\documentclass{article} 
\usepackage{iclr2021_conference,times}

\usepackage{amsmath,amsfonts,bm}

\def\eqref#1{equation~\ref{#1}}

\def\1{\bm{1}}

\DeclareMathAlphabet{\mathsfit}{\encodingdefault}{\sfdefault}{m}{sl}
\SetMathAlphabet{\mathsfit}{bold}{\encodingdefault}{\sfdefault}{bx}{n}

\usepackage{hyperref}
\usepackage{url}

\title{Inequality, Crime and Public Health: A Survey of Emerging Trends in Urban Data Science}

\iclrfinalcopy

\author{Massimiliano Luca$^{1,2}$, Gian Maria Campedelli$^{3}$, Simone Centellegher$^{1}$, \\ \textbf{Michele Tizzoni$^{3}$, Bruno Lepri$^{1}$} \\ \\
 $^{1}$ Mobile and Social Computing Lab, Bruno Kessler Foundation \\
 $^{2}$ Faculty of Computer Science, Free University of Bolzano \\ 
 $^{3}$ Department of Sociology and Social Research, University of Trento \\ \\ 
\texttt{\{mluca,centellegher,lepri\}@fbk.eu} \\ \texttt{\{gianmaria.campedelli, michele.tizzoni\}@unitn.it} 
}

\begin{document}

\maketitle
\begin{abstract}
Urban agglomerations are constantly and rapidly evolving ecosystems, with globalization and increasing urbanization posing new challenges in sustainable urban development well summarized in the United Nations' Sustainable Development Goals (SDGs). The advent of the digital age generated by modern alternative data sources provides new tools to tackle these challenges with spatio-temporal scales that were previously unavailable with census statistics. In this review, we present how new digital data sources are employed to provide data-driven insights to study and track (i) urban crime and public safety; (ii) socioeconomic inequalities and segregation; and (iii) public health, with a particular focus on the city scale. 

\end{abstract}

\section{Introduction}
Cities occupy only 3\% of the global surface but are inhabited by more than 50\% of the world's population\footnote{https://www.un.org/sustainabledevelopment/cities/}. Timely and accurate data are thus becoming fundamental for policymakers and municipalities to control cities' dynamics and respond to multiple societal challenges. In 2015, the United Nations set out 17 Sustainable Development Goals (SDGs)\footnote{https://sdgs.un.org/goals}  that summarize the new challenges we have to face to guarantee everyone a better and more sustainable future. Examples of such goals are about guaranteeing good quality of (accessible) health and well-being, reduction of inequalities, and design of sustainable and safe cities and communities. 

It is clear that big urban agglomerations have a pivotal role in the accomplishment of such goals as many of them are fundamentally related to human movements, displacement, and interactions \citep{glaeser2012triumph, sassen2019}. More in general, it is known that human dynamics are related to the diffusion of viral diseases \citep{eubank2004modelling, colizza2007modeling, perkins2014theory}, to the behavioral responses in case of natural disasters \citep{bohorquez2009, bagrow2011}, to the optimization of traffic volumes \citep{batty2013, mazzoli2019}, to the economic growth, innovation and social integration \citep{bettencourt2007growth, pan2013, schlapfer2014scaling}, to the severity of air pollution and the consumption of energy, water and other resources \citep{bettencourt2007growth, bettencourt2010unified}. 

To monitor the progress towards the aforementioned societal challenges, it is fundamental to have an always up-to-date picture of cities. In the past, institutions had to rely almost exclusively on census data and official statistics. However, both these data sources have some intrinsic limitations including (i) the time gap between the data collection and the actual availability of the data, and (ii) the frequencies and costs of the data collection campaigns \citep{lazer2009}. Luckily, we are in the middle of a digital sensing revolution with billions of data that are generated every second and that can be employed to have an almost real-time picture of cities' dynamics at low costs. Examples of such data include tracks from GPS devices embedded in smartphones, vehicles, or boats, records produced by the communication between phones and the cellular network, and geotagged posts from social media platforms \citep{gonzalez2008understanding, zheng2010geolife, moreira2013predicting, spinsanti2013mobility, blondel2015survey, cui2018social}. Finally, other data sources like satellite images, street networks and points of interest can provide precious information to integrate with human dynamics data in order to capture socio-economic aspects \citep{ deville2014dynamic, jean2016combining, tatem2017worldpop, lepri2022understanding, weber2018census, yeh2020using}

In this review paper, we showcase and discuss how alternative data sources have been employed by researchers to study the relationship between human dynamics and three SDGs: (i) crime diffusion and public safety, (ii) socio-economic inequalities and segregation, and (iii) public health and disease diffusion. Also, we have decided to focus on the studies that investigate such dynamics in urban agglomerations. Thus, excluding studies that, for example, investigate the relationship between international mobility and the global diffusion of diseases, or map the socio-economic inequalities across countries.

The paper is structured as follows. We start in Section \ref{sec:data_sources} by first describing the new sources of data used in these lines of research. Then, in Section \ref{sec:crime}, we discuss the implication of using computational methods for studying urban crime and public safety. After showing how data sources like mobile phone data, social media, and others have been used by researchers for a variety of aspects related to crime and security, we conclude the Section with a critical reflection. In Section \ref{sec:inequalities} and Section \ref{sec:public_health}, we follow a similar structure covering works related to socio-economic inequalities, segregation, and public health. Finally, in Section \ref{sec:conclusions} we conclude the paper with a brief discussion.

\section{Data sources}\label{sec:data_sources}
The digital age has revolutionized the world we live in, and simple actions such as clicking on a website, sending an email, paying with a credit card or making a phone call generate so-called digital traces. Digital traces track information about our daily behaviors and in the last decades this rich and vast amount of information created new research opportunities to better understand and study human behavior \citep{lazer2009,lazer2020}.

In this section we are going to describe the most commonly type of data used in this line of research.

\subsection{Call Detail Records}
Telecommunication companies collect information regarding people's exchanges by means of Call Detail Records (CDRs), which contain real-world observations on how, when and with whom a person communicates \citep{blondel2015survey, luca2021leveraging}.
A Call Detail Record is a tuple $(\mbox{u}_o, \mbox{u}_i, t, d, A_o, A_i)$ which contains privacy-enhanced metadata about the caller $\mbox{u}_o$, the callee $\mbox{u}_i$, the timestamp $t$ of when the call took place and its duration $d$. $A_o$ and $A_i$ represent respectively the outgoing and incoming Radio Base Stations (RBSs), namely the antennas that delivered the communication through the network. 
Mobile phone data may cover a large sample size on a national scale and aggregated mobility flows have been inferred by counting the number of users that move between RBSs or administrative units such as neighborhoods and municipalities \citep{calabrese2011estimating}. 
Since the position and the coverage area of each RBS are known, a user's telecommunication event represents a proxy of the user's geographic location. The precision of this location can vary from 100 meters in urban areas to kilometers in rural areas. 
This approximation implies that the user's spatial and temporal resolution is given by where and when a user makes a call or sends an SMS leading to sparse and incomplete mobility trajectories.
Nevertheless, CDRs have been proven valuable in studying and understanding human mobility \citep{gonzalez2008understanding, simini2012universal,csaji2013exploring,blondel2015survey,pappalardo2015returners}.

A less common type of data used in the literature are the eXtended Detail Records (XDRs), which are generated by telecommunication companies when a user uploads or downloads data from the Internet using their phone's connection. A single event is a privacy-enhanced record $(\mbox{u}, t, A, k)$ where $t$ is the timestamp of the event, $A$ is the RBS that managed the connection, and $k$ the amount of uploaded/downloaded information. Given the higher frequency of mobile internet connections, XDRs reduce the problem of sparsity characterizing CDRs \citep{chen2019complete, luca2022modeling}.

\subsection{GPS location data}
Location intelligence companies collect GPS location data of opt-in individuals from third-party mobile apps through a Software Development Kit (SDK) that captures user locations through GPS signals in Android and iOS devices.
In general, a data point contains privacy-enhanced information like the \textit{user identifier}, the \textit{timestamp}, and geographic information such as \textit{longitude} and \textit{latitude}. 
In the last years, to help a prompt response to the COVID-19 pandemic, location intelligence companies such as Cuebiq\footnote{https://www.cuebiq.com/}, Unacast\footnote{https://www.unacast.com/}, and Safegraph\footnote{https://www.safegraph.com/} made available several datasets for research purposes \citep{chang2021mobility,hunter2021effect,lucchini2021living,aleta2022quantifying}. The collected GPS location data generally provide more precise location information than CDRs. Unfortunately, location intelligence companies do not share details either on how the data is collected or from which mobile apps, potentially compromising their population representativeness.

On the same line, Big Tech companies such as Facebook (Data4Good\footnote{https://dataforgood.facebook.com/dfg/covid-19}) and Google (Community Mobility Reports\footnote{https://www.google.com/covid19/mobility/}) have provided GPS location data collected directly from their platforms. However, they have shared these data in an aggregated fashion both in time and space. Thus, the shared data have the advantage of covering all the countries the Big Tech companies operate in but are less precise than the GPS location data provided by location intelligence companies.

\subsection{Social Media data}
Social media platforms such as Facebook, Instagram and Twitter facilitate the creation and sharing of content with posts that can contain text, videos, photos, etc. together with a timestamp and an (optional) geographic location. 
For example on Twitter, people can share geo-located tweets with either their precise geographical position (latitude and longitude) or the location suggested by the platform (e.g., restaurant, landmarks, etc.) where the user was located when the tweet was published.
Platforms like Foursquare are location-based social networking websites where users share their locations by checking in at points of interest (POIs), such as restaurants, pubs, shops, museums, etc. The users' location is thus available given that such venues have a geographical location (latitude and longitude).
For most social media platforms, geotagged posts are downloadable through their Application Programming Interfaces (APIs). 
From spatial and temporal information present in social media posts is thus possible to infer users' mobility trajectories from their posts' history. 
Such data suffer from data sparsity problems with respect to datasets collected through mobile phones. 
Nevertheless, social media data have been proven valuable in modeling and dealing with different societal challenges such as urban crime, public health, unemployment, etc. \citep{WangAutomaticCrimePrediction2012a, broniatowski2013national,llorente2015social}.

\subsection{Other data sources}

\subsubsection*{Credit card transactions}
Credit cards are universal across the world, but they have received relatively little attention to date. Since people’s spending has become increasingly digitized, it is possible to capture consumer behavior at an unprecedented scale. 
Each credit card transaction generally consists of privacy-enhanced information such as a \textit{user identifier}, the \textit{timestamp} of the transaction, and the \textit{transaction type} represented by the Merchant Category Code (MCC).
Recent research has begun to use transaction records to provide insights on financial well-being \citep{singh2015money}, individual traits \citep{gladstone2019can,tovanich2021inferring}, purchase behavior in urban populations \citep{di2018sequences,dong2017social} and segregation \citep{dong2020}.

\subsubsection*{Satellite imagery}
There exist several types of satellite imagery collected by governments and private companies and they can be mainly divided by their spatial, spectral, temporal, radiometric and geometric resolutions \citep{campbell2011introduction}. As an example, the Landsat  Program represents the longest-running project for the acquisition of satellite imagery of Earth: they provide freely downloadable repeated (average return period of 16 days) imagery with a geometric resolution of 30 meters for the entire planet. Satellite data has been proven useful for different tasks such as tracking urbanization \citep{tatem2017worldpop,strano2021agglomeration} and forecasting diseases \citep{dister1997landscape,rogers2002satellite,ford2009using}.

\subsubsection*{Wearables}
In the last few years, wearable sensors such as smartwatches have steadily grown in availability. These devices collect physiological and activity data such as heart rate, sleep, step count and calories burnt. This information can be exploited to track in almost real-time a person’s health. As an example, recently smartwatches were used to investigate changes in physiological parameters in response to a COVID-19 infection or COVID-19 vaccines \citep{wiedermann2022evidence,guan2022higher}.

\subsubsection*{Census}
Census data is collected by governments to monitor and gather information about the population of a country. The data is then used to have a reliable picture of the current population, including important information such as demographics and socio-economic conditions.
Despite the vast amount of data produced in the digital age, census data remains widely used since it can be jointly used (at an aggregate level) with newer sources of data such as CDRs and provide valuable insights on the general population.

\section{Urban Crime and Security}
\label{sec:crime}
\subsection{The Computational Contamination in Research on Crime}
Over the last two decades, scholars from various fields and disciplines have focused on crime and public safety by leveraging the potential of computational methods and novel big data sources. This wave of methodological innovation transcended the traditional boundaries of criminology as a discipline, fostering the interest of social scientists as well as computer scientists, statisticians, applied mathematicians and physicists.  
As a result, the study of crime has been invested by contamination of approaches, techniques, and viewpoints \citep{BrantinghamComputerSimulationTool2004,GroffSimulatedexperimentstheir2008,BogomolovMovesStreetClassifying2015,DOrsognaStatisticalphysicscrime2015b, BouchardSocialNetworkAnalysis2016,FaustSocialNetworksCrime2019,DeNadaiSocioeconomicbuiltenvironment2020,HaywardArtificialintelligencecrime2021,CampedelliMachinelearningcriminology2022a}. 

Interestingly, the link between computational methods and the study of crime is not as recent as many scholars portray. For instance, \citet{CampedelliMachinelearningcriminology2022a} noted how, despite attempts to rebrand such a relationship in terms of novelty, the dialogue between Artificial Intelligence (AI) and research on crime has roots that date back to the 1980s. The relationship in fact emerged decades ago as the result of two processes: the use of AI-based approaches for predictive purposes \citep{IcoveAutomatedCrimeProfiling1986} and the exploration of AI as a tool for aiding sociological theorizing \citep{BrentThereRoleArtificial1988, Andersonartificialintelligencetheory1989a, WoolgarWhynotsociology1989}.

Hence, while it is limiting to describe the link between computational methods and the study of crime only by focusing on the recent past, it is nevertheless true that recent years have led to an acceleration in this dialogue, at least in terms of scientific productivity. The reasons behind this fact are four-fold. First, administrative data in digital format have become more and more ubiquitous and easy to access. Second, the democratization of programming languages made it easier for criminologists and crime researchers without a computer science background to explore the potential of algorithmic methods. Third, the availability of other digital sources such as social media data, GPS data, and mobile phone data enriched the information horizon available to study crime. Fourth, following a trend that was already in place, governments and institutions in many Western countries pushed for data-driven solutions to reduce crime, thus increasing funding opportunities in academia as well as business opportunities in the digital and technological sectors.
All these factors together made it easier for scholars to gather, process, and analyze data related to crime and security issues, substantially increasing the number of publications and projects over the years \citep{CampedelliWherearewe2020a}. 

Methodology-wise, crime has been investigated through a plethora of different techniques and frameworks. Besides traditional statistical approaches that target either correlational or causal outcomes, geospatial modeling, network science, agent-based modeling, and machine learning have been the four main areas on which scholars have focused their attention. Virtually every area of criminology and crime research has been --- to some extent --- explored by computational approaches: from white collar crime \citep{Ribeirodynamicalstructurepolitical2018b, Luna-PlaCorruptioncomplexityscientific2020a, KerteszComplexityscienceapproach2021} to terrorism \citep{MoonModelingSimulatingTerrorist2007d, ChuangLocalalliancesrivalries2019b,CampedelliLearningfutureterrorist2021b}, from illicit drugs \citep{MackeySolutionDetectClassify2018,SarkerMachineLearningNatural2019,MaglioccaModelingcocainetraffickers2019} to organized crime \citep{NardinSimulatingprotectionrackets2016,TroitzschCanagentbasedsimulation2017a, CalderoniRecruitmentOrganizedCrime2021}, from gun violence \citep{MohlerMarkedpointprocess2014a, GreenModelingContagionSocial2017a, LoefflerGunViolenceContagious2018a} to cyber-crime \citep{ShalaginovCybercrimeinvestigations2017,DuxburyNetworkStructureOpioid2018, Duxburyresponsivenesscriminalnetworks2020}, from recidivism \citep{TollenaarWhichmethodpredicts2013,DuweOutOldNew2017, BerkAlmostpoliticallyacceptable2020} to predictive policing \citep{MohlerSelfExcitingPointProcess2011a,CaplanRiskTerrainModeling2011a,PerryPredictivePolicingRole2013}. Particularly, the dialogue between computational methods and the study of recidivism and predictive policing not only focused on technical innovations to optimize forecasting and predictive models, but also provoked vivid debates regarding critical issues of algorithmic accountability, fairness, and transparency \citep{Lumpredictserve2016, Dresselaccuracyfairnesslimits2018c,RichardsonDirtyDataBad2019,Akpinareffectdifferentialvictim2021, PurvesFairnessAlgorithmicPolicing2022}. In fact, although the computational analysis of crime has remained chiefly confined to the academic sphere, in some cases --- such as predictive policing and criminal justice risk assessment tools --- algorithmic solutions have been deployed by courts and law enforcement agencies. In the United States, where this transition from academia to the public and private sectors has been faster, data-driven tools to aid police agencies and courts have a long history \citep{BerkMachineLearningRisk2019b}. Yet, the rapid diffusion of novel tools, coupled with their secrecy, pushed scholars, activists and journalists to scrutinize the effects that these software have on high-stake settings, showing that these instruments often lead to disparate and unfair treatment against minorities, reinforcing discrimination and over-policing in policing and criminal justice. Two sides hence emerged: one populated by those defending the benefits and potential of computational approaches for predicting crime and recidivism (among other things), and those calling for either the elimination of such tools or their heavy regulation. 

Within the kaleidoscope of areas in which the computational wave has spread, the study of urban crime has certainly fostered significant scholarly interest. Urban crime trivially embraces all those deviant and criminal behaviors occurring in urban settings and, therefore, can be seen as a higher-level category containing some of those previously mentioned, as the study of illicit drugs (when distributed or consumed in urban settings), the study of violent crime (when perpetrated in urban settings) or predictive policing itself, which by definition targets a certain urban area.

\subsection{Computational Methods, Big Data and Urban Crime}

\subsubsection{The advantages in studying urban crime today}
There are some specific reasons behind the fact that urban crime has attracted so much scholarly attention. First and foremost, one of the most popular regularities in the empirical study of crime is the so-called ``law of crime concentration" \citep{WeisburdLawCrimeConcentration2015}. Inspired by the theoretical tradition on crime and place \citep{ShawJuveniledelinquencyurban1942, CohenSocialChangeCrime1979, BrantinghamComputerSimulationTool2004, EckCrimePlaceCrime1995}, the ``law of crime concentration" states that most crimes in a city are concentrated in specific small areas, such as blocks, streets or neighborhoods. In other words, crime clusters spatially \citep{Johnsonbriefhistoryanalysis2010b}. The dawn of this empirical finding dates back to the early seminal cartographic works of \citet{QueteletResearchPropensityCrime1831} in the Nineteenth century. Over the decades, scores of studies emerged in the context of routine activity \citep{CohenSocialChangeCrime1979} and crime pattern theories \citep{BrantinghamPatternsCrime1984} have verified this finding not only in the United States but in many other countries all around the world \citep{YeSpacetimeinteraction2015, deMeloCrimeconcentrationssimilarities2015a, MazeikaCrimeHotSpots2017, Breetzkeconcentrationurbancrime2018, FavarinThismustbe2018a, UmarAssessingSpatialConcentration2020}. Second, not only does crime cluster spatially, it also clusters temporally. It is in fact well known that the probability that a crime occurs is not homogeneous across time windows \citep{AaltonenShortTermTemporalClustering2018,PiatkowskaTemporalClusteringHate2021, HolbrookScalableBayesianinference2021}. In general, crime has its own higher-level seasonalities and these temporal dynamics also vary across crime types \citep{YanSeasonalityPropertyCrime2004, LinningCrimeSeasonalityExamining2017, AaltonenShortTermTemporalClustering2018}. 
In many cases, these two layers intersect --- especially in the case of urban crime --- creating spatio-temporal regularities that allow for deeper analytical scrutiny. In general, spatial and temporal patterns create the conditions for deploying statistical and mathematical models taking advantage of the non-random data of criminal phenomena for forecasting and predictive purposes. A third reason behind the strong relationship between computational methods and urban crime is that both traditional and more novel data sources for studying crime both make spatial and temporal information available. In the United States and other Western countries, for instance, data on reported crimes or calls for service are digitally recorded and easily accessible at the city level, combining information on the type of crime with information on the time and location of the offense. At the same time, data providers and tech companies sell or offer data on social media activity, mobile usage, public transportation, point of interest attendance, and GPS tracking. Overall, the availability of digital data beyond traditional administrative records has allowed scholars to expand the typical analytical frame in which crime --- patterned but highly dynamic --- is studied considering only fixed or static factors, variables, and conditions, such as the built environment or socio-economic characteristics. In light of this, the study of urban crime is aided by an arsenal of information that is rich --- often way richer than the one available to the study of other crime contexts --- and can be connected to other human phenomena that are known to be patterned, such as mobility flows. Fourth, the computational analysis of urban crime has straightforward practical consequences that transcend the pure research dimension. While the gap between empirical evidence and policy solutions may be wide for other areas of inquiry, the translation of empirical evidence to crime reduction strategies has always been much faster in the study of urban crime.  

Scholars have taken advantage of these conditions and amassed a relevant number of studies with mainly two goals: disentangling crime correlates and forecasting or predicting crime trends and locations. The two goals are interrelated, as optimal forecasting can be achieved only through the selection of relevant correlates and, in turn, the study of correlates cannot be deemed independent from the need to optimize predictive performance.

\subsubsection{Mobility, urban crime and ecological networks}
Taxi flows and mobility patterns, as proxied for instance by the analysis of activity at point of interest (POI) locations, have been critical components of recent studies targeting urban crime. 
Some of the works emerging in this area have been framed using agent-based models (ABMs). ABM refers to generative models that allow research to simulate social and criminal phenomena by incorporating empirical or artificial data to investigate research questions that are impractical to be investigated in the real world (e.g., for ethical or monetary reasons). Although ABMs pose several major limitations to the reliability of findings when simulations are not appropriately designed and cannot be validated \citep{GroffStateArtAgentBased2019c,CampedelliCriminologyCrossroadsComputational2022}, when models are carried out properly they offer a compelling set of benefits for criminologists and crime researchers, including theory testing, scenario exploration, and long-term forecasting. 

Within this line of research, \citet{RosesSimulatingOffenderMobility2020} propose a simulation model for offender mobility in New York City (NYC) using open data to simulate urban structure, location-based information to proxy human activity and taxi flow data to proxy human mobility between different areas of the city. By comparing 35 different mobility patterns, the authors highlight the benefits of integrating taxi flow data with previous crime data and popular activity nodes to simulate offenders' mobility meaningfully. In another example, \citet{Rosesdatadrivenagentbasedsimulation2021} designed a model aimed at identifying drivers of relevant crime patterns through openly available static and dynamic geographical and temporal features, and proposed a data-driven decision-making process based on machine learning to allow artificial agents to decide whether to engage in crime based on their perception of the surrounding environment. Focusing again on NYC and targeting crime counts at the street level, the authors indicate the stability of high crime areas, in line with the criminological literature, and highlight the importance of the spatial environment in predicting crime hotspots. Agent-based modeling, however, is far from being the only methodological framework utilized in the study of urban crime through mobility data.

\citet{WangCrimeRateInference2016}, for instance, focused on Chicago to infer crime rates at the neighborhood level using POI and taxi flow data through more traditional statistical approaches like linear and negative binomial regression, indicating that including these information sources reduces prediction error by 17.6 percentage points. In a subsequent extension of the work, \citet{WangNonStationaryModelCrime2019} propose a graphically weighted regression approach for crime rate inference that aims at capturing the non-stationarity nature of crime across neighborhoods in the same urban context, i.e., Chicago, from 2010 to 2014. The assumption behind the analysis is that the same features may have different relationships to crime across different spatial contexts, thus involving a further layer of complexity in the dynamic nature of criminal phenomena. 
Chicago has been historically one of the US cities that attracted the highest scholarly attention in the study of deviance and crime \citep{SampsonGreatAmericancity2012}. In recent years, Chicago has also been the focus of several papers investigating crime from a computational perspective. Besides the articles mentioned above, others have explored the promises of sophisticated statistical techniques to shed light on the city's crime dynamics. \citet{PapachristosConnectedCrimeEnduring2018a}, for instance, studied how criminal co-offending (measured via co-arrests data) generates pathways between neighborhoods in Chicago, creating a spatial network that facilitates the diffusion of crime in time and space. Their statistical analyses demonstrate that these ``neighborhood networks" are stable over time, generated by various processes, including structural characteristics and social dynamics. Their work fits into a growing body of literature that assess the interdependency of neighborhoods within an urban context \citep{TitaCrimeNeighborhoodsUnits2009, PetersonSegregatedSpatialLocations2009, GraifUrbanPovertyNeighborhood2014}, unfolding the connectivity of communities within cities, despite the belief that, given urban segregation and crime clustering, co-offending patterns should also be clustered. Relying on the interdependence of communities via spatial network representations, \citet{GraifNetworkSpilloversNeighborhood2021} study the relationship between crime and commuting patterns across Chicago communities by concentrating on mobility patterns between job and home locations of the city residents. They mainly investigate whether exposure to workplaces characterized by higher disadvantage leads to an increase in local crime, suggesting that this relationship exists. In other words, \citet{GraifNetworkSpilloversNeighborhood2021} shows that disadvantage in the extra-local network of communities where citizens work is associated with higher crime levels in the communities where the same citizens live. \citet{SampsonEnduringNeighborhoodEffect2022} depart from a similar theoretical perspective to show that a neighborhood's well-being is statistically dependent upon the well-being of the communities that their residents visit or the communities from which visitors come. Remarkably, their analysis outlines that mobility-based socioeconomic disadvantage explains rates of violence and homicide in Chicago neighborhoods. The combination of low scores of residential socioeconomic conditions of residents, visited communities, and visitors constitutes what the authors label  ``triple disadvantage", elaborating on how this concept is theoretically and technically valuable for explaining crime dynamics in Chicago \citep{LevyTripleDisadvantageNeighborhood2020}.

\subsubsection{Urban crime and mobile phone data}
Mobility and people dynamics within urban contexts have also been investigated by means of mobile phone data. One of the first notable examples is the work by \citet{BogomolovOnceCrimeCrime2014a} in which mobile network infrastructure data on London, UK, is combined with traditional demographic information and geo-localized open data to show that human behavioral data significantly improve the prediction of crime hotspots. London has been the focus of another early work by \citet{TraunmuellerMiningMobilePhone2014} in which anonymized mobile telecommunication data are used to investigate urban crime theories. From such telecommunication data, authors extract quantitative proxies for mapping the presence of people in a given area and find that the age diversity and the ratio of visitors in a given area are negatively related to crime, in line with theoretical concepts proposed by Jane Jacobs \citep{jacobs1961} such as the one of ``natural surveillance" and by Felson and Clarke \citep{felson1998}, i.e., ratio of young people.
\citet{SongCrimeFeedsLegal2019} utilized geocoded tracks of mobile phones to analyze if the intensity of population mobility among pairs of communities in a large Chinese city can help shed light on offenders' decision-making processes. The study explicitly considers thefts, and its outcomes suggest that such a measure of mobility leads to a higher predictive performance of theft locations compared to the traditional analysis of crime generators. By leveraging mobile phone data and fine-grained spatio-temporal data on violent crime in Manchester, UK, \citet{HaleemExposedPopulationViolent2021} proposed the use of the ``exposed population-at-risk" concept to shed light on public crime hotspots on Saturday nights.  
\citet{DeNadaiSocioeconomicbuiltenvironment2020} sought instead to examine the link of socioeconomic conditions, built environment and mobility patterns with violent and property crime across multiple cities. The authors identify the focus on single urban contexts as one of the main shortcomings of the existing literature. They hence focus on four contexts with very different social, cultural, and urban characteristics --- i.e., Bogotá, Boston, Chicago, and Los Angeles --- to provide higher external generalizability of their findings. Mobility flows are proxied through the use of mobile phone data in the form of CDRs. The work shows that combining information on people, crime, places, and human mobility produces better-performing models in terms of descriptive and predictive accuracy. 

\subsubsection{The role of social media}
In the thriving literature focusing on mobility and crime, recent studies have also sought to unfold the potential of social media to capture the dynamic dimension of human behavior in urban contexts, in line with the hypothesis emerging from other studies in the same area of research, namely that resident population does not explain the complexity of the ecology of crime. Among social media platforms, Twitter has certainly received greater attention from scholars interested in urban crime. \citet{WangUsingTwitterNextPlace2015} proposed the use of Twitter data for solving the problem of ``next-place prediction", thus seeking to estimate people's individual trajectories. According to the authors, Twitter posts provide rich contextual information in the form of text that can be used to construct such individual trajectories even in cases when no direct reference to geospatial information is available. They hence present two models designed to extract geographic information from general texts allowing them to predict the type of venue a user will visit and the distance between the user and a given type of venue in the future. By leveraging this computational approach, they apply their methodology to test the correlation between next-place prediction and crime. \citet{YangCrimeTelescopecrimehotspot2018}  included data from Twitter (and particularly data on the sentiment and topic of tweets) in their CrimeTelescope platform, a software intended to provide optimized crime hotspots prediction in New York. Besides Twitter data, CrimeTelescope also included information on urban infrastructure via POIs from Foursquare and historical crime data. The statistical outcomes of the study suggested that this multi-modal combination of data leads to better predictive performance (up to 5.2\%) compared to traditional approaches only using data on past crimes.   \citet{Mallesonimpactusingsocial2015} highlighted that there is a relationship between the density of tweets in a given area and shifts in crime concentration. Similarly, \citet{HippUsingSocialMedia2019} integrated geocoded Twitter data into models to capture the temporal ambient population in Southern California, arguing that social media data can be promising to test routine activity and crime pattern theories. \citet{WoRecreatingHumanMobility2022} examines the potential of four Twitter-derived measures to predict crime counts across more than 2,300 block groups in the city of Los Angeles. The aims of the study are specifically two. First, the authors seek to represent local human activity distinguishing between insider and outsider occupants of a neighborhood. Second, they analyze whether statistical relationships exist between property and violent crime and Twitter-derived measures of the ambient population in Los Angeles. Wo and co-authors conclude that Twitter is powerful in aiding research on ambient population and crime distributions at the spatial level. 
However, not all studies using Twitter data reached the same positive and promising conclusions. \citet{TuckerWhoTweetsWhere2021}, in fact, critically tested whether geotagged Twitter data correlate with events of public violence and private conflict during weekday days, weekday nights, weekend days, and weekend nights in the city of Boston. The authors indicate that Twitter works as a proxy of human dynamics only for particular types of locations and activities, thus recommending caution in the use of tweets as comprehensive sources for mapping the ambient population within a city.

\subsubsection{Critical Reflections}

The study of crime has been impacted by the vast array of computational methodologies that have spread across social sciences in the last two decades, and urban crime in particular has benefited from this methodological contamination and the increasing availability of digital data sources coming from mobile phones, social media, transportation information, and other geolocalized trace data. This availability has opened many new possibilities to test criminological theories and improve predictive accuracy in terms of crime hotspots and crime patterns in space and time. Nonetheless, it should be noted that important caveats should be considered in critically evaluating the potential and relevance of novel data sources for the study of urban crime. Particularly, as noted by \citet{BrowningHumanMobilityCrime2021}, representativeness and generalizability of both mobility at the place- and person-level is a problem. Browning and colleagues, for instance, argue that representativeness can be an issue when considering data that are collected based on voluntary choices of users and that this representativeness trivially poses a risk to the generalizability of findings across urban contexts (or even across areas within the same urban context). Hence, scholars must recognize this limitation and adopt strategies to mitigate it. Strategies may include methodological innovations in terms of weighting and result validation. Furthermore, the reliance on novel digital data may lead to an increase in the scholarly unbalance towards Western urban contexts. In fact, while accessibility to digital communication technologies is very high in the Western world, the scenario is very different for countries in other regions of the planet, reinforcing the abovementioned issue of representativity and generalizability when deploying these data sources outside the Western context. Finally, it is worth considering the societal implications of governmental decisions to incorporate mobile phones, GPS, POIs, and social media data into software designed for crime prediction, especially in non-democratic countries. In political contexts in which civil, political, and human rights are not sufficiently protected and guaranteed, the exploitation of multi-modal data sources may significantly increase the state of surveillance over citizens, causing detrimental effects on their liberties and well-being. Scholars should thus engage more in the ethical consequences of information systems engineered to collect as much data as possible to protect public safety and crime control allegedly.

\section{Socioeconomic Inequalities and Segregation}
\label{sec:inequalities}

\subsection{The Computational Contamination in Research on Inequalities and Segregation}
Reducing inequalities is of crucial importance to guarantee a more sustainable and just future for our cities and societies. Indeed, socioeconomic inequalities and income segregation threaten access to health and negatively impact health population levels \citep{wilkinson2006,pickett2015}, prevent equal access to educational opportunities \citep{quillian2014,logan2016}, and hinder social and economic development \citep{neves2016}. Moreover, they are are intimately related to opportunities offered by neighborhoods \citep{SampsonGreatAmericancity2012,chetty2016effects,manduca2019,sampson2019,hedefak2020} and human movements and interactions \citep{eagle2010,chetty2016effects,wang2018,dong2020,chetty2022a,chetty2022b}. 
Before the advent of the digital era, socioeconomic inequalities were studied through surveys and census data. However, while census data provide a large-scale representativeness of the population, it lacks the ability to capture and provide an up-to-date picture of cities' dynamics and citizens' behaviors, routines, and habits \citep{lazer2009}. As previously discussed, census data are collected every few years and are made publicly available several months after they were collected \citep{lazer2009}. Instead, digital data provide alternative data sources that allow capturing different facets of human behavior: human interactions, human movements, and human encounters are just a few examples of human behavior which may play an essential role in investigating inequalities and segregation and that nowadays can be studied by means of mobile phone data (i.e., CDRs, GPS traces, etc.) and other digital traces (e.g., credit card transactions) \citep{lazer2009,lazer2020}.

In what follows, we discuss how several researchers, from economics and computational social science, have started using alternative data sources to study the daily behaviors and routines associated with socioeconomic inequalities and segregation in cities.

\subsection{Computational Methods, Big Data and Inequalities}
As more people are moving to cities, governments have to deal with novel challenges like gentrification, unaffordability, segregation, and inequality \citep{glaeser2001,florida2017new}. The place where a person lives can have substantial impacts on health \citep{wilkinson2006}, economic opportunities \citep{chetty2014land,chetty2016effects}, infrastructure and services accessibility \citep{glaeser2001,ewing2016,florida2017new}, and many other aspects, both at a city and national scale \citep{chetty2014land,shelton2015social}. Thus, measuring inequalities and segregation with timely and accurate data is of paramount importance, and alternative data sources and ubiquitous technologies are starting to play a central role in deeply analyzing factors and behaviors associated to inequalities such as environmental inequalities \citep{dass2022strategies,brazil2022environmental}, social mixing and income segregation \citep{fan2022diversity,shelton2015social, moro2021mobility}, and community resilience \citep{hong2021measuring}.

\subsubsection{GPS and mobile phone data}
\citet{athey2020experienced} developed a measure of experienced isolation (by race) to capture individuals’ exposure to other (diverse) individuals using GPS data in the US. They found that the isolation individuals experience in their daily life is lower than the one measured by standard residential isolation metrics, especially in cities with higher levels of public transportation, density, education and income.
\citet{jarv2015ethnic} moved beyond residential segregation to explore individuals' activity spaces, namely the locations visited by an individual because of their regular activities and routines. They exploited CDRs in the city of Tallin in Estonia to measure ethnic segregation (Estonian versus Russian). They found that, for example, activity locations of Russian speakers tend to be more concentrated in regions with a prevalence of Russian-speaking communities.
\citet{xu2019quantifying} leveraged multiple urban datasets (e.g., CDRs records, housing prices and income data) to study citizens' segregation by their socio-economic status and its evolution in both physical and social (communication) spaces in Singapore. They found relatively higher levels of segregation in wealthier classes for both social and physical space.
\citet{hong2021measuring} leveraged mobile phone data to measure the inequalities in community resilience to the Harvey hurricane in Texas. By measuring the mobility behavior of the individuals, the authors highlighted socio-economical and racial disparities in resilience capacity and evacuation patterns, suggesting the adoption of novel data-driven policies to prioritize equal allocations of resources to vulnerable neighborhoods.
Another study \citep{dass2022strategies} used mobile phone data, socio-demographic data, and infection rates information to measure accessibility to green spaces in Boston, at the beginning of the COVID-19 pandemic. The authors discovered inequalities, where communities with higher infections and higher prevalence of black residents experienced greater infection exposure per visit.
\citet{fan2022diversity} employed mobile phone data of half a million people located in three different metropolitan areas in the US to study how people experienced social mixing in urban streets. The authors found that the density of people's street visits only explains the 26\% of street-level diversity (e.g., social mixing), while the adjacent amenities, residential diversity, and income level explain the 44\% of the designed diversity score. Also, \citet{fan2022diversity} shows that while streets densely visited tend to have more crime, diverse streets have fewer crimes.
\citet{moro2021mobility} leveraged high-resolution mobility data of more than 4.5 million users in eleven big US cities to study income segregation. Previously, income segregation was studied using static residential patterns with high spatial resolutions. Thanks to the fine-grained mobility data, the authors found that the income segregation associated with places and individuals may significantly vary even for places that are close to each other. The authors proposed a model and showed that the experienced income segregation of individuals is associated with the exploration of new locations and places visited by visitors from different income groups. In general, \citet{moro2021mobility} highlights the importance of considering mobility patterns when we aim at measuring income segregation.
\citet{yabe2022behavioral} investigated how social interactions (e.g., encounters) changed during the COVID-19 pandemic with respect to income diversity. The authors relied on a dataset of millions of mobile phone users in multiple US cities for a period of three years before and during the pandemic. Overall, \citet{yabe2022behavioral} found that the diversity of individual-level urban encounters decreased significantly despite in 2021, indices related to aggregated mobility recovered to pre-pandemic levels. The authors argued that the pandemic could have long-lasting implications on urban income diversity.
\citet{brazil2022environmental} utilize mobile phone data to study the mobility behavior of individuals to uncover environmental inequalities. The author found that people from minority groups and poorer neighborhoods tend to travel to areas with greater levels of air pollution with respect to white and richer neighborhoods. 

\subsubsection{The role of other data sources}
Using hundreds of millions of geotagged tweets, \citet{wang2018} have measured neighborhood isolation for 50 American cities finding that residents of black and Hispanic neighborhoods, despite their socio-economic status, are less exposed to either non-poor or white neighborhoods than residents of white neighborhoods.
Using the same dataset, \cite{phillips2021social} have computed the social integration of 50 American cities with indices that measure the extent to which residents in each neighborhood travel to other neighborhoods in a city. They have shown that cities with greater population densities and less racial segregation have higher levels of structural connectedness. 
Using Twitter data and geo-located credit card transactions \citet{morales2019segregation} investigated segregation in the physical and online space together with their relationship. They show that physical and online interactions in urban areas are segregated by income and that information does not flow homogeneously across social classes in either the physical or the virtual space. In a follow-up study, \citet{dong2020} found that segregation in urban and online interactions seems stronger than the residential ones. Indeed, while residential neighborhoods sometimes might consist of a mix of different socioeconomic groups, purchase activities and online interactions seem to take place more often between neighborhoods whose economic conditions are similar. Additionally, \citet{dong2020} have shown that segregation increases with differences in socioeconomic status but this effect is asymmetric for shopping behaviors. In fact, the number of movements from poorer to wealthier neighborhoods is larger than vice versa.
\cite{hilman2022socioeconomic} instead leveraged check-in datasets collected from location-based social media and census information to study segregation levels in 20 cities in the US. The authors found an upwards bias for which people of a certain socioeconomic class mostly visited places in the same class with rare visits to locations from higher classes. Furthermore, this bias increases with socioeconomic status and correlates with metrics for estimating racial residential segregation. In recent work, \citet{shelton2015social} showed that geotagged Twitter data can be used to capture the socio-spatial relations of territories dynamically. In particular, the authors analyzed the data for the city of Louisville in Kentucky, US, to show that analyzing the segregation of the city using static data is not sufficient to understand its dynamics.

\citet{nicoletti2022disadvantaged} leveraged a combination of census data and geographical information from OpenStreetMap (OSM) to measure the accessibility to different points of interest (POIs). The authors defined a metric of accessibility and found that, in more than 50 cities, their metric suggests that inequalities appear proportional to growth processes. Also, for 10 of the cities, low accessibility scores were associated with communities with a larger share of minorities and with lower income levels.

Inequalities can also be related to the possibilities of accessing transportation. 
For example, by analyzing Boston's BlueBikers program mobility data, \citet{fraser4076776cycling} showed that some neighborhoods use bike-sharing programs more than others. However, by considering the underlying socio-demographic characteristics, it emerged that there are significantly different adoption rates with respect to race and income level. \citet{fraser4076776cycling} also pointed out how, by analyzing the mobility network over time (e.g., 2011 to 2021), Boston's program is gradually reaching a broader range of neighborhoods. 

In a couple of recent works, \citet{chetty2022a, chetty2022b} have leveraged data from Facebook on 21 billion friendships. In particular, in a first study \citet{chetty2022a} they have measured three types of social capital by postal code in the US: (i) connectedness, namely friendship between people with different characteristics (i.e., high income vs low income), (ii) social cohesion, which is the extent to which networks of friends are clustered in cliques, and (iii) civic engagement, which measures as rates of volunteering or participation in civic organizations. Interestingly, the share of high-income friends among people with low income (called economic connectedness by the authors) is one of the more powerful predictors of upward economic (e.g., income) mobility. Moreover, \citet{chetty2022a} have also found that differences in economic connectedness can explain upward income mobility, even when controlling for other strong neighborhood-level predictors such as poverty rates, racial segregation, and inequality. In a companion study and again using Facebook data, \citet{chetty2022b} have shown that about half of the social disconnection across different socioeconomic groups is explained by differences in exposure to people with high socioeconomic status in places such as schools. Instead, the other half is explained by a friendship bias, namely a lower tendency of low-income people to establish friendships with high-income individuals. This ability to disentangle differences in exposures and friendship bias is of paramount relevance for building effective interventions and strategies to increase economic connectedness and thus decrease income segregation and inequalities.

\subsubsection{Critical Reflections}
New methodologies and novel insights on segregation and inequalities in urban environments have been developed thanks to novel data sources that enabled the study of different facets of human behavior and interaction at spatio-temporal scales that were previously unavailable.
While these new approaches have brought substantial advantages for a timely study of human dynamics, it is still difficult to use these new methodologies and data sources to effectively measure the real impact of policies and interventions for reducing inequalities. In particular, this second aspect will require the development of a simulation framework able to generate different scenarios enabled by specific interventions. Moreover, given the proprietary nature of these novel data sources, it is often difficult to have access to longitudinal data that span multiple years and therefore it is problematic to track whether an intervention had an effective impact over time. An example of the potential of a study based on a longitudinal dataset is represented by \citep{yabe2022behavioral}, which using three years of data found that the COVID-19 pandemic still has long-lasting implications on urban income diversity in US cities.
Furthermore, several studies are based on aggregated data and not on individual data for privacy reasons. This means that pieces of information such as poverty and deprivation aren't directly available at the individual level and scholars have to develop proxies of such measures at an aggregated level (i.e., neighborhood) which may hinder the understanding of actual inequalities. As an example, \citet{gundougdu2019bridging} had to move the two definitions of bridging and bonding social capital from an individual level to an aggregated level due to the unavailability of individual-level data.

\section{Public Health}
\label{sec:public_health}

\subsection{The Computational Contamination in Research on Public Health}

Public health is an inherently interdisciplinary field, whose key focus, preventing disease and promoting health, largely benefits from the contribution of different scholarly expertise, ranging from medicine to the social sciences, psychology and economics \cite{gavens2018interdisciplinary}.
The past decades have seen an ever-increasing adoption of computational and digital methods in the research on public health, and many have been advocating for increasingly closer collaboration between computer scientists and public health scholars \cite{epstein2013collaborations}. 

In particular, major advances in the use of computational methods for public health have been witnessed in the area of infectious disease epidemiology, specifically to model and understand the patterns of disease dynamics and related causes.
The use of mathematical models to study the spread of infectious diseases dates back to the seminal works of~\cite{ross1916application} and~\cite{kermack1927contribution} at the beginning of the XX century, who first introduced the law of mass action in epidemiology. 
Over the years, epidemic models became increasingly complex, pursuing a higher level of realism by introducing additional interacting components, such as spatially defined structures~ \cite{rvachev1985mathematical,sattenspiel1995structured}, age-stratified contact patterns~\cite{fumanelli2012inferring}, human movements on long and short scales~ \cite{colizza2007modeling,balcan2009multiscale} and, more in general, human behavior~\cite{perra2011towards,funk2010modelling}.  
At the same time, the development of such models has been possible thanks to the increasing availability of computing power, thus allowing the \textit{in-silico} recreation of populations with unprecedented levels of detail.
If, in the origins of mathematical epidemiology, models were based on the assumption of a single, closed and well-mixed population~ \cite{grassly2008mathematical}, modern epidemic models are usually structured as Agent-Based Models (ABMs), simulating the daily routines of up to hundreds of millions of individuals and their close contacts, in households, schools, and workplaces, requiring large-scale computational infrastructures~\cite{ajelli2010comparing,merler2010role,ferguson2006strategies}.

The growth of computing power has been matched by even faster growth in data availability. 
With the diffusion of ubiquitous technologies and the rise of the Internet era, the field of epidemiology has been rapidly contaminated by digital approaches leading to a newly defined ``digital epidemiology"~\cite{salathe2012digital}. 
Digital epidemiology, in the definition given by~\cite{salathe2018digital}, refers to epidemiology that uses data that was generated outside the public health system, data that were not collected with a specific public health-related purpose. 
The first study that brought to worldwide attention the potential use of a novel digital data source in epidemiology described Google Flu Trends, a system to monitor flu activity in more than 25 countries based on search query data~\cite{ginsberg2009detecting}.
The service was shut down in 2015, but historical data are still available. Also, the same study sparked a significant controversy on the accuracy of such emerging models and their potential biases~\cite{lazer2014google,lazer2014parable}.
Soon thereafter, studies on digital epidemiology started growing exponentially, using a variety of digital sources to track disease prevalence and design public health interventions~\cite{bansal2016big,althouse2015enhancing}. 
Many studies followed the seminal example of Google Flu Trends by integrating different web data sources to forecast flu activity~\cite{polgreen2008using,shaman2012forecasting,shaman2013real,lampos2015advances,yuan2013monitoring}. 
As other data sources became rapidly available, their use has been explored in a wide range of epidemiological applications.
Mobile phone data have been used to measure human movements and inform both spatially structured epidemic models~\cite{wesolowski2016connecting, wesolowski2012quantifying,tatem2009use}, and surveillance systems \cite{barlacchi2017you}.
Other studies have further advanced epidemic modeling and forecasting by combining additional data streams such as social media data~\cite{lampos2010flu,zhang2015social,zhang2017forecasting}, internet media reports~\cite{freifeld2008healthmap}, wearable sensors~\cite{viboud2020fitbit, isella2011close}, satellite imagery~\cite{castro2021using,bharti2016measuring}.   
Finally, the opportunity provided by the Web to directly engage users in scientific research has opened the path to participatory surveillance systems, moving beyond the initial paradigm of passively collected data sources~\cite{paolotti2014web,smolinski2015flu,brownstein2017combining}. 

In this context, the COVID-19 pandemic has marked a turning point for digital epidemiology.
While before 2020, digital epidemiology has been mainly studied as a proof-of-concept with a few real-time applications, since the early days of the COVID-19 outbreak in China, digital approaches have played a crucial role across the whole pandemic life cycle~\cite{oliver2020mobile}, ranging from predictive modeling~\cite{poletto2020applications} to the population-scale deployment of digital contact tracing apps~\cite{colizza2021time}.

In the next sections, we discuss in more detail the role of the different alternative data sources and modeling techniques in computational epidemiology with a specific focus on applications in the urban context. First, we briefly highlight the general advantages of studying urban public health using digital approaches and big data. After that, we highlight the roles of mobile phone data, social media data, and other novel data streams. 

\subsection{Computational Methods, Big Data, and Urban Public Health}

\subsubsection{The advantages of studying urban public health today}

Modern epidemiology traces its roots to the foundational work of John Snow, who first identified the Broad Street pump as the source of the 1854 London cholera outbreak by mapping disease prevalence in Soho, and showing how cases occurred around this pump \cite{shiode2015mortality}.
Since then, cartography and mapping spatial disease patterns have represented a fundamental tool for epidemiologists \cite{koch2011disease}.
In particular, the analysis of disease patterns in cities has attracted significant attention from scholars, as large metropolitan areas represent the main hubs of disease emergence and spreading \cite{ali2011networked,connolly2021extended}, a fact that has been well exemplified by the most recent global pandemics. 
Due to the high density of people and close proximity of living and working spaces in cities, infectious diseases can easily spread from person to person.
Additionally, the crowded and often unsanitary living conditions in many cities can provide a conducive environment for the spread of infectious diseases. For example, inadequate access to clean water and sanitation facilities can lead to the proliferation of waterborne diseases such as cholera.
Furthermore, the rapid pace of urbanization and population growth in many cities can put a strain on existing healthcare systems, making it more difficult to effectively detect and respond to outbreaks of infectious diseases \cite{neiderud2015urbanization}.

In recent years, scholars have investigated the role of urban features in the spread of infectious diseases, through a number of computational methods. 
On the one hand, computational epidemic models have been developed to capture the complexities of human behavior in the urban environment, from fine-scale human movements \cite{perkins2014theory} to contact networks \cite{eubank2004modelling}.
On the other hand, many studies have explored the effect of city characteristics, such as urban population scaling laws \cite{bettencourt2007growth}, on health outcomes \cite{rocha2015non, bilal2021scaling} through a mix of theoretical and computational approaches \cite{tizzoni2015scaling, schlapfer2014scaling}. 
As digital trace data have become pervasive, providing researchers with a tool to investigate human behavior at a high spatial resolution, several studies have advanced our understanding of the role of urban structures in the spread of epidemics.
By combining novel data sources with spatially resolved records of disease incidence, researchers have shown that variations in mobility patterns and the associated spatio-temporal fluctuations in population size can predict variations in the dynamics of seasonal flu epidemics \cite{dalziel2013human, dalziel2018urbanization, zachreson2018urbanization}.
Similarly, the hierarchical structure of cities, and their different organization as single-center or multi-center systems, has been shown to predict inter-city variations in the spreading dynamics of respiratory infections \cite{brizuela2021understanding,aguilar2022impact,rader2020crowding}.

Furthermore, the availability of high-resolution digital sources has also enabled the study of determinants of non-communicable diseases and chronic health conditions in urban areas.
In particular, the analysis of digital trace data has allowed the characterization of neighborhoods based on novel behavioral indicators, thus providing new metrics to explain the observed residents' health outcomes \cite{sadilek2013modeling}.
Mobile phone data, social media data, and remote sensing have been used to model the pulse of urban life at a scale and granularity that would be hard to achieve with traditional methods.
Overall, research in this area has demonstrated that novel digital sources represent an invaluable tool to monitor the health conditions of cities, understand their dynamics and inform public health policies.
In the following sections, we provide an overview of some relevant contributions, based on different data sources, to address public health issues in large metropolitan areas.

\subsubsection{The role of mobile phone data}

Location data generated by mobile phones have played a pivotal role in the modeling of human mobility and population settlements at international \cite{kraemer2020mapping}, national \cite{deville2014dynamic}, and smaller spatial scales \cite{alessandretti2022human}.
Since the early days of digital epidemiology, mobile phone data have represented an invaluable data source to connect empirical human mobility patterns and the spatial spread of infectious diseases. They have been used to calibrate epidemic models, understand disease spreading patterns, and evaluate intervention strategies against them \cite{wesolowski2016connecting}.
Initial efforts to incorporate mobile phone derived mobility metrics into epidemic models have been mostly focused on large spatial scales, such as country-wide movements. This is the case, for instance of seminal work by \cite{tatem2009worldwide}, who leveraged mobile phone data collected in Zanzibar to estimate the relation between human mobility flows and parasite carrier movements and rates of malaria importation. The authors found that most of the people in Zanzibar traveled low-risk short distances but risk groups visiting higher-risk regions for extended periods could be identified. Similarly, \cite{wesolowski2012quantifying} used mobile phone data and malaria prevalence information to estimate how people's movements were related to parasite importation between different regions. With their study, the authors were able to identify sources and sinks of imported infections and also to identify critical travel routes. 
Applications to city-scale epidemic scenarios have been generally more scarce, however, until the COVID-19 pandemic. 

The COVID-19 pandemic has represented a defining moment for the use of mobile phone-derived data in epidemic modeling in cities. 
For the first time, high-resolution temporally resolved positioning data, collected from millions of users, became available to researchers. 
Such an unprecedented amount of information has fostered the development of a new generation of ABMs, with the ability to recreate synthetic populations of large urban areas with extraordinary realism, which was deemed impossible until a few years ago. 
While, in 2011, \cite{cooley2011role} built their ABM of 7 million individuals living in New York City, only based on the most recent census surveys, 10 years later, \cite{aleta2020modelling} could explicitly model the time-varying interactions of 100,000 people in the Boston Metropolitan Area, accounting for more than 5 million interactions in schools, workplaces, and households, derived from empirical co-location events. 
Their study showed that a response system based on enhanced testing and contact tracing could be an important tool to mitigate the spread of COVID-19, once social distancing measures were relaxed. 
In the following paper, \cite{aleta2022quantifying} developed a similar model, integrating individual-level mobility data with socio-demographic information, to generate synthetic populations in New York City and Seattle, and simulate transmission events in more than 400,000 locations within the two cities. Such a detailed model allowed them to characterize the risk of COVID-19 transmission in different venues, identifying the most likely locations of superspreading events.        
In a similar effort, \cite{chang2021mobility} developed an epidemic ABM describing the mobility networks of ten of the largest US metropolitan areas. These networks mapped the hourly movements of 98 million people from census block groups, resulting in 5 billion dynamic edges.
Simulations of COVID-19 spread on these large spatially-resolved networks demonstrated how higher infection rates among disadvantaged racial and socioeconomic groups were solely a result of differences in mobility in response to non-pharmaceutical interventions (NPIs). 
Studies of intra-city mobility during the pandemic were not limited to the USA or Europe. 
As an example, \cite{gozzi2021estimating} investigated the dynamics of COVID-19 in the metro area of Santiago de Chile, using anonymized mobile phone data from 1.4 million users.
By combining mobility traces and a compartmental epidemic model, they found that mobility responses to the lockdown were highly unequal in the city, with most deprived areas experiencing higher levels of mobility and, as a consequence, higher infection rates. 

Other studies have used mobile phone data to investigate the socio-economic effects of NPIs in cities.  
\cite{bonaccorsi2020economic} investigated the impact of the first lockdown measures in Italy, using a large-scale mobility dataset provided by Meta. They found that the impact on mobility was stronger in municipalities with higher fiscal capacity, while, at the same time, mobility reductions were larger in municipalities with higher income inequalities. Their results prompted fiscal interventions targeting the unequal effects of COVID-19 mitigation measures. 
On a similar note, \cite{gauvin2021socio} used anonymized individual location data to study the mobility responses to COVID-19 in the neighborhoods of 3 major Italian cities.
Their analysis uncovered the desertification of historic city centers, which persisted after the end of the first lockdown. Such a center-periphery gradient was mainly associated with differences in educational attainment. Similar results were found by \cite{glodeanu2021social} who evaluated socio-economic disparities of mobility responses in the neighborhoods of Madrid. 

As mobility restrictions were removed, and social life returned to normal, other studies focused on persistent changes in human behavior, that followed the pandemic response. 
For instance, \cite{lucchini2021living} used location data to analyze how people changed their mobility patterns and person-to-person contacts in response to NPIs in the US. Interestingly, they found a persistent reduction in close contacts and in the number of venues visited, even after the lifting of COVID-19 mandates.   
Using crowdsourced mobility data from 45 million devices, \cite{li2022aggravated} found evidence of aggravated social segregation in the 12 largest US metropolitan areas, as a consequence of the COVID-19 mobility restrictions.
Other studies, instead, have investigated the effects of the pandemic on lifestyles and individual habits. 
A notable example is a work by \cite{hunter2021effect}, who investigated the effects of the COVID-19 pandemic on walking habits in 10 major metropolitan areas of the US. The authors used individual-level mobility data to identify changes in the walking behavior of more than 1.6 million anonymized mobile phone users. Their findings highlighted a dramatic decline in walking habits during the first wave of the pandemic. Moreover, they found that once restrictions were lifted, walking levels recovered to pre-COVID-19 measures in high-income areas, whereas low-income areas were still well below pre-COVID-19 levels.     

Finally, recent modeling advances have further developed the field of large-scale ABMs by combining high-resolution mobility data with detailed information on the economic role of individuals, as workers and consumers. Recent work by \cite{pangallo2022unequal} developed an ABM of the New York-Newark-Jersey City Metro Area that is representative of the real population across multiple socio-economic characteristics, including their employment status, the industry they work in, and their ability to work from home. Parameterizing the model with privacy-enhanced location data, the authors could explore the complex tradeoff between health and economy with an unprecedented level of realism.

\subsubsection{The role of social media}

In digital epidemiology, social media data have always represented an important source of information to infer disease prevalence from health-related behaviors or symptoms reported by users \cite{aiello2020social, brownstein2009digital}. 
Among social media, Twitter is the one that has attracted the most attention from scholars, thanks to the public availability, and machine readability, of basically all its content~\cite{mejova2015twitter}. 
The most typical use of Twitter data involves the automatic identification of relevant tweets, through either keyword search or natural-language processing, to identify posts whose content is related to some health condition \cite{paul2011you}. For instance, tweets posted by users who report Influenza-Like Illness (ILI) symptoms. Collected tweets are then used as input to predictive models that aim at reproducing some known baseline, such as the ILI trends reported by official public health surveillance.  

While initial efforts in this direction were mostly focused on measuring aggregated statistics of disease prevalence at the national level \cite{broniatowski2013national, gesualdo2013influenza}, the availability of geo-tagged social media data with GPS accuracy, in particular from Twitter, has allowed mapping users' health conditions at a very high spatial resolution, reaching the scale of a city \cite{sadilek2012predicting}. 
In a seminal paper, \cite{nagar2014case} used geo-referenced city-level Twitter data as a means of forecasting real-time ILI emergency department visits in New York City. 
They demonstrated that, at that spatial resolution, Twitter data could effectively capture the dynamics of flu in the boroughs of NYC. They also found that a model using the number of infection-related tweets outperformed one based on the number of web searches in predicting the number of ILI-related visits to emergency departments.
Similarly, \cite{lu2018accurate} used Twitter data to model seasonal flu epidemics in the Boston Metropolitan Area.  

Social media also represents a valuable data source to monitor non-communicable diseases and health habits in cities.
A comprehensive study by \cite{nguyen2016building} developed a publicly available neighborhood-level data set with indicators related to health behaviors and wellbeing in the USA.
Interestingly, the authors found that greater happiness, and positivity toward physical activity and toward healthy foods, assessed via tweets, were associated with lower all-cause mortality and prevalence of chronic conditions such as obesity and diabetes and lower physical inactivity, and smoking.
Similarly, Twitter data has been proven useful to map dietary habits in US cities, down to the level of census tract. A study by \cite{gore2015you} investigated how the obesity rate of an urban geographic area correlates with the contents of geo-tagged tweets in that area. 
In recent work by \cite{sigalo2022using}, the authors analyzed about sixty-thousand geolocated food-related tweets collected across 25 cities, in the USA. They found associations between a census tract being classified as a food desert and an increase in the number of tweets in a census tract that mentioned unhealthy foods.
Instagram, a video and photo sharing social network with more than 1 billion users worldwide, represents another relevant data source to investigate health habits, and in particular dietary choices, at scale. As an example, \cite{de2016characterizing} showed that the textual content of Instagram posts predicts with high accuracy food deserts in the metropolitan areas of the US, while \cite{mejova2015foodporn} combined data from Instagram, Twitter, and Foursquare to correlate dietary choices and the prevalence of obesity across the USA. 
Another public health-relevant use of social media, which has been extensively explored by health departments in the US, is monitoring reports of foodborne illnesses. Two notable studies investigated the potential use of Twitter posts \cite{harris2014health} and Yelp reviews \cite{harrison2014using} to track food poisoning outbreaks in Chicago and New York City, respectively. Both studies demonstrated the high impact of using social media data to improve surveillance in collaboration with city public health authorities. 
Finally, \cite{aiello2016chatty} used a random sample of 17 million geo-referenced Flickr photos taken within Greater London between 2010 and 2015 to create a high-resolution map of the sound landscape of the city. They further leveraged such dataset to quantify the effects of noise on population health, correlating noise exposure levels with hypertension rates, at a very high spatial granularity \cite{gasco2020social}.

\subsubsection{The role of other data sources}

Beyond mobile phone location data and social media, several studies have demonstrated the potential use of other digital sources for public health research.   
As already mentioned, multiple studies have leveraged search query data of services like Google Search \cite{ginsberg2009detecting, lampos2015advances}, Baidu \cite{yuan2013monitoring}, or Bing \cite{lampos2021tracking}, to monitor epidemics at scale.
Internet search queries have not only been used to track the spread of infectious diseases but also to monitor other health conditions, for instance, mental health.
As an example, \cite{adler2019search} combined official demographic statistics with data generated from Bing queries to gain insight into suicide rates per state in India as reported by the official census.

Another useful, but mostly untapped, data source to monitor disease incidence is Wikipedia pageview data. 
A landmark paper by \cite{mciver2014wikipedia} showed that the number of Wikipedia article views of specific health-related pages was a good predictor of ILI activity in the US. 
However, even though Wikipedia pageview data are geolocated, their availability with geo-encoded information is limited due to privacy reasons. Thus, city-scale studies are scarce. A notable example is a work by \cite{tizzoni2020impact}, who measured changes in awareness in the United States during the 2016 Zika epidemic through geo-localized Wikipedia pageview data, at the level of US city. 
They examined the attention to Zika in 788 cities in the United States with a population larger than 40,000 and found clear and distinct patterns of attention, varying with the exposure to the virus and the volume of media coverage.  

Electronic records of retail market purchases represent a novel and interesting data stream, whose potential has been recently explored.
\cite{miliou2021predicting} proposed to use retail market data to improve the forecasting of seasonal flu. In particular, the authors showed that by identifying some specific co-purchases of products, by specific customers, it is possible to model seasonal flu incidence in Italy, 4 weeks in advance, with improved accuracy with respect to an autoregressive baseline.
\cite{aiello2019large} collected and analyzed a similar dataset, reaching an unprecedented level of spatial granularity.
By analyzing 1.6 billion food item purchases and 1.1B medical prescriptions for the entire city of London over the course of one year, they showed that nutrient diversity and amount of calories are the two strongest predictors of the prevalence of three chronic health conditions: hypertension, high cholesterol, and diabetes.

\subsection{Critical Reflections}

The future of urban public health is undoubtedly going to be more and more digital. 
It is clear, however, that several challenges lie ahead, as has been evidenced by the adoption of computational and digital technologies during the COVID-19 pandemic.
Thanks to increasing computing power and the availability of high-resolution behavioral data, computational models of epidemics are able to capture key determinants of transmission with impressive detail. However, they often lack a structural integration with socioeconomic dimensions that are known to affect epidemic outcomes. 
Recent studies have pointed out the need for equitable approaches in digital epidemiology, to address socioeconomic gaps in disease surveillance and modeling \cite{tizzoni2022addressing, buckee2021thinking}. 
Future work should aim at reducing health disparities during health emergencies through closer collaboration between epidemiologists and social scientists, psychologists, and economists.  

Of course, the use of passively collected digital traces in public health comes with significant privacy and ethical concerns.
In an urban context, measuring behaviors at a high spatial granularity represents a key advantage with respect to traditional data sources. However, reaching a high granularity may imply a higher risk of data re-identification, especially with small sample sizes, thus putting individual privacy at risk. 
In the future, it will important to understand what privacy-preserving mechanisms can be most effective in minimizing such risks while preserving the potential of data analysis, even at a high spatial resolution. 

Finally, the use of novel digital sources requires a careful understanding of their limitations and their scope.
For instance, mobile phone-derived mobility metrics have been proven useful to understand the dynamics of COVID-19 in the early phase of the outbreak, however, the relationship between mobility indicators and epidemic outcomes is not straightforward \cite{kishore2021mobility}. 
While mobile phone data are clearly useful to measure changes in human behavior and link them with epidemic dynamics, such link often varies over time, and understanding this varying relationship poses significant challenges to scholars and policymakers who may want to use mobile phone data to evaluate the effectiveness of interventions or forecast future epidemic trajectories. 
Further work is needed to define methods that can systematically assess the quality of mobile phone-derived mobility metrics and make them comparable across different settings and data providers.

\section{Conclusions}
\label{sec:conclusions}
Cities are the beating heart of our modern societies. With more than half of the world's population living there, multiple emerging societal challenges require modern solutions. In particular, measuring the efficiency of deployed policies and progress towards specific SDGs, it is fundamental to have an always up-to-date picture of human dynamics in cities (e.g., how people move, how they interact with each other). In this context, it is clear that a pivotal role is played by the data collected from alternatives (ubiquitous) data sources like mobile phones, social media, GPS traces, satellite images, wearable devices and many others. In our review paper, we showcase how such alternative data are employed to monitor signs of progress toward some specific United Nations' Sustainable Development Goals. In particular, after a discussion about the different alternative data sources, we review how such information has been used to monitor urban crime and public safety. After that, we highlight the role of such data in reducing socioeconomic inequalities and segregation. Finally, we showed how they impacted research about public health. In all the sections, we start with a brief discussion about the advantages of using big data with respect to other techniques. Afterwards, we describe how different studies use such information. Finally, we conclude every section with some critical reflections about limitations and potential future directions.  
\bibliography{iclr2021_conference}

\begin{thebibliography}{272}
\providecommand{\natexlab}[1]{#1}
\providecommand{\url}[1]{\texttt{#1}}
\expandafter\ifx\csname urlstyle\endcsname\relax
  \providecommand{\doi}[1]{doi: #1}\else
  \providecommand{\doi}{doi: \begingroup \urlstyle{rm}\Url}\fi

\bibitem[Aaltonen et~al.(2018)Aaltonen, Kivivuori, and
  Kuitunen]{AaltonenShortTermTemporalClustering2018}
Mikko Aaltonen, Janne Kivivuori, and Laura Kuitunen.
\newblock Short-{Term} {Temporal} {Clustering} of {Police}-{Reported} {Violent}
  {Offending} and {Victimization}: {Examining} {Timing} and the {Role} of
  {Revenge}.
\newblock \emph{Criminal Justice Review}, 43\penalty0 (3):\penalty0 309--324,
  September 2018.
\newblock ISSN 0734-0168, 1556-3839.
\newblock \doi{10.1177/0734016818761100}.
\newblock URL \url{http://journals.sagepub.com/doi/10.1177/0734016818761100}.

\bibitem[Adler et~al.(2019)Adler, Cattuto, Kalimeri, Paolotti, Tizzoni,
  Verhulst, Yom-Tov, Young, et~al.]{adler2019search}
Natalia Adler, Ciro Cattuto, Kyriaki Kalimeri, Daniela Paolotti, Michele
  Tizzoni, Stefaan Verhulst, Elad Yom-Tov, Andrew Young, et~al.
\newblock How search engine data enhance the understanding of determinants of
  suicide in india and inform prevention: observational study.
\newblock \emph{Journal of medical internet research}, 21\penalty0
  (1):\penalty0 e10179, 2019.

\bibitem[Aguilar et~al.(2022)Aguilar, Bassolas, Ghoshal, Hazarie, Kirkley,
  Mazzoli, Meloni, Mimar, Nicosia, Ramasco, et~al.]{aguilar2022impact}
Javier Aguilar, Aleix Bassolas, Gourab Ghoshal, Surendra Hazarie, Alec Kirkley,
  Mattia Mazzoli, Sandro Meloni, Sayat Mimar, Vincenzo Nicosia, Jos{\'e}~J
  Ramasco, et~al.
\newblock Impact of urban structure on infectious disease spreading.
\newblock \emph{Scientific reports}, 12\penalty0 (1):\penalty0 1--13, 2022.

\bibitem[Aiello et~al.(2020)Aiello, Renson, and Zivich]{aiello2020social}
Allison~E Aiello, Audrey Renson, and Paul Zivich.
\newblock Social media-and internet-based disease surveillance for public
  health.
\newblock \emph{Annual review of public health}, 41:\penalty0 101, 2020.

\bibitem[Aiello et~al.(2016)Aiello, Schifanella, Quercia, and
  Aletta]{aiello2016chatty}
Luca~Maria Aiello, Rossano Schifanella, Daniele Quercia, and Francesco Aletta.
\newblock Chatty maps: constructing sound maps of urban areas from social media
  data.
\newblock \emph{Royal Society open science}, 3\penalty0 (3):\penalty0 150690,
  2016.

\bibitem[Aiello et~al.(2019)Aiello, Schifanella, Quercia, and
  Del~Prete]{aiello2019large}
Luca~Maria Aiello, Rossano Schifanella, Daniele Quercia, and Lucia Del~Prete.
\newblock Large-scale and high-resolution analysis of food purchases and health
  outcomes.
\newblock \emph{EPJ Data Science}, 8\penalty0 (1):\penalty0 1--22, 2019.

\bibitem[Ajelli et~al.(2010)Ajelli, Gon{\c{c}}alves, Balcan, Colizza, Hu,
  Ramasco, Merler, and Vespignani]{ajelli2010comparing}
Marco Ajelli, Bruno Gon{\c{c}}alves, Duygu Balcan, Vittoria Colizza, Hao Hu,
  Jos{\'e}~J Ramasco, Stefano Merler, and Alessandro Vespignani.
\newblock Comparing large-scale computational approaches to epidemic modeling:
  agent-based versus structured metapopulation models.
\newblock \emph{BMC infectious diseases}, 10\penalty0 (1):\penalty0 1--13,
  2010.

\bibitem[Akpinar et~al.(2021)Akpinar, De-Arteaga, and
  Chouldechova]{Akpinareffectdifferentialvictim2021}
Nil-Jana Akpinar, Maria De-Arteaga, and Alexandra Chouldechova.
\newblock The effect of differential victim crime reporting on predictive
  policing systems.
\newblock In \emph{Proceedings of the 2021 {ACM} {Conference} on {Fairness},
  {Accountability}, and {Transparency}}, {FAccT} '21, pp.\  838--849, New York,
  NY, USA, March 2021. Association for Computing Machinery.
\newblock ISBN 978-1-4503-8309-7.
\newblock \doi{10.1145/3442188.3445877}.
\newblock URL \url{https://doi.org/10.1145/3442188.3445877}.

\bibitem[Alessandretti(2022)]{alessandretti2022human}
Laura Alessandretti.
\newblock What human mobility data tell us about covid-19 spread.
\newblock \emph{Nature Reviews Physics}, 4\penalty0 (1):\penalty0 12--13, 2022.

\bibitem[Aleta et~al.(2020)Aleta, Martin-Corral, Pastore~y Piontti, Ajelli,
  Litvinova, Chinazzi, Dean, Halloran, Longini~Jr, Merler,
  et~al.]{aleta2020modelling}
Alberto Aleta, David Martin-Corral, Ana Pastore~y Piontti, Marco Ajelli, Maria
  Litvinova, Matteo Chinazzi, Natalie~E Dean, M~Elizabeth Halloran, Ira~M
  Longini~Jr, Stefano Merler, et~al.
\newblock Modelling the impact of testing, contact tracing and household
  quarantine on second waves of covid-19.
\newblock \emph{Nature Human Behaviour}, 4\penalty0 (9):\penalty0 964--971,
  2020.

\bibitem[Aleta et~al.(2022)Aleta, Mart{\'\i}n-Corral, Bakker, Pastore~y
  Piontti, Ajelli, Litvinova, Chinazzi, Dean, Halloran, Longini~Jr,
  et~al.]{aleta2022quantifying}
Alberto Aleta, David Mart{\'\i}n-Corral, Michiel~A Bakker, Ana Pastore~y
  Piontti, Marco Ajelli, Maria Litvinova, Matteo Chinazzi, Natalie~E Dean,
  M~Elizabeth Halloran, Ira~M Longini~Jr, et~al.
\newblock Quantifying the importance and location of sars-cov-2 transmission
  events in large metropolitan areas.
\newblock \emph{Proceedings of the National Academy of Sciences}, 119\penalty0
  (26):\penalty0 e2112182119, 2022.

\bibitem[Ali \& Keil(2011)Ali and Keil]{ali2011networked}
S~Harris Ali and Roger Keil.
\newblock \emph{Networked disease: emerging infections in the global city}.
\newblock John Wiley \& Sons, 2011.

\bibitem[Althouse et~al.(2015)Althouse, Scarpino, Meyers, Ayers, Bargsten,
  Baumbach, Brownstein, Castro, Clapham, Cummings,
  et~al.]{althouse2015enhancing}
Benjamin~M Althouse, Samuel~V Scarpino, Lauren~Ancel Meyers, John~W Ayers,
  Marisa Bargsten, Joan Baumbach, John~S Brownstein, Lauren Castro, Hannah
  Clapham, Derek~AT Cummings, et~al.
\newblock Enhancing disease surveillance with novel data streams: challenges
  and opportunities.
\newblock \emph{EPJ data science}, 4\penalty0 (1):\penalty0 1--8, 2015.

\bibitem[Anderson(1989)]{Andersonartificialintelligencetheory1989a}
Bo~Anderson.
\newblock On artificial intelligence and theory construction in sociology.
\newblock \emph{Journal of Mathematical Sociology}, 14\penalty0 (2-3):\penalty0
  209--216, 1989.

\bibitem[Athey et~al.(2020)Athey, Ferguson, Gentzkow, and
  Schmidt]{athey2020experienced}
Susan Athey, Billy~A Ferguson, Matthew Gentzkow, and Tobias Schmidt.
\newblock Experienced segregation.
\newblock Technical report, National Bureau of Economic Research, 2020.

\bibitem[Bagrow et~al.(2011)Bagrow, Wang, and Barabasi]{bagrow2011}
J.~P. Bagrow, D.~Wang, and A.-L. Barabasi.
\newblock Collective response of human populations to large-scale emergencies.
\newblock \emph{PLos One}, 6\penalty0 (3):\penalty0 e17680, 2011.

\bibitem[Balcan et~al.(2009)Balcan, Colizza, Gon{\c{c}}alves, Hu, Ramasco, and
  Vespignani]{balcan2009multiscale}
Duygu Balcan, Vittoria Colizza, Bruno Gon{\c{c}}alves, Hao Hu, Jos{\'e}~J
  Ramasco, and Alessandro Vespignani.
\newblock Multiscale mobility networks and the spatial spreading of infectious
  diseases.
\newblock \emph{Proceedings of the National Academy of Sciences}, 106\penalty0
  (51):\penalty0 21484--21489, 2009.

\bibitem[Bansal et~al.(2016)Bansal, Chowell, Simonsen, Vespignani, and
  Viboud]{bansal2016big}
Shweta Bansal, Gerardo Chowell, Lone Simonsen, Alessandro Vespignani, and
  C{\'e}cile Viboud.
\newblock Big data for infectious disease surveillance and modeling.
\newblock \emph{The Journal of infectious diseases}, 214\penalty0
  (suppl\_4):\penalty0 S375--S379, 2016.

\bibitem[Barlacchi et~al.(2017)Barlacchi, Perentis, Mehrotra, Musolesi, and
  Lepri]{barlacchi2017you}
Gianni Barlacchi, Christos Perentis, Abhinav Mehrotra, Mirco Musolesi, and
  Bruno Lepri.
\newblock Are you getting sick? predicting influenza-like symptoms using human
  mobility behaviors.
\newblock \emph{EPJ data science}, 6:\penalty0 1--15, 2017.

\bibitem[Batty(2013)]{batty2013}
M.~Batty.
\newblock \emph{The new science of cities}.
\newblock MIT Press, 2013.

\bibitem[Berk(2019)]{BerkMachineLearningRisk2019b}
Richard Berk.
\newblock \emph{Machine {Learning} {Risk} {Assessments} in {Criminal} {Justice}
  {Settings}}.
\newblock Springer International Publishing, Cham, 2019.
\newblock ISBN 978-3-030-02271-6 978-3-030-02272-3.
\newblock \doi{10.1007/978-3-030-02272-3}.
\newblock URL \url{http://link.springer.com/10.1007/978-3-030-02272-3}.

\bibitem[Berk \& Elzarka(2020)Berk and
  Elzarka]{BerkAlmostpoliticallyacceptable2020}
Richard Berk and Ayya~A. Elzarka.
\newblock Almost politically acceptable criminal justice risk assessment.
\newblock \emph{Criminology \& Public Policy}, 19\penalty0 (4):\penalty0
  1231--1257, 2020.
\newblock ISSN 1745-9133.
\newblock \doi{10.1111/1745-9133.12500}.
\newblock URL
  \url{https://onlinelibrary.wiley.com/doi/abs/10.1111/1745-9133.12500}.
\newblock \_eprint:
  https://onlinelibrary.wiley.com/doi/pdf/10.1111/1745-9133.12500.

\bibitem[Bettencourt \& West(2010)Bettencourt and West]{bettencourt2010unified}
Luis Bettencourt and Geoffrey West.
\newblock A unified theory of urban living.
\newblock \emph{Nature}, 467\penalty0 (7318):\penalty0 912--913, 2010.

\bibitem[Bettencourt et~al.(2007)Bettencourt, Lobo, Helbing, K{\"u}hnert, and
  West]{bettencourt2007growth}
Lu{\'\i}s~MA Bettencourt, Jos{\'e} Lobo, Dirk Helbing, Christian K{\"u}hnert,
  and Geoffrey~B West.
\newblock Growth, innovation, scaling, and the pace of life in cities.
\newblock \emph{Proceedings of the national academy of sciences}, 104\penalty0
  (17):\penalty0 7301--7306, 2007.

\bibitem[Bharti et~al.(2016)Bharti, Djibo, Tatem, Grenfell, and
  Ferrari]{bharti2016measuring}
Nita Bharti, Ali Djibo, Andrew~J Tatem, Bryan~T Grenfell, and Matthew~J
  Ferrari.
\newblock Measuring populations to improve vaccination coverage.
\newblock \emph{Scientific reports}, 6\penalty0 (1):\penalty0 1--10, 2016.

\bibitem[Bilal et~al.(2021)Bilal, de~Castro, Alfaro, Barrientos-Gutierrez,
  Barreto, Leveau, Martinez-Folgar, Miranda, Montes, Mullachery,
  et~al.]{bilal2021scaling}
Usama Bilal, Caio~P de~Castro, Tania Alfaro, Tonatiuh Barrientos-Gutierrez,
  Mauricio~L Barreto, Carlos~M Leveau, Kevin Martinez-Folgar, J~Jaime Miranda,
  Felipe Montes, Pricila Mullachery, et~al.
\newblock Scaling of mortality in 742 metropolitan areas of the americas.
\newblock \emph{Science advances}, 7\penalty0 (50):\penalty0 eabl6325, 2021.

\bibitem[Blondel et~al.(2015)Blondel, Decuyper, and Krings]{blondel2015survey}
Vincent~D Blondel, Adeline Decuyper, and Gautier Krings.
\newblock A survey of results on mobile phone datasets analysis.
\newblock \emph{EPJ data science}, 4\penalty0 (1):\penalty0 10, 2015.

\bibitem[Bogomolov et~al.(2014)Bogomolov, Lepri, Staiano, Oliver, Pianesi, and
  Pentland]{BogomolovOnceCrimeCrime2014a}
Andrey Bogomolov, Bruno Lepri, Jacopo Staiano, Nuria Oliver, Fabio Pianesi, and
  Alex Pentland.
\newblock Once {Upon} a {Crime}: {Towards} {Crime} {Prediction} from
  {Demographics} and {Mobile} {Data}.
\newblock In \emph{Proceedings of the 16th {International} {Conference} on
  {Multimodal} {Interaction}}, pp.\  427--434, Istanbul Turkey, November 2014.
  ACM.
\newblock ISBN 978-1-4503-2885-2.
\newblock \doi{10.1145/2663204.2663254}.
\newblock URL \url{https://dl.acm.org/doi/10.1145/2663204.2663254}.

\bibitem[Bogomolov et~al.(2015)Bogomolov, Lepri, Staiano, Letouzé, Oliver,
  Pianesi, and Pentland]{BogomolovMovesStreetClassifying2015}
Andrey Bogomolov, Bruno Lepri, Jacopo Staiano, Emmanuel Letouzé, Nuria Oliver,
  Fabio Pianesi, and Alex Pentland.
\newblock Moves on the {Street}: {Classifying} {Crime} {Hotspots} {Using}
  {Aggregated} {Anonymized} {Data} on {People} {Dynamics}.
\newblock \emph{Big Data}, 3\penalty0 (3):\penalty0 148--158, September 2015.
\newblock ISSN 2167-6461, 2167-647X.
\newblock \doi{10.1089/big.2014.0054}.
\newblock URL \url{http://www.liebertpub.com/doi/10.1089/big.2014.0054}.

\bibitem[Bohorquez et~al.(2009)Bohorquez, Gourley, Dixon, Spagat, and
  Johnson]{bohorquez2009}
J.~Bohorquez, S.~Gourley, A.~Dixon, M.~Spagat, and N.~Johnson.
\newblock Common ecology quantifies human insurgency.
\newblock \emph{Nature}, 462:\penalty0 911--914, 2009.

\bibitem[Bonaccorsi et~al.(2020)Bonaccorsi, Pierri, Cinelli, Flori, Galeazzi,
  Porcelli, Schmidt, Valensise, Scala, Quattrociocchi,
  et~al.]{bonaccorsi2020economic}
Giovanni Bonaccorsi, Francesco Pierri, Matteo Cinelli, Andrea Flori, Alessandro
  Galeazzi, Francesco Porcelli, Ana~Lucia Schmidt, Carlo~Michele Valensise,
  Antonio Scala, Walter Quattrociocchi, et~al.
\newblock Economic and social consequences of human mobility restrictions under
  covid-19.
\newblock \emph{Proceedings of the National Academy of Sciences}, 117\penalty0
  (27):\penalty0 15530--15535, 2020.

\bibitem[Bouchard \& Malm(2016)Bouchard and
  Malm]{BouchardSocialNetworkAnalysis2016}
Martin Bouchard and Aili Malm.
\newblock \emph{Social {Network} {Analysis} and {Its} {Contribution} to
  {Research} on {Crime} and {Criminal} {Justice}}, volume~1.
\newblock Oxford University Press, November 2016.
\newblock \doi{10.1093/oxfordhb/9780199935383.013.21}.
\newblock URL
  \url{http://oxfordhandbooks.com/view/10.1093/oxfordhb/9780199935383.001.0001/oxfordhb-9780199935383-e-21}.

\bibitem[Brantingham \& Brantingham(1984)Brantingham and
  Brantingham]{BrantinghamPatternsCrime1984}
P.~J. Brantingham and P.~L. Brantingham.
\newblock \emph{Patterns in {Crime}}.
\newblock Macmillan, 1984.
\newblock ISBN 978-0-02-313520-0.

\bibitem[Brantingham \& Brantingham(2004)Brantingham and
  Brantingham]{BrantinghamComputerSimulationTool2004}
Patricia~L Brantingham and Paul~J Brantingham.
\newblock Computer {Simulation} as a {Tool} for {Environmental}
  {Criminologists}.
\newblock \emph{Security Journal}, 17\penalty0 (1):\penalty0 21--30, January
  2004.
\newblock ISSN 1743-4645.
\newblock \doi{10.1057/palgrave.sj.8340159}.
\newblock URL \url{https://doi.org/10.1057/palgrave.sj.8340159}.

\bibitem[Brazil(2022)]{brazil2022environmental}
Noli Brazil.
\newblock Environmental inequality in the neighborhood networks of urban
  mobility in us cities.
\newblock \emph{Proceedings of the National Academy of Sciences}, 119\penalty0
  (17):\penalty0 e2117776119, 2022.

\bibitem[Breetzke(2018)]{Breetzkeconcentrationurbancrime2018}
Gregory~D. Breetzke.
\newblock The concentration of urban crime in space by race: evidence from
  {South} {Africa}.
\newblock \emph{Urban Geography}, 39\penalty0 (8):\penalty0 1195--1220,
  September 2018.
\newblock ISSN 0272-3638, 1938-2847.
\newblock \doi{10.1080/02723638.2018.1440127}.
\newblock URL
  \url{https://www.tandfonline.com/doi/full/10.1080/02723638.2018.1440127}.

\bibitem[Brent(1988)]{BrentThereRoleArtificial1988}
Edward Brent.
\newblock Is {There} a {Role} for {Artificial} {Intelligence} in {Sociological}
  {Theorizing}?
\newblock \emph{The American Sociologist}, 19\penalty0 (2):\penalty0 158--166,
  1988.
\newblock ISSN 0003-1232.
\newblock URL \url{https://www.jstor.org/stable/27698419}.
\newblock Publisher: Springer.

\bibitem[Brizuela et~al.(2021)Brizuela, Garc{\'\i}a-Chan, Gutierrez~Pulido, and
  Chowell]{brizuela2021understanding}
Noel~G Brizuela, N{\'e}stor Garc{\'\i}a-Chan, Humberto Gutierrez~Pulido, and
  Gerardo Chowell.
\newblock Understanding the role of urban design in disease spreading.
\newblock \emph{Proceedings of the Royal Society A}, 477\penalty0
  (2245):\penalty0 20200524, 2021.

\bibitem[Broniatowski et~al.(2013)Broniatowski, Paul, and
  Dredze]{broniatowski2013national}
David~A Broniatowski, Michael~J Paul, and Mark Dredze.
\newblock National and local influenza surveillance through twitter: an
  analysis of the 2012-2013 influenza epidemic.
\newblock \emph{PloS one}, 8\penalty0 (12):\penalty0 e83672, 2013.

\bibitem[Browning et~al.(2021)Browning, Pinchak, and
  Calder]{BrowningHumanMobilityCrime2021}
Christopher~R. Browning, Nicolo~P. Pinchak, and Catherine~A. Calder.
\newblock Human {Mobility} and {Crime}: {Theoretical} {Approaches} and {Novel}
  {Data} {Collection} {Strategies}.
\newblock \emph{Annual Review of Criminology}, 4\penalty0 (1):\penalty0
  99--123, 2021.
\newblock \doi{10.1146/annurev-criminol-061020-021551}.
\newblock URL \url{https://doi.org/10.1146/annurev-criminol-061020-021551}.
\newblock \_eprint: https://doi.org/10.1146/annurev-criminol-061020-021551.

\bibitem[Brownstein et~al.(2009)Brownstein, Freifeld, and
  Madoff]{brownstein2009digital}
John~S Brownstein, Clark~C Freifeld, and Lawrence~C Madoff.
\newblock Digital disease detection—harnessing the web for public health
  surveillance.
\newblock \emph{The New England journal of medicine}, 360\penalty0
  (21):\penalty0 2153, 2009.

\bibitem[Brownstein et~al.(2017)Brownstein, Chu, Marathe, Marathe, Nguyen,
  Paolotti, Perra, Perrotta, Santillana, Swarup,
  et~al.]{brownstein2017combining}
John~S Brownstein, Shuyu Chu, Achla Marathe, Madhav~V Marathe, Andre~T Nguyen,
  Daniela Paolotti, Nicola Perra, Daniela Perrotta, Mauricio Santillana,
  Samarth Swarup, et~al.
\newblock Combining participatory influenza surveillance with modeling and
  forecasting: Three alternative approaches.
\newblock \emph{JMIR public health and surveillance}, 3\penalty0 (4):\penalty0
  e7344, 2017.

\bibitem[Buckee et~al.(2021)Buckee, Noor, and Sattenspiel]{buckee2021thinking}
Caroline Buckee, Abdisalan Noor, and Lisa Sattenspiel.
\newblock Thinking clearly about social aspects of infectious disease
  transmission.
\newblock \emph{Nature}, 595\penalty0 (7866):\penalty0 205--213, 2021.

\bibitem[Calabrese et~al.(2011)Calabrese, Di~Lorenzo, Liu, and
  Ratti]{calabrese2011estimating}
Francesco Calabrese, Giusy Di~Lorenzo, Liang Liu, and Carlo Ratti.
\newblock Estimating origin-destination flows using opportunistically collected
  mobile phone location data from one million users in boston metropolitan
  area.
\newblock 2011.

\bibitem[Calderoni et~al.(2021)Calderoni, Campedelli, Szekely, Paolucci, and
  Andrighetto]{CalderoniRecruitmentOrganizedCrime2021}
Francesco Calderoni, Gian~Maria Campedelli, Aron Szekely, Mario Paolucci, and
  Giulia Andrighetto.
\newblock Recruitment into {Organized} {Crime}: {An} {Agent}-{Based} {Approach}
  {Testing} the {Impact} of {Different} {Policies}.
\newblock \emph{Journal of Quantitative Criminology}, February 2021.
\newblock ISSN 0748-4518, 1573-7799.
\newblock \doi{10.1007/s10940-020-09489-z}.
\newblock URL \url{http://link.springer.com/10.1007/s10940-020-09489-z}.

\bibitem[Campbell \& Wynne(2011)Campbell and Wynne]{campbell2011introduction}
James~B Campbell and Randolph~H Wynne.
\newblock \emph{Introduction to remote sensing}.
\newblock Guilford Press, 2011.

\bibitem[Campedelli(2020)]{CampedelliWherearewe2020a}
Gian~Maria Campedelli.
\newblock Where are we? {Using} {Scopus} to map the literature at the
  intersection between artificial intelligence and research on crime.
\newblock \emph{Journal of Computational Social Science}, September 2020.
\newblock ISSN 2432-2717, 2432-2725.
\newblock \doi{10.1007/s42001-020-00082-9}.
\newblock URL \url{http://link.springer.com/10.1007/s42001-020-00082-9}.

\bibitem[Campedelli(2022{\natexlab{a}})]{CampedelliCriminologyCrossroadsComputational2022}
Gian~Maria Campedelli.
\newblock Criminology {At} the {Crossroads}? {Computational} {Perspectives}.
\newblock In \emph{Machine learning for criminology and crime research: at the
  crossroads}. Routledge, New York, 2022{\natexlab{a}}.
\newblock ISBN 978-1-00-321773-2.

\bibitem[Campedelli(2022{\natexlab{b}})]{CampedelliMachinelearningcriminology2022a}
Gian~Maria Campedelli.
\newblock \emph{Machine learning for criminology and crime research: at the
  crossroads}.
\newblock Routledge advances in criminology. Routledge, New York, NY, first
  edition edition, 2022{\natexlab{b}}.
\newblock ISBN 978-1-03-210919-0 978-1-03-210928-2.

\bibitem[Campedelli et~al.(2021)Campedelli, Bartulovic, and
  Carley]{CampedelliLearningfutureterrorist2021b}
Gian~Maria Campedelli, Mihovil Bartulovic, and Kathleen~M. Carley.
\newblock Learning future terrorist targets through temporal meta-graphs.
\newblock \emph{Scientific Reports}, 11\penalty0 (1):\penalty0 8533, April
  2021.
\newblock ISSN 2045-2322.
\newblock \doi{10.1038/s41598-021-87709-7}.
\newblock URL \url{https://www.nature.com/articles/s41598-021-87709-7}.
\newblock Number: 1 Publisher: Nature Publishing Group.

\bibitem[Caplan et~al.(2011)Caplan, Kennedy, and
  Miller]{CaplanRiskTerrainModeling2011a}
Joel~M. Caplan, Leslie~W. Kennedy, and Joel Miller.
\newblock Risk {Terrain} {Modeling}: {Brokering} {Criminological} {Theory} and
  {GIS} {Methods} for {Crime} {Forecasting}.
\newblock \emph{Justice Quarterly}, 28\penalty0 (2):\penalty0 360--381, April
  2011.
\newblock ISSN 0741-8825, 1745-9109.
\newblock \doi{10.1080/07418825.2010.486037}.
\newblock URL
  \url{http://www.tandfonline.com/doi/abs/10.1080/07418825.2010.486037}.

\bibitem[Castro et~al.(2021)Castro, Generous, Luo, Pastore~y Piontti, Martinez,
  Gomes, Osthus, Fairchild, Ziemann, Vespignani, et~al.]{castro2021using}
Lauren~A Castro, Nicholas Generous, Wei Luo, Ana Pastore~y Piontti, Kaitlyn
  Martinez, Marcelo~FC Gomes, Dave Osthus, Geoffrey Fairchild, Amanda Ziemann,
  Alessandro Vespignani, et~al.
\newblock Using heterogeneous data to identify signatures of dengue outbreaks
  at fine spatio-temporal scales across brazil.
\newblock \emph{PLoS neglected tropical diseases}, 15\penalty0 (5):\penalty0
  e0009392, 2021.

\bibitem[Chang et~al.(2021)Chang, Pierson, Koh, Gerardin, Redbird, Grusky, and
  Leskovec]{chang2021mobility}
Serina Chang, Emma Pierson, Pang~Wei Koh, Jaline Gerardin, Beth Redbird, David
  Grusky, and Jure Leskovec.
\newblock Mobility network models of covid-19 explain inequities and inform
  reopening.
\newblock \emph{Nature}, 589\penalty0 (7840):\penalty0 82--87, 2021.

\bibitem[Chen et~al.(2019)Chen, Viana, Fiore, and Sarraute]{chen2019complete}
Guangshuo Chen, Aline~Carneiro Viana, Marco Fiore, and Carlos Sarraute.
\newblock Complete trajectory reconstruction from sparse mobile phone data.
\newblock \emph{EPJ Data Science}, 8\penalty0 (1):\penalty0 30, 2019.

\bibitem[Chetty et~al.(2014)Chetty, Hendren, Kline, and Saez]{chetty2014land}
Raj Chetty, Nathaniel Hendren, Patrick Kline, and Emmanuel Saez.
\newblock Where is the land of opportunity? the geography of intergenerational
  mobility in the united states.
\newblock \emph{The Quarterly Journal of Economics}, 129\penalty0 (4):\penalty0
  1553--1623, 2014.

\bibitem[Chetty et~al.(2016)Chetty, Hendren, and Katz]{chetty2016effects}
Raj Chetty, Nathaniel Hendren, and Lawrence~F Katz.
\newblock The effects of exposure to better neighborhoods on children: New
  evidence from the moving to opportunity experiment.
\newblock \emph{American Economic Review}, 106\penalty0 (4):\penalty0 855--902,
  2016.

\bibitem[Chetty et~al.(2022a)Chetty, Jackson, Kuchler, Stroebel, Hendren,
  Fluegge, Gong, Gonzalez, Grondin, Jacob, Johnston, Koenen, Laguna-Muggenburg,
  Mudekereza, Rutter, Thor, Townsend, Ruby, Bailey, Barberá, Bhole, and
  Wernerfelt]{chetty2022a}
Raj Chetty, Matthew~O Jackson, Theresa Kuchler, Johannes Stroebel, Nathaniel
  Hendren, Robert~B Fluegge, Sara Gong, Federico Gonzalez, Armelle Grondin,
  Matthew Jacob, Drew Johnston, Martin Koenen, Eduardo Laguna-Muggenburg,
  Florian Mudekereza, Tom Rutter, Nicolaj Thor, Wilbur Townsend, Zhang Ruby,
  Mike Bailey, Pablo Barberá, Monica Bhole, and Nils Wernerfelt.
\newblock Social capital i: measurement and associations with economic
  mobility.
\newblock \emph{Nature}, 608:\penalty0 108–121, 2022a.

\bibitem[Chetty et~al.(2022b)Chetty, Jackson, Kuchler, Stroebel, Hendren,
  Fluegge, Gong, Gonzalez, Grondin, Jacob, Johnston, Koenen, Laguna-Muggenburg,
  Mudekereza, Rutter, Thor, Townsend, Ruby, Bailey, Barberá, Bhole, and
  Wernerfelt]{chetty2022b}
Raj Chetty, Matthew~O Jackson, Theresa Kuchler, Johannes Stroebel, Nathaniel
  Hendren, Robert~B Fluegge, Sara Gong, Federico Gonzalez, Armelle Grondin,
  Matthew Jacob, Drew Johnston, Martin Koenen, Eduardo Laguna-Muggenburg,
  Florian Mudekereza, Tom Rutter, Nicolaj Thor, Wilbur Townsend, Zhang Ruby,
  Mike Bailey, Pablo Barberá, Monica Bhole, and Nils Wernerfelt.
\newblock Social capital i: measurement and associations with economic
  mobility.
\newblock \emph{Nature}, 608:\penalty0 122–134, 2022b.

\bibitem[Chuang et~al.(2019)Chuang, Ben-Asher, and
  D’Orsogna]{ChuangLocalalliancesrivalries2019b}
Yao-Li Chuang, Noam Ben-Asher, and Maria~R. D’Orsogna.
\newblock Local alliances and rivalries shape near-repeat terror activity of
  al-{Qaeda}, {ISIS}, and insurgents.
\newblock \emph{Proceedings of the National Academy of Sciences}, 116\penalty0
  (42):\penalty0 20898--20903, October 2019.
\newblock \doi{10.1073/pnas.1904418116}.
\newblock URL \url{https://www.pnas.org/doi/abs/10.1073/pnas.1904418116}.
\newblock Publisher: Proceedings of the National Academy of Sciences.

\bibitem[Cohen \& Felson(1979)Cohen and Felson]{CohenSocialChangeCrime1979}
Lawrence~E. Cohen and Marcus Felson.
\newblock Social {Change} and {Crime} {Rate} {Trends}: {A} {Routine} {Activity}
  {Approach}.
\newblock \emph{American Sociological Review}, 44\penalty0 (4):\penalty0
  588--608, 1979.
\newblock ISSN 0003-1224.
\newblock \doi{10.2307/2094589}.
\newblock URL \url{https://www.jstor.org/stable/2094589}.
\newblock Publisher: [American Sociological Association, Sage Publications,
  Inc.].

\bibitem[Colizza et~al.(2007)Colizza, Barrat, Barthelemy, Valleron, and
  Vespignani]{colizza2007modeling}
Vittoria Colizza, Alain Barrat, Marc Barthelemy, Alain-Jacques Valleron, and
  Alessandro Vespignani.
\newblock Modeling the worldwide spread of pandemic influenza: baseline case
  and containment interventions.
\newblock \emph{PLoS medicine}, 4\penalty0 (1):\penalty0 e13, 2007.

\bibitem[Colizza et~al.(2021)Colizza, Grill, Mikolajczyk, Cattuto, Kucharski,
  Riley, Kendall, Lythgoe, Bonsall, Wymant, et~al.]{colizza2021time}
Vittoria Colizza, Eva Grill, Rafael Mikolajczyk, Ciro Cattuto, Adam Kucharski,
  Steven Riley, Michelle Kendall, Katrina Lythgoe, David Bonsall, Chris Wymant,
  et~al.
\newblock Time to evaluate covid-19 contact-tracing apps.
\newblock \emph{Nature Medicine}, 27\penalty0 (3):\penalty0 361--362, 2021.

\bibitem[Connolly et~al.(2021)Connolly, Keil, and Ali]{connolly2021extended}
Creighton Connolly, Roger Keil, and S~Harris Ali.
\newblock Extended urbanisation and the spatialities of infectious disease:
  Demographic change, infrastructure and governance.
\newblock \emph{Urban studies}, 58\penalty0 (2):\penalty0 245--263, 2021.

\bibitem[Cooley et~al.(2011)Cooley, Brown, Cajka, Chasteen, Ganapathi,
  Grefenstette, Hollingsworth, Lee, Levine, Wheaton, et~al.]{cooley2011role}
Philip Cooley, Shawn Brown, James Cajka, Bernadette Chasteen, Laxminarayana
  Ganapathi, John Grefenstette, Craig~R Hollingsworth, Bruce~Y Lee, Burton
  Levine, William~D Wheaton, et~al.
\newblock The role of subway travel in an influenza epidemic: a new york city
  simulation.
\newblock \emph{Journal of urban health}, 88\penalty0 (5):\penalty0 982--995,
  2011.

\bibitem[Cs{\'a}ji et~al.(2013)Cs{\'a}ji, Browet, Traag, Delvenne, Huens,
  Van~Dooren, Smoreda, and Blondel]{csaji2013exploring}
Bal{\'a}zs~Cs Cs{\'a}ji, Arnaud Browet, Vincent~A Traag, Jean-Charles Delvenne,
  Etienne Huens, Paul Van~Dooren, Zbigniew Smoreda, and Vincent~D Blondel.
\newblock Exploring the mobility of mobile phone users.
\newblock \emph{Physica A: statistical mechanics and its applications},
  392\penalty0 (6):\penalty0 1459--1473, 2013.

\bibitem[Cui et~al.(2018)Cui, Xie, and Liu]{cui2018social}
Yilan Cui, Xing Xie, and Yi~Liu.
\newblock Social media and mobility landscape: Uncovering spatial patterns of
  urban human mobility with multi source data.
\newblock \emph{Frontiers of Environmental Science \& Engineering}, 12\penalty0
  (5):\penalty0 7, 2018.

\bibitem[Dalziel et~al.(2013)Dalziel, Pourbohloul, and
  Ellner]{dalziel2013human}
Benjamin~D Dalziel, Babak Pourbohloul, and Stephen~P Ellner.
\newblock Human mobility patterns predict divergent epidemic dynamics among
  cities.
\newblock \emph{Proceedings of the Royal Society B: Biological Sciences},
  280\penalty0 (1766):\penalty0 20130763, 2013.

\bibitem[Dalziel et~al.(2018)Dalziel, Kissler, Gog, Viboud, Bj{\o}rnstad,
  Metcalf, and Grenfell]{dalziel2018urbanization}
Benjamin~D Dalziel, Stephen Kissler, Julia~R Gog, Cecile Viboud, Ottar~N
  Bj{\o}rnstad, C~Jessica~E Metcalf, and Bryan~T Grenfell.
\newblock Urbanization and humidity shape the intensity of influenza epidemics
  in us cities.
\newblock \emph{Science}, 362\penalty0 (6410):\penalty0 75--79, 2018.

\bibitem[Dass et~al.(2022)Dass, O’Brien, and Ristea]{dass2022strategies}
Sarina Dass, Daniel~T O’Brien, and Alina Ristea.
\newblock Strategies and inequities in balancing recreation and covid exposure
  when visiting green spaces.
\newblock \emph{Environment and Planning B: Urban Analytics and City Science},
  pp.\  23998083221114645, 2022.

\bibitem[De~Choudhury et~al.(2016)De~Choudhury, Sharma, and
  Kiciman]{de2016characterizing}
Munmun De~Choudhury, Sanket Sharma, and Emre Kiciman.
\newblock Characterizing dietary choices, nutrition, and language in food
  deserts via social media.
\newblock In \emph{Proceedings of the 19th acm conference on computer-supported
  cooperative work \& social computing}, pp.\  1157--1170, 2016.

\bibitem[De~Melo et~al.(2015)De~Melo, Matias, and
  Andresen]{deMeloCrimeconcentrationssimilarities2015a}
Silas~Nogueira De~Melo, Lindon~Fonseca Matias, and Martin~A. Andresen.
\newblock Crime concentrations and similarities in spatial crime patterns
  in a {Brazilian} context.
\newblock \emph{Applied Geography}, 62:\penalty0 314--324, August 2015.
\newblock ISSN 0143-6228.
\newblock \doi{10.1016/j.apgeog.2015.05.012}.
\newblock URL
  \url{https://www.sciencedirect.com/science/article/pii/S0143622815001320}.

\bibitem[De~Nadai et~al.(2020)De~Nadai, Xu, Letouzé, González, and
  Lepri]{DeNadaiSocioeconomicbuiltenvironment2020}
Marco De~Nadai, Yanyan Xu, Emmanuel Letouzé, Marta~C. González, and Bruno
  Lepri.
\newblock Socio-economic, built environment, and mobility conditions associated
  with crime: a study of multiple cities.
\newblock \emph{Scientific Reports}, 10\penalty0 (1):\penalty0 13871, December
  2020.
\newblock ISSN 2045-2322.
\newblock \doi{10.1038/s41598-020-70808-2}.
\newblock URL \url{http://www.nature.com/articles/s41598-020-70808-2}.

\bibitem[Deville et~al.(2014)Deville, Linard, Martin, Gilbert, Stevens,
  Gaughan, Blondel, and Tatem]{deville2014dynamic}
Pierre Deville, Catherine Linard, Samuel Martin, Marius Gilbert, Forrest~R
  Stevens, Andrea~E Gaughan, Vincent~D Blondel, and Andrew~J Tatem.
\newblock Dynamic population mapping using mobile phone data.
\newblock \emph{Proceedings of the National Academy of Sciences}, 111\penalty0
  (45):\penalty0 15888--15893, 2014.

\bibitem[Di~Clemente et~al.(2018)Di~Clemente, Luengo-Oroz, Travizano, Xu,
  Vaitla, and Gonz{\'a}lez]{di2018sequences}
Riccardo Di~Clemente, Miguel Luengo-Oroz, Matias Travizano, Sharon Xu, Bapu
  Vaitla, and Marta~C Gonz{\'a}lez.
\newblock Sequences of purchases in credit card data reveal lifestyles in urban
  populations.
\newblock \emph{Nature communications}, 9\penalty0 (1):\penalty0 1--8, 2018.

\bibitem[Dister et~al.(1997)Dister, Fish, Bros, Frank, and
  Wood]{dister1997landscape}
Sheri~W Dister, Durland Fish, Shannon~M Bros, Denise~H Frank, and Byron~L Wood.
\newblock Landscape characterization of peridomestic risk for lyme disease
  using satellite imagery.
\newblock \emph{The American journal of tropical medicine and hygiene},
  57\penalty0 (6):\penalty0 687--692, 1997.

\bibitem[Dong et~al.(2017)Dong, Suhara, Bozkaya, Singh, Lepri, and
  Pentland]{dong2017social}
Xiaowen Dong, Yoshihiko Suhara, Bur{\c{c}}in Bozkaya, Vivek~K Singh, Bruno
  Lepri, and Alex~‘Sandy’ Pentland.
\newblock Social bridges in urban purchase behavior.
\newblock \emph{ACM Transactions on Intelligent Systems and Technology (TIST)},
  9\penalty0 (3):\penalty0 1--29, 2017.

\bibitem[Dong et~al.(2020)Dong, Morales, Jahani, Moro, Lepri, Bozkaya,
  Sarraute, Bar-Yam, and Pentland]{dong2020}
Xiaowen Dong, Alfredo~J Morales, Eaman Jahani, Esteban Moro, Bruno Lepri,
  Burcin Bozkaya, Carlos Sarraute, Yaneer Bar-Yam, and Alex Pentland.
\newblock Segregated interactions in urban and online space.
\newblock \emph{EPJ Data Science}, 9\penalty0 (20), 2020.

\bibitem[D'Orsogna \& Perc(2015)D'Orsogna and
  Perc]{DOrsognaStatisticalphysicscrime2015b}
Maria~R. D'Orsogna and Matjaž Perc.
\newblock Statistical physics of crime: {A} review.
\newblock \emph{Physics of Life Reviews}, 12:\penalty0 1--21, March 2015.
\newblock ISSN 15710645.
\newblock \doi{10.1016/j.plrev.2014.11.001}.
\newblock URL
  \url{https://linkinghub.elsevier.com/retrieve/pii/S1571064514001730}.

\bibitem[Dressel \& Farid(2018)Dressel and
  Farid]{Dresselaccuracyfairnesslimits2018c}
Julia Dressel and Hany Farid.
\newblock The accuracy, fairness, and limits of predicting recidivism.
\newblock \emph{Science Advances}, 4\penalty0 (1):\penalty0 eaao5580, January
  2018.
\newblock ISSN 2375-2548.
\newblock \doi{10.1126/sciadv.aao5580}.
\newblock URL \url{https://www.science.org/doi/10.1126/sciadv.aao5580}.

\bibitem[Duwe \& Kim(2017)Duwe and Kim]{DuweOutOldNew2017}
Grant Duwe and KiDeuk Kim.
\newblock Out {With} the {Old} and in {With} the {New}? {An} {Empirical}
  {Comparison} of {Supervised} {Learning} {Algorithms} to {Predict}
  {Recidivism}.
\newblock \emph{Criminal Justice Policy Review}, 28\penalty0 (6):\penalty0
  570--600, July 2017.
\newblock ISSN 0887-4034, 1552-3586.
\newblock \doi{10.1177/0887403415604899}.
\newblock URL \url{http://journals.sagepub.com/doi/10.1177/0887403415604899}.

\bibitem[Duxbury \& Haynie(2020)Duxbury and
  Haynie]{Duxburyresponsivenesscriminalnetworks2020}
Scott Duxbury and Dana~L. Haynie.
\newblock The responsiveness of criminal networks to intentional attacks:
  {Disrupting} darknet drug trade.
\newblock \emph{PLOS ONE}, 15\penalty0 (9):\penalty0 e0238019, September 2020.
\newblock ISSN 1932-6203.
\newblock \doi{10.1371/journal.pone.0238019}.
\newblock URL
  \url{https://journals.plos.org/plosone/article?id=10.1371/journal.pone.0238019}.
\newblock Publisher: Public Library of Science.

\bibitem[Duxbury \& Haynie(2018)Duxbury and
  Haynie]{DuxburyNetworkStructureOpioid2018}
Scott~W. Duxbury and Dana~L. Haynie.
\newblock The {Network} {Structure} of {Opioid} {Distribution} on a {Darknet}
  {Cryptomarket}.
\newblock \emph{Journal of Quantitative Criminology}, 34\penalty0 (4):\penalty0
  921--941, December 2018.
\newblock ISSN 1573-7799.
\newblock \doi{10.1007/s10940-017-9359-4}.
\newblock URL \url{https://doi.org/10.1007/s10940-017-9359-4}.

\bibitem[Eagle et~al.(2010)Eagle, Macy, and Claxton]{eagle2010}
Nathan Eagle, Michael Macy, and Rob Claxton.
\newblock Network diversity and economic development.
\newblock \emph{Science}, 328\penalty0 (5981):\penalty0 1029--1031, 2010.

\bibitem[Eck \& Weisburd(1995)Eck and Weisburd]{EckCrimePlaceCrime1995}
John~E. Eck and David Weisburd (eds.).
\newblock \emph{Crime and Place: Crime Prevention Studies}.
\newblock Willow Tree Pr, Monsey, NY, December 1995.
\newblock ISBN 978-1-881798-05-7.

\bibitem[Epstein(2013)]{epstein2013collaborations}
Jennifer~A Epstein.
\newblock Collaborations between public health and computer science: a path
  worth pursuing.
\newblock \emph{Am J Public Health Res}, 1\penalty0 (7):\penalty0 166--170,
  2013.

\bibitem[Eubank et~al.(2004)Eubank, Guclu, Anil~Kumar, Marathe, Srinivasan,
  Toroczkai, and Wang]{eubank2004modelling}
Stephen Eubank, Hasan Guclu, VS~Anil~Kumar, Madhav~V Marathe, Aravind
  Srinivasan, Zoltan Toroczkai, and Nan Wang.
\newblock Modelling disease outbreaks in realistic urban social networks.
\newblock \emph{Nature}, 429\penalty0 (6988):\penalty0 180--184, 2004.

\bibitem[Fan et~al.(2022)Fan, Su, Sun, Noyman, Zhang, Pentland, and
  Moro]{fan2022diversity}
Zhuangyuan Fan, Tianyu Su, Maoran Sun, Ariel Noyman, Fan Zhang, Alex~Sandy
  Pentland, and Esteban Moro.
\newblock Diversity beyond density: experienced social mixing of urban streets.
\newblock \emph{arXiv preprint arXiv:2209.07041}, 2022.

\bibitem[Faust \& Tita(2019)Faust and Tita]{FaustSocialNetworksCrime2019}
Katherine Faust and George~E. Tita.
\newblock Social {Networks} and {Crime}: {Pitfalls} and {Promises} for
  {Advancing} the {Field}.
\newblock \emph{Annual Review of Criminology}, 2\penalty0 (1):\penalty0
  99--122, January 2019.
\newblock ISSN 2572-4568.
\newblock \doi{10.1146/annurev-criminol-011518-024701}.
\newblock URL
  \url{https://www.annualreviews.org/doi/10.1146/annurev-criminol-011518-024701}.

\bibitem[Favarin(2018)]{FavarinThismustbe2018a}
Serena Favarin.
\newblock This must be the place (to commit a crime). {Testing} the law of
  crime concentration in {Milan}, {Italy}.
\newblock \emph{European Journal of Criminology}, 15\penalty0 (6):\penalty0
  702--729, November 2018.
\newblock ISSN 1477-3708, 1741-2609.
\newblock \doi{10.1177/1477370818757700}.
\newblock URL \url{http://journals.sagepub.com/doi/10.1177/1477370818757700}.

\bibitem[Felson \& Clarke(1998)Felson and Clarke]{felson1998}
M.~Felson and R.V. Clarke.
\newblock Opportunity makes the thief: Practical theory for crime prevention.
\newblock \emph{Police Research Series, Paper 98}, 1998.

\bibitem[Ferguson et~al.(2006)Ferguson, Cummings, Fraser, Cajka, Cooley, and
  Burke]{ferguson2006strategies}
Neil~M Ferguson, Derek~AT Cummings, Christophe Fraser, James~C Cajka, Philip~C
  Cooley, and Donald~S Burke.
\newblock Strategies for mitigating an influenza pandemic.
\newblock \emph{Nature}, 442\penalty0 (7101):\penalty0 448--452, 2006.

\bibitem[Florida(2017)]{florida2017new}
Richard Florida.
\newblock \emph{The new urban crisis: How our cities are increasing inequality,
  deepening segregation, and failing the middle class-and what we can do about
  it}.
\newblock Hachette UK, 2017.

\bibitem[Ford et~al.(2009)Ford, Colwell, Rose, Morse, Rogers, and
  Yates]{ford2009using}
Timothy~E Ford, Rita~R Colwell, Joan~B Rose, Stephen~S Morse, David~J Rogers,
  and Terry~L Yates.
\newblock Using satellite images of environmental changes to predict infectious
  disease outbreaks.
\newblock \emph{Emerging infectious diseases}, 15\penalty0 (9):\penalty0 1341,
  2009.

\bibitem[Fraser et~al.(2022)Fraser, Van~Woert, Olivieri, Baron, Buckley, and
  Lalli]{fraser4076776cycling}
Timothy Fraser, Katherine Van~Woert, Sophia Olivieri, Jonathan Baron, Katelyn
  Buckley, and Pamela Lalli.
\newblock Cycling cities: Measuring transportation equity in bikeshare
  networks.
\newblock \emph{Available at SSRN 4076776}, 2022.

\bibitem[Freifeld et~al.(2008)Freifeld, Mandl, Reis, and
  Brownstein]{freifeld2008healthmap}
Clark~C Freifeld, Kenneth~D Mandl, Ben~Y Reis, and John~S Brownstein.
\newblock Healthmap: global infectious disease monitoring through automated
  classification and visualization of internet media reports.
\newblock \emph{Journal of the American Medical Informatics Association},
  15\penalty0 (2):\penalty0 150--157, 2008.

\bibitem[Fumanelli et~al.(2012)Fumanelli, Ajelli, Manfredi, Vespignani, and
  Merler]{fumanelli2012inferring}
Laura Fumanelli, Marco Ajelli, Piero Manfredi, Alessandro Vespignani, and
  Stefano Merler.
\newblock Inferring the structure of social contacts from demographic data in
  the analysis of infectious diseases spread.
\newblock 2012.

\bibitem[Funk et~al.(2010)Funk, Salath{\'e}, and Jansen]{funk2010modelling}
Sebastian Funk, Marcel Salath{\'e}, and Vincent~AA Jansen.
\newblock Modelling the influence of human behaviour on the spread of
  infectious diseases: a review.
\newblock \emph{Journal of the Royal Society Interface}, 7\penalty0
  (50):\penalty0 1247--1256, 2010.

\bibitem[Gasco et~al.(2020)Gasco, Schifanella, Aiello, Quercia, Asensio, and
  de~Arcas]{gasco2020social}
Luis Gasco, Rossano Schifanella, Luca~Maria Aiello, Daniele Quercia, Cesar
  Asensio, and Guillermo de~Arcas.
\newblock Social media and open data to quantify the effects of noise on
  health.
\newblock \emph{Frontiers in Sustainable Cities}, 2:\penalty0 41, 2020.

\bibitem[Gauvin et~al.(2021)Gauvin, Bajardi, Pepe, Lake, Privitera, and
  Tizzoni]{gauvin2021socio}
Laetitia Gauvin, Paolo Bajardi, Emanuele Pepe, Brennan Lake, Filippo Privitera,
  and Michele Tizzoni.
\newblock Socio-economic determinants of mobility responses during the first
  wave of covid-19 in italy: from provinces to neighbourhoods.
\newblock \emph{Journal of The Royal Society Interface}, 18\penalty0
  (181):\penalty0 20210092, 2021.

\bibitem[Gavens et~al.(2018)Gavens, Holmes, B{\"u}hringer, McLeod, Neumann,
  Lingford-Hughes, Hock, and Meier]{gavens2018interdisciplinary}
Lucy Gavens, Joanne Holmes, Gerhard B{\"u}hringer, Jordache McLeod, Maike
  Neumann, A~Lingford-Hughes, Emma~S Hock, and Petra~Sylvia Meier.
\newblock Interdisciplinary working in public health research: a proposed good
  practice checklist.
\newblock \emph{Journal of Public Health}, 40\penalty0 (1):\penalty0 175--182,
  2018.

\bibitem[Gesualdo et~al.(2013)Gesualdo, Stilo, Agricola, Gonfiantini, Pandolfi,
  Velardi, and Tozzi]{gesualdo2013influenza}
Francesco Gesualdo, Giovanni Stilo, Eleonora Agricola, Michaela~V Gonfiantini,
  Elisabetta Pandolfi, Paola Velardi, and Alberto~E Tozzi.
\newblock Influenza-like illness surveillance on twitter through automated
  learning of na{\"\i}ve language.
\newblock \emph{PLoS One}, 8\penalty0 (12):\penalty0 e82489, 2013.

\bibitem[Ginsberg et~al.(2009)Ginsberg, Mohebbi, Patel, Brammer, Smolinski, and
  Brilliant]{ginsberg2009detecting}
Jeremy Ginsberg, Matthew~H Mohebbi, Rajan~S Patel, Lynnette Brammer, Mark~S
  Smolinski, and Larry Brilliant.
\newblock Detecting influenza epidemics using search engine query data.
\newblock \emph{Nature}, 457\penalty0 (7232):\penalty0 1012--1014, 2009.

\bibitem[Gladstone et~al.(2019)Gladstone, Matz, and Lemaire]{gladstone2019can}
Joe~J Gladstone, Sandra~C Matz, and Alain Lemaire.
\newblock Can psychological traits be inferred from spending? evidence from
  transaction data.
\newblock \emph{Psychological science}, 30\penalty0 (7):\penalty0 1087--1096,
  2019.

\bibitem[Glaeser(2012)]{glaeser2012triumph}
Edward Glaeser.
\newblock \emph{Triumph of the city: How our greatest invention makes us
  richer, smarter, greener, healthier, and happier}.
\newblock Penguin, 2012.

\bibitem[Glaeser et~al.(2001)Glaeser, Kolko, and Saiz]{glaeser2001}
E.L. Glaeser, J.~Kolko, and A.~Saiz.
\newblock Consumer city.
\newblock \emph{Journal of Economic Geography}, 1:\penalty0 27--50, 2001.

\bibitem[Glodeanu et~al.(2021)Glodeanu, Gull{\'o}n, and
  Bilal]{glodeanu2021social}
Adri{\'a}n Glodeanu, Pedro Gull{\'o}n, and Usama Bilal.
\newblock Social inequalities in mobility during and following the covid-19
  associated lockdown of the madrid metropolitan area in spain.
\newblock \emph{Health \& Place}, 70:\penalty0 102580, 2021.

\bibitem[Gonzalez et~al.(2008)Gonzalez, Hidalgo, and
  Barabasi]{gonzalez2008understanding}
Marta~C Gonzalez, Cesar~A Hidalgo, and Albert-Laszlo Barabasi.
\newblock Understanding individual human mobility patterns.
\newblock \emph{nature}, 453\penalty0 (7196):\penalty0 779--782, 2008.

\bibitem[Gore et~al.(2015)Gore, Diallo, and Padilla]{gore2015you}
Ross~Joseph Gore, Saikou Diallo, and Jose Padilla.
\newblock You are what you tweet: connecting the geographic variation in
  america’s obesity rate to twitter content.
\newblock \emph{PloS one}, 10\penalty0 (9):\penalty0 e0133505, 2015.

\bibitem[Gozzi et~al.(2021)Gozzi, Tizzoni, Chinazzi, Ferres, Vespignani, and
  Perra]{gozzi2021estimating}
Nicol{\`o} Gozzi, Michele Tizzoni, Matteo Chinazzi, Leo Ferres, Alessandro
  Vespignani, and Nicola Perra.
\newblock Estimating the effect of social inequalities on the mitigation of
  covid-19 across communities in santiago de chile.
\newblock \emph{Nature communications}, 12\penalty0 (1):\penalty0 1--9, 2021.

\bibitem[Graif et~al.(2014)Graif, Gladfelter, and
  Matthews]{GraifUrbanPovertyNeighborhood2014}
Corina Graif, Andrew~S. Gladfelter, and Stephen~A. Matthews.
\newblock Urban {Poverty} and {Neighborhood} {Effects} on {Crime}:
  {Incorporating} {Spatial} and {Network} {Perspectives}: {Neighborhood}
  {Poverty}, {Crime}, and {Exposure} {Networks}.
\newblock \emph{Sociology Compass}, 8\penalty0 (9):\penalty0 1140--1155,
  September 2014.
\newblock ISSN 17519020.
\newblock \doi{10.1111/soc4.12199}.
\newblock URL \url{https://onlinelibrary.wiley.com/doi/10.1111/soc4.12199}.

\bibitem[Graif et~al.(2021)Graif, Freelin, Kuo, Wang, Li, and
  Kifer]{GraifNetworkSpilloversNeighborhood2021}
Corina Graif, Brittany~N. Freelin, Yu-Hsuan Kuo, Hongjian Wang, Zhenhui Li, and
  Daniel Kifer.
\newblock Network {Spillovers} and {Neighborhood} {Crime}: {A} {Computational}
  {Statistics} {Analysis} of {Employment}-{Based} {Networks} of
  {Neighborhoods}.
\newblock \emph{Justice Quarterly}, 38\penalty0 (2):\penalty0 344--374,
  February 2021.
\newblock ISSN 0741-8825, 1745-9109.
\newblock \doi{10.1080/07418825.2019.1602160}.
\newblock URL
  \url{https://www.tandfonline.com/doi/full/10.1080/07418825.2019.1602160}.

\bibitem[Grassly \& Fraser(2008)Grassly and Fraser]{grassly2008mathematical}
Nicholas~C Grassly and Christophe Fraser.
\newblock Mathematical models of infectious disease transmission.
\newblock \emph{Nature Reviews Microbiology}, 6\penalty0 (6):\penalty0
  477--487, 2008.

\bibitem[Green et~al.(2017)Green, Horel, and
  Papachristos]{GreenModelingContagionSocial2017a}
Ben Green, Thibaut Horel, and Andrew~V. Papachristos.
\newblock Modeling {Contagion} {Through} {Social} {Networks} to {Explain} and
  {Predict} {Gunshot} {Violence} in {Chicago}, 2006 to 2014.
\newblock \emph{JAMA Internal Medicine}, 177\penalty0 (3):\penalty0 326--333,
  March 2017.
\newblock ISSN 2168-6106.
\newblock \doi{10.1001/jamainternmed.2016.8245}.
\newblock URL \url{https://doi.org/10.1001/jamainternmed.2016.8245}.

\bibitem[Groff \& Mazerolle(2008)Groff and
  Mazerolle]{GroffSimulatedexperimentstheir2008}
Elizabeth Groff and Lorraine Mazerolle.
\newblock Simulated experiments and their potential role in criminology and
  criminal justice.
\newblock \emph{Journal of Experimental Criminology}, 4\penalty0 (3):\penalty0
  187, August 2008.
\newblock ISSN 1572-8315.
\newblock \doi{10.1007/s11292-008-9058-0}.
\newblock URL \url{https://doi.org/10.1007/s11292-008-9058-0}.

\bibitem[Groff et~al.(2019)Groff, Johnson, and
  Thornton]{GroffStateArtAgentBased2019c}
Elizabeth Groff, Shane~D. Johnson, and Amy Thornton.
\newblock State of the {Art} in {Agent}-{Based} {Modeling} of {Urban} {Crime}:
  {An} {Overview}.
\newblock \emph{Journal of Quantitative Criminology}, 35\penalty0 (1):\penalty0
  155--193, March 2019.
\newblock ISSN 0748-4518, 1573-7799.
\newblock \doi{10.1007/s10940-018-9376-y}.
\newblock URL \url{http://link.springer.com/10.1007/s10940-018-9376-y}.

\bibitem[Guan et~al.(2022)Guan, Mofaz, Qian, Patalon, Shmueli, Yamin, and
  Brandeau]{guan2022higher}
Grace Guan, Merav Mofaz, Gary Qian, Tal Patalon, Erez Shmueli, Dan Yamin, and
  Margaret~L Brandeau.
\newblock Higher sensitivity monitoring of reactions to covid-19 vaccination
  using smartwatches.
\newblock \emph{NPJ Digital Medicine}, 5\penalty0 (1):\penalty0 1--9, 2022.

\bibitem[G{\"u}ndo{\u{g}}du et~al.(2019)G{\"u}ndo{\u{g}}du, Panzarasa, Oliver,
  and Lepri]{gundougdu2019bridging}
Didem G{\"u}ndo{\u{g}}du, Pietro Panzarasa, Nuria Oliver, and Bruno Lepri.
\newblock The bridging and bonding structures of place-centric networks:
  Evidence from a developing country.
\newblock \emph{PloS one}, 14\penalty0 (9):\penalty0 e0221148, 2019.

\bibitem[Haleem et~al.(2021)Haleem, Do~Lee, Ellison, and
  Bannister]{HaleemExposedPopulationViolent2021}
Muhammad~Salman Haleem, Won Do~Lee, Mark Ellison, and Jon Bannister.
\newblock The ‘{Exposed}’ {Population}, {Violent} {Crime} in {Public}
  {Space} and the {Night}-time {Economy} in {Manchester}, {UK}.
\newblock \emph{European Journal on Criminal Policy and Research}, 27\penalty0
  (3):\penalty0 335--352, September 2021.
\newblock ISSN 1572-9869.
\newblock \doi{10.1007/s10610-020-09452-5}.
\newblock URL \url{https://doi.org/10.1007/s10610-020-09452-5}.

\bibitem[Harris et~al.(2014)Harris, Mansour, Choucair, Olson, Nissen, and
  Bhatt]{harris2014health}
Jenine~K Harris, Raed Mansour, Bechara Choucair, Joe Olson, Cory Nissen, and
  Jay Bhatt.
\newblock Health department use of social media to identify foodborne
  illness—chicago, illinois, 2013--2014.
\newblock \emph{Morbidity and Mortality Weekly Report}, 63\penalty0
  (32):\penalty0 681, 2014.

\bibitem[Harrison et~al.(2014)Harrison, Jorder, Stern, Stavinsky, Reddy,
  Hanson, Waechter, Lowe, Gravano, and Balter]{harrison2014using}
Cassandra Harrison, Mohip Jorder, Henri Stern, Faina Stavinsky, Vasudha Reddy,
  Heather Hanson, HaeNa Waechter, Luther Lowe, Luis Gravano, and Sharon Balter.
\newblock Using online reviews by restaurant patrons to identify unreported
  cases of foodborne illness—new york city, 2012--2013.
\newblock \emph{Morbidity and Mortality Weekly Report}, 63\penalty0
  (20):\penalty0 441, 2014.

\bibitem[Hayward \& Maas(2021)Hayward and
  Maas]{HaywardArtificialintelligencecrime2021}
Keith~J Hayward and Matthijs~M Maas.
\newblock Artificial intelligence and crime: {A} primer for criminologists.
\newblock \emph{Crime, Media, Culture: An International Journal}, 17\penalty0
  (2):\penalty0 209--233, August 2021.
\newblock ISSN 1741-6590, 1741-6604.
\newblock \doi{10.1177/1741659020917434}.
\newblock URL \url{http://journals.sagepub.com/doi/10.1177/1741659020917434}.

\bibitem[Hedefalk \& Dribe(2020)Hedefalk and Dribe]{hedefak2020}
Finn Hedefalk and Martin Dribe.
\newblock The social context of nearest neighbors shapes educational attainment
  regardless of class origin.
\newblock \emph{Proc. Natl Acad. Sci.}, 117\penalty0 (26):\penalty0
  14918--14925, 2020.

\bibitem[Hilman et~al.(2022)Hilman, I{\~n}iguez, and
  Karsai]{hilman2022socioeconomic}
Rafiazka~Millanida Hilman, Gerardo I{\~n}iguez, and M{\'a}rton Karsai.
\newblock Socioeconomic biases in urban mixing patterns of us metropolitan
  areas.
\newblock \emph{EPJ data science}, 11\penalty0 (1):\penalty0 32, 2022.

\bibitem[Hipp et~al.(2019)Hipp, Bates, Lichman, and
  Smyth]{HippUsingSocialMedia2019}
John~R. Hipp, Christopher Bates, Moshe Lichman, and Padhraic Smyth.
\newblock Using {Social} {Media} to {Measure} {Temporal} {Ambient}
  {Population}: {Does} it {Help} {Explain} {Local} {Crime} {Rates}?
\newblock \emph{Justice Quarterly}, 36\penalty0 (4):\penalty0 718--748, June
  2019.
\newblock ISSN 0741-8825, 1745-9109.
\newblock \doi{10.1080/07418825.2018.1445276}.
\newblock URL
  \url{https://www.tandfonline.com/doi/full/10.1080/07418825.2018.1445276}.

\bibitem[Holbrook et~al.(2021)Holbrook, Loeffler, Flaxman, and
  Suchard]{HolbrookScalableBayesianinference2021}
Andrew~J. Holbrook, Charles~E. Loeffler, Seth~R. Flaxman, and Marc~A. Suchard.
\newblock Scalable {Bayesian} inference for self-excitatory stochastic
  processes applied to big {American} gunfire data.
\newblock \emph{Statistics and Computing}, 31\penalty0 (1):\penalty0 4, January
  2021.
\newblock ISSN 0960-3174, 1573-1375.
\newblock \doi{10.1007/s11222-020-09980-4}.
\newblock URL \url{http://link.springer.com/10.1007/s11222-020-09980-4}.

\bibitem[Hong et~al.(2021)Hong, Bonczak, Gupta, and
  Kontokosta]{hong2021measuring}
Boyeong Hong, Bartosz~J Bonczak, Arpit Gupta, and Constantine~E Kontokosta.
\newblock Measuring inequality in community resilience to natural disasters
  using large-scale mobility data.
\newblock \emph{Nature communications}, 12\penalty0 (1):\penalty0 1--9, 2021.

\bibitem[Hunter et~al.(2021)Hunter, Garcia, de~Sa, Zapata-Diomedi, Millett,
  Woodcock, Pentland, and Moro]{hunter2021effect}
Ruth~F Hunter, Leandro Garcia, Thiago~Herick de~Sa, Belen Zapata-Diomedi,
  Christopher Millett, James Woodcock, Alex’Sandy’ Pentland, and Esteban
  Moro.
\newblock Effect of covid-19 response policies on walking behavior in us
  cities.
\newblock \emph{Nature communications}, 12\penalty0 (1):\penalty0 1--9, 2021.

\bibitem[Icove(1986)]{IcoveAutomatedCrimeProfiling1986}
David~J. Icove.
\newblock Automated {Crime} {Profiling}.
\newblock \emph{FBI Law Enforcement Bulletin}, 55\penalty0 (27), 1986.

\bibitem[Isella et~al.(2011)Isella, Romano, Barrat, Cattuto, Colizza, Van~den
  Broeck, Gesualdo, Pandolfi, Rav{\`a}, Rizzo, et~al.]{isella2011close}
Lorenzo Isella, Mariateresa Romano, Alain Barrat, Ciro Cattuto, Vittoria
  Colizza, Wouter Van~den Broeck, Francesco Gesualdo, Elisabetta Pandolfi,
  Lucilla Rav{\`a}, Caterina Rizzo, et~al.
\newblock Close encounters in a pediatric ward: measuring face-to-face
  proximity and mixing patterns with wearable sensors.
\newblock \emph{PloS one}, 6\penalty0 (2):\penalty0 e17144, 2011.

\bibitem[Jacobs(1961)]{jacobs1961}
J.~Jacobs.
\newblock \emph{The death and life of great American cities}.
\newblock New York: Random House, 1961.

\bibitem[J{\"a}rv et~al.(2015)J{\"a}rv, M{\"u}{\"u}risepp, Ahas, Derudder, and
  Witlox]{jarv2015ethnic}
Olle J{\"a}rv, Kerli M{\"u}{\"u}risepp, Rein Ahas, Ben Derudder, and Frank
  Witlox.
\newblock Ethnic differences in activity spaces as a characteristic of
  segregation: A study based on mobile phone usage in tallinn, estonia.
\newblock \emph{Urban Studies}, 52\penalty0 (14):\penalty0 2680--2698, 2015.

\bibitem[Jean et~al.(2016)Jean, Burke, Xie, Davis, Lobell, and
  Ermon]{jean2016combining}
Neal Jean, Marshall Burke, Michael Xie, W~Matthew Davis, David~B Lobell, and
  Stefano Ermon.
\newblock Combining satellite imagery and machine learning to predict poverty.
\newblock \emph{Science}, 353\penalty0 (6301):\penalty0 790--794, 2016.

\bibitem[Johnson(2010)]{Johnsonbriefhistoryanalysis2010b}
Shane~D. Johnson.
\newblock A brief history of the analysis of crime concentration.
\newblock \emph{European Journal of Applied Mathematics}, 21\penalty0
  (4-5):\penalty0 349--370, October 2010.
\newblock ISSN 1469-4425, 0956-7925.
\newblock \doi{10.1017/S0956792510000082}.
\newblock URL
  \url{https://www.cambridge.org/core/journals/european-journal-of-applied-mathematics/article/abs/brief-history-of-the-analysis-of-crime-concentration/4DD75FAB576E54B318DE883E55E6CA6A}.
\newblock Publisher: Cambridge University Press.

\bibitem[Kermack \& McKendrick(1927)Kermack and
  McKendrick]{kermack1927contribution}
William~Ogilvy Kermack and Anderson~G McKendrick.
\newblock A contribution to the mathematical theory of epidemics.
\newblock \emph{Proceedings of the royal society of london. Series A,
  Containing papers of a mathematical and physical character}, 115\penalty0
  (772):\penalty0 700--721, 1927.

\bibitem[Kertész \& Wachs(2021)Kertész and
  Wachs]{KerteszComplexityscienceapproach2021}
János Kertész and Johannes Wachs.
\newblock Complexity science approach to economic crime.
\newblock \emph{Nature Reviews Physics}, 3\penalty0 (2):\penalty0 70--71,
  February 2021.
\newblock ISSN 2522-5820.
\newblock \doi{10.1038/s42254-020-0238-9}.
\newblock URL \url{https://www.nature.com/articles/s42254-020-0238-9}.
\newblock Bandiera\_abtest: a Cg\_type: Nature Research Journals Number: 2
  Primary\_atype: Comments \& Opinion Publisher: Nature Publishing Group
  Subject\_term: Complex networks Subject\_term\_id: complex-networks.

\bibitem[Kishore(2021)]{kishore2021mobility}
Nishant Kishore.
\newblock Mobility data as a proxy for epidemic measures.
\newblock \emph{Nature Computational Science}, 1\penalty0 (9):\penalty0
  567--568, 2021.

\bibitem[Koch(2011)]{koch2011disease}
Tom Koch.
\newblock \emph{Disease maps: epidemics on the ground}.
\newblock University of Chicago Press, 2011.

\bibitem[Kraemer et~al.(2020)Kraemer, Sadilek, Zhang, Marchal, Tuli, Cohn,
  Hswen, Perkins, Smith, Reiner, et~al.]{kraemer2020mapping}
Moritz~UG Kraemer, Adam Sadilek, Qian Zhang, Nahema~A Marchal, Gaurav Tuli,
  Emily~L Cohn, Yulin Hswen, T~Alex Perkins, David~L Smith, Robert~C Reiner,
  et~al.
\newblock Mapping global variation in human mobility.
\newblock \emph{Nature Human Behaviour}, 4\penalty0 (8):\penalty0 800--810,
  2020.

\bibitem[Lampos et~al.(2010)Lampos, Bie, and Cristianini]{lampos2010flu}
Vasileios Lampos, Tijl~De Bie, and Nello Cristianini.
\newblock Flu detector-tracking epidemics on twitter.
\newblock In \emph{Joint European conference on machine learning and knowledge
  discovery in databases}, pp.\  599--602. Springer, 2010.

\bibitem[Lampos et~al.(2015)Lampos, Miller, Crossan, and
  Stefansen]{lampos2015advances}
Vasileios Lampos, Andrew~C Miller, Steve Crossan, and Christian Stefansen.
\newblock Advances in nowcasting influenza-like illness rates using search
  query logs.
\newblock \emph{Scientific reports}, 5\penalty0 (1):\penalty0 1--10, 2015.

\bibitem[Lampos et~al.(2021)Lampos, Majumder, Yom-Tov, Edelstein, Moura,
  Hamada, Rangaka, McKendry, and Cox]{lampos2021tracking}
Vasileios Lampos, Maimuna~S Majumder, Elad Yom-Tov, Michael Edelstein, Simon
  Moura, Yohhei Hamada, Molebogeng~X Rangaka, Rachel~A McKendry, and Ingemar~J
  Cox.
\newblock Tracking covid-19 using online search.
\newblock \emph{NPJ digital medicine}, 4\penalty0 (1):\penalty0 1--11, 2021.

\bibitem[Lazer et~al.(2009)Lazer, Pentland, Adamic, Aral, Barabasi, Brewer,
  Christakis, Contractor, Fowler, Gutmann, Jebara, King, Macy, Roy, and
  Van~Alstyne]{lazer2009}
D.~Lazer, A.~Pentland, L.~Adamic, S.~Aral, A.~L. Barabasi, D.~Brewer,
  N.~Christakis, N.~Contractor, J.~Fowler, M.~Gutmann, T.~Jebara, G.~King,
  M.~Macy, D.~Roy, and M.~Van~Alstyne.
\newblock Computational social science.
\newblock \emph{Science}, 323\penalty0 (5915):\penalty0 721, 2009.

\bibitem[Lazer et~al.(2020)Lazer, Pentland, Watts, Aral, Athey, Contractor,
  Freelon, Gonzalez-Bailon, King, Margetts, Nelson, Salganik, Strohmaier,
  Vespignani, and Wagner]{lazer2020}
D.~Lazer, A.~Pentland, D.~Watts, S.~Aral, S.~Athey, N.~Contractor, D.~Freelon,
  S.~Gonzalez-Bailon, G.~King, H.~Margetts, A.~Nelson, M.~J. Salganik,
  M.~Strohmaier, A.~Vespignani, and C.~Wagner.
\newblock Computational social science: Obstacles and opportunities.
\newblock \emph{Science}, 369\penalty0 (6507):\penalty0 1060--1062, 2020.

\bibitem[Lazer et~al.(2014{\natexlab{a}})Lazer, Kennedy, King, and
  Vespignani]{lazer2014google}
David Lazer, Ryan Kennedy, Gary King, and Alessandro Vespignani.
\newblock Google flu trends still appears sick: An evaluation of the 2013-2014
  flu season.
\newblock \emph{Available at SSRN 2408560}, 2014{\natexlab{a}}.

\bibitem[Lazer et~al.(2014{\natexlab{b}})Lazer, Kennedy, King, and
  Vespignani]{lazer2014parable}
David Lazer, Ryan Kennedy, Gary King, and Alessandro Vespignani.
\newblock The parable of google flu: traps in big data analysis.
\newblock \emph{Science}, 343\penalty0 (6176):\penalty0 1203--1205,
  2014{\natexlab{b}}.

\bibitem[Lepri et~al.(2022)Lepri, Centellegher, and
  Nadai]{lepri2022understanding}
Bruno Lepri, Simone Centellegher, and Marco~De Nadai.
\newblock Understanding and rewiring cities.
\newblock In \emph{European Conference on Advances in Databases and Information
  Systems}, pp.\  3--10. Springer, 2022.

\bibitem[Levy et~al.(2020)Levy, Phillips, and
  Sampson]{LevyTripleDisadvantageNeighborhood2020}
Brian~L. Levy, Nolan~E. Phillips, and Robert~J. Sampson.
\newblock Triple {Disadvantage}: {Neighborhood} {Networks} of {Everyday}
  {Urban} {Mobility} and {Violence} in {U}.{S}. {Cities}.
\newblock \emph{American Sociological Review}, 85\penalty0 (6):\penalty0
  925--956, December 2020.
\newblock ISSN 0003-1224.
\newblock \doi{10.1177/0003122420972323}.
\newblock URL \url{https://doi.org/10.1177/0003122420972323}.
\newblock Publisher: SAGE Publications Inc.

\bibitem[Li et~al.(2022)Li, Huang, Li, and Xu]{li2022aggravated}
Xiao Li, Xiao Huang, Dongying Li, and Yang Xu.
\newblock Aggravated social segregation during the covid-19 pandemic: Evidence
  from crowdsourced mobility data in twelve most populated us metropolitan
  areas.
\newblock \emph{Sustainable Cities and Society}, 81:\penalty0 103869, 2022.

\bibitem[Linning et~al.(2017)Linning, Andresen, and
  Brantingham]{LinningCrimeSeasonalityExamining2017}
Shannon~J. Linning, Martin~A. Andresen, and Paul~J. Brantingham.
\newblock Crime {Seasonality}: {Examining} the {Temporal} {Fluctuations} of
  {Property} {Crime} in {Cities} {With} {Varying} {Climates}.
\newblock \emph{International Journal of Offender Therapy and Comparative
  Criminology}, 61\penalty0 (16):\penalty0 1866--1891, December 2017.
\newblock ISSN 0306-624X, 1552-6933.
\newblock \doi{10.1177/0306624X16632259}.
\newblock URL \url{http://journals.sagepub.com/doi/10.1177/0306624X16632259}.

\bibitem[Llorente et~al.(2015)Llorente, Garcia-Herranz, Cebrian, and
  Moro]{llorente2015social}
Alejandro Llorente, Manuel Garcia-Herranz, Manuel Cebrian, and Esteban Moro.
\newblock Social media fingerprints of unemployment.
\newblock \emph{PloS one}, 10\penalty0 (5):\penalty0 e0128692, 2015.

\bibitem[Loeffler \& Flaxman(2018)Loeffler and
  Flaxman]{LoefflerGunViolenceContagious2018a}
Charles Loeffler and Seth Flaxman.
\newblock Is {Gun} {Violence} {Contagious}? {A} {Spatiotemporal} {Test}.
\newblock \emph{Journal of Quantitative Criminology}, 34\penalty0 (4):\penalty0
  999--1017, December 2018.
\newblock ISSN 1573-7799.
\newblock \doi{10.1007/s10940-017-9363-8}.
\newblock URL \url{https://doi.org/10.1007/s10940-017-9363-8}.

\bibitem[Logan \& Burdick-Will(2016)Logan and Burdick-Will]{logan2016}
John~R Logan and Julia Burdick-Will.
\newblock School segregation, charter schools, and access to quality education.
\newblock \emph{Journal of Urban Affairs}, 38:\penalty0 323--343, 2016.

\bibitem[Lu et~al.(2018)Lu, Hou, Baltrusaitis, Shah, Leskovec, Hawkins,
  Brownstein, Conidi, Gunn, Gray, et~al.]{lu2018accurate}
Fred~Sun Lu, Suqin Hou, Kristin Baltrusaitis, Manan Shah, Jure Leskovec, Jared
  Hawkins, John Brownstein, Giuseppe Conidi, Julia Gunn, Josh Gray, et~al.
\newblock Accurate influenza monitoring and forecasting using novel internet
  data streams: a case study in the boston metropolis.
\newblock \emph{JMIR public health and surveillance}, 4\penalty0 (1):\penalty0
  e8950, 2018.

\bibitem[Luca et~al.(2021)Luca, Barlacchi, Oliver, and
  Lepri]{luca2021leveraging}
Massimiliano Luca, Gianni Barlacchi, Nuria Oliver, and Bruno Lepri.
\newblock Leveraging mobile phone data for migration flows.
\newblock \emph{arXiv preprint arXiv:2105.14956}, 2021.

\bibitem[Luca et~al.(2022)Luca, Lepri, Frias-Martinez, and
  Lutu]{luca2022modeling}
Massimiliano Luca, Bruno Lepri, Enrique Frias-Martinez, and Andra Lutu.
\newblock Modeling international mobility using roaming cell phone traces
  during covid-19 pandemic.
\newblock \emph{EPJ Data Science}, 11\penalty0 (1):\penalty0 22, 2022.

\bibitem[Lucchini et~al.(2021)Lucchini, Centellegher, Pappalardo, Gallotti,
  Privitera, Lepri, and De~Nadai]{lucchini2021living}
Lorenzo Lucchini, Simone Centellegher, Luca Pappalardo, Riccardo Gallotti,
  Filippo Privitera, Bruno Lepri, and Marco De~Nadai.
\newblock Living in a pandemic: changes in mobility routines, social activity
  and adherence to covid-19 protective measures.
\newblock \emph{Scientific reports}, 11\penalty0 (1):\penalty0 1--12, 2021.

\bibitem[Lum \& Isaac(2016)Lum and Isaac]{Lumpredictserve2016}
Kristian Lum and William Isaac.
\newblock To predict and serve?
\newblock \emph{Significance}, 13\penalty0 (5):\penalty0 14--19, 2016.
\newblock ISSN 1740-9713.
\newblock \doi{10.1111/j.1740-9713.2016.00960.x}.
\newblock URL
  \url{https://rss.onlinelibrary.wiley.com/doi/abs/10.1111/j.1740-9713.2016.00960.x}.
\newblock \_eprint:
  https://rss.onlinelibrary.wiley.com/doi/pdf/10.1111/j.1740-9713.2016.00960.x.

\bibitem[Luna-Pla \& Nicolás-Carlock(2020)Luna-Pla and
  Nicolás-Carlock]{Luna-PlaCorruptioncomplexityscientific2020a}
Issa Luna-Pla and José~R. Nicolás-Carlock.
\newblock Corruption and complexity: a scientific framework for the analysis of
  corruption networks.
\newblock \emph{Applied Network Science}, 5\penalty0 (1):\penalty0 13, February
  2020.
\newblock ISSN 2364-8228.
\newblock \doi{10.1007/s41109-020-00258-2}.
\newblock URL \url{https://doi.org/10.1007/s41109-020-00258-2}.

\bibitem[Mackey et~al.(2018)Mackey, Kalyanam, Klugman, Kuzmenko, and
  Gupta]{MackeySolutionDetectClassify2018}
Tim Mackey, Janani Kalyanam, Josh Klugman, Ella Kuzmenko, and Rashmi Gupta.
\newblock Solution to {Detect}, {Classify}, and {Report} {Illicit} {Online}
  {Marketing} and {Sales} of {Controlled} {Substances} via {Twitter}: {Using}
  {Machine} {Learning} and {Web} {Forensics} to {Combat} {Digital} {Opioid}
  {Access}.
\newblock \emph{Journal of Medical Internet Research}, 20\penalty0
  (4):\penalty0 e10029, April 2018.
\newblock ISSN 1438-8871.
\newblock \doi{10.2196/10029}.
\newblock URL \url{http://www.jmir.org/2018/4/e10029/}.

\bibitem[Magliocca et~al.(2019)Magliocca, McSweeney, Sesnie, Tellman, Devine,
  Nielsen, Pearson, and Wrathall]{MaglioccaModelingcocainetraffickers2019}
Nicholas~R. Magliocca, Kendra McSweeney, Steven~E. Sesnie, Elizabeth Tellman,
  Jennifer~A. Devine, Erik~A. Nielsen, Zoe Pearson, and David~J. Wrathall.
\newblock Modeling cocaine traffickers and counterdrug interdiction forces as a
  complex adaptive system.
\newblock \emph{Proceedings of the National Academy of Sciences}, 116\penalty0
  (16):\penalty0 7784--7792, April 2019.
\newblock ISSN 0027-8424, 1091-6490.
\newblock \doi{10.1073/pnas.1812459116}.
\newblock URL \url{http://www.pnas.org/lookup/doi/10.1073/pnas.1812459116}.

\bibitem[Malleson \& Andresen(2015)Malleson and
  Andresen]{Mallesonimpactusingsocial2015}
Nick Malleson and Martin~A. Andresen.
\newblock The impact of using social media data in crime rate calculations:
  shifting hot spots and changing spatial patterns.
\newblock \emph{Cartography and Geographic Information Science}, 42, 2015.

\bibitem[Manduca \& Sampson(2019)Manduca and Sampson]{manduca2019}
Robert Manduca and Robert~J. Sampson.
\newblock Punishing and toxic neighborhood environments independently predict
  the intergenerational social mobility of black and white children.
\newblock \emph{Proc. Natl Acad. Sci.}, 116\penalty0 (16):\penalty0
  7772–7777, 2019.

\bibitem[Mazeika \& Kumar(2017)Mazeika and Kumar]{MazeikaCrimeHotSpots2017}
David~M. Mazeika and Sumit Kumar.
\newblock Do crime hot spots exist in developing countries? evidence from
  india.
\newblock \emph{Journal of Quantitative Criminology}, 33\penalty0 (1):\penalty0
  45--61, March 2017.
\newblock ISSN 1573-7799.
\newblock \doi{10.1007/s10940-016-9280-2}.
\newblock URL \url{https://doi.org/10.1007/s10940-016-9280-2}.

\bibitem[Mazzoli et~al.(2019)Mazzoli, Molas, Bassolas, Lenormand, Colet, and
  Ramasco]{mazzoli2019}
M.~Mazzoli, A.~Molas, A.~Bassolas, M.~Lenormand, P.~Colet, and J.~J. Ramasco.
\newblock Field theory for recurrent mobility.
\newblock \emph{Nature Communications}, 10:\penalty0 3895, 2019.

\bibitem[McIver \& Brownstein(2014)McIver and Brownstein]{mciver2014wikipedia}
David~J McIver and John~S Brownstein.
\newblock Wikipedia usage estimates prevalence of influenza-like illness in the
  united states in near real-time.
\newblock \emph{PLoS computational biology}, 10\penalty0 (4):\penalty0
  e1003581, 2014.

\bibitem[Mejova et~al.(2015{\natexlab{a}})Mejova, Haddadi, Noulas, and
  Weber]{mejova2015foodporn}
Yelena Mejova, Hamed Haddadi, Anastasios Noulas, and Ingmar Weber.
\newblock \# foodporn: Obesity patterns in culinary interactions.
\newblock In \emph{Proceedings of the 5th International Conference on Digital
  Health 2015}, pp.\  51--58, 2015{\natexlab{a}}.

\bibitem[Mejova et~al.(2015{\natexlab{b}})Mejova, Weber, and
  Macy]{mejova2015twitter}
Yelena Mejova, Ingmar Weber, and Michael~W Macy.
\newblock \emph{Twitter: a digital socioscope}.
\newblock Cambridge University Press, 2015{\natexlab{b}}.

\bibitem[Merler \& Ajelli(2010)Merler and Ajelli]{merler2010role}
Stefano Merler and Marco Ajelli.
\newblock The role of population heterogeneity and human mobility in the spread
  of pandemic influenza.
\newblock \emph{Proceedings of the Royal Society B: Biological Sciences},
  277\penalty0 (1681):\penalty0 557--565, 2010.

\bibitem[Miliou et~al.(2021)Miliou, Xiong, Rinzivillo, Zhang, Rossetti,
  Giannotti, Pedreschi, and Vespignani]{miliou2021predicting}
Ioanna Miliou, Xinyue Xiong, Salvatore Rinzivillo, Qian Zhang, Giulio Rossetti,
  Fosca Giannotti, Dino Pedreschi, and Alessandro Vespignani.
\newblock Predicting seasonal influenza using supermarket retail records.
\newblock \emph{PLOS Computational Biology}, 17\penalty0 (7):\penalty0
  e1009087, 2021.

\bibitem[Mohler et~al.(2011)Mohler, Short, Brantingham, Schoenberg, and
  Tita]{MohlerSelfExcitingPointProcess2011a}
G.~O. Mohler, M.~B. Short, P.~J. Brantingham, F.~P. Schoenberg, and G.~E. Tita.
\newblock Self-{Exciting} {Point} {Process} {Modeling} of {Crime}.
\newblock \emph{Journal of the American Statistical Association}, 106\penalty0
  (493):\penalty0 100--108, March 2011.
\newblock ISSN 0162-1459.
\newblock \doi{10.1198/jasa.2011.ap09546}.
\newblock URL \url{https://doi.org/10.1198/jasa.2011.ap09546}.
\newblock Publisher: Taylor \& Francis \_eprint:
  https://doi.org/10.1198/jasa.2011.ap09546.

\bibitem[Mohler(2014)]{MohlerMarkedpointprocess2014a}
George Mohler.
\newblock Marked point process hotspot maps for homicide and gun crime
  prediction in {Chicago}.
\newblock \emph{International Journal of Forecasting}, 30\penalty0
  (3):\penalty0 491--497, July 2014.
\newblock ISSN 0169-2070.
\newblock \doi{10.1016/j.ijforecast.2014.01.004}.
\newblock URL
  \url{https://www.sciencedirect.com/science/article/pii/S0169207014000284}.

\bibitem[Moon \& Carley(2007)Moon and
  Carley]{MoonModelingSimulatingTerrorist2007d}
Il-Chul Moon and Kathleen~M. Carley.
\newblock Modeling and {Simulating} {Terrorist} {Networks} in {Social} and
  {Geospatial} {Dimensions}.
\newblock \emph{IEEE Intelligent Systems}, 22\penalty0 (5):\penalty0 40--49,
  September 2007.
\newblock ISSN 1541-1672.
\newblock \doi{10.1109/MIS.2007.4338493}.
\newblock URL \url{http://ieeexplore.ieee.org/document/4338493/}.

\bibitem[Morales et~al.(2019)Morales, Dong, Bar-Yam, and
  ‘Sandy’Pentland]{morales2019segregation}
Alfredo~J Morales, Xiaowen Dong, Yaneer Bar-Yam, and Alex ‘Sandy’Pentland.
\newblock Segregation and polarization in urban areas.
\newblock \emph{Royal Society Open Science}, 6\penalty0 (10):\penalty0 190573,
  2019.

\bibitem[Moreira-Matias et~al.(2013)Moreira-Matias, Gama, Ferreira,
  Mendes-Moreira, and Damas]{moreira2013predicting}
Luis Moreira-Matias, Joao Gama, Michel Ferreira, Joao Mendes-Moreira, and Luis
  Damas.
\newblock Predicting taxi--passenger demand using streaming data.
\newblock \emph{IEEE Transactions on Intelligent Transportation Systems},
  14\penalty0 (3):\penalty0 1393--1402, 2013.

\bibitem[Moro et~al.(2021)Moro, Calacci, Dong, and Pentland]{moro2021mobility}
Esteban Moro, Dan Calacci, Xiaowen Dong, and Alex Pentland.
\newblock Mobility patterns are associated with experienced income segregation
  in large us cities.
\newblock \emph{Nature communications}, 12\penalty0 (1):\penalty0 1--10, 2021.

\bibitem[Nagar et~al.(2014)Nagar, Yuan, Freifeld, Santillana, Nojima, Chunara,
  Brownstein, et~al.]{nagar2014case}
Ruchit Nagar, Qingyu Yuan, Clark~C Freifeld, Mauricio Santillana, Aaron Nojima,
  Rumi Chunara, John~S Brownstein, et~al.
\newblock A case study of the new york city 2012-2013 influenza season with
  daily geocoded twitter data from temporal and spatiotemporal perspectives.
\newblock \emph{Journal of medical Internet research}, 16\penalty0
  (10):\penalty0 e3416, 2014.

\bibitem[Nardin et~al.(2016)Nardin, Andrighetto, Conte, Sz\'{e}kely, Anzola,
  Elsenbroich, Lotzmann, Neumann, Punzo, and
  Troitzsch]{NardinSimulatingprotectionrackets2016}
Luis~G. Nardin, Giulia Andrighetto, Rosaria Conte, \'{A}ron Sz\'{e}kely, David
  Anzola, Corinna Elsenbroich, Ulf Lotzmann, Martin Neumann, Valentina Punzo,
  and Klaus~G. Troitzsch.
\newblock Simulating protection rackets: a case study of the {Sicilian}
  {Mafia}.
\newblock \emph{Autonomous Agents and Multi-Agent Systems}, 30\penalty0
  (6):\penalty0 1117--1147, November 2016.
\newblock ISSN 1387-2532, 1573-7454.
\newblock \doi{10.1007/s10458-016-9330-z}.
\newblock URL \url{http://link.springer.com/10.1007/s10458-016-9330-z}.

\bibitem[Neiderud(2015)]{neiderud2015urbanization}
Carl-Johan Neiderud.
\newblock How urbanization affects the epidemiology of emerging infectious
  diseases.
\newblock \emph{Infection ecology \& epidemiology}, 5\penalty0 (1):\penalty0
  27060, 2015.

\bibitem[Neves et~al.(2016)Neves, Afonso, and Silva]{neves2016}
Pedro~Cunha Neves, Oscar Afonso, and Sandra~Tavares Silva.
\newblock A meta-analytic reassessment of the effects of inequality on growth.
\newblock \emph{World Development}, 78:\penalty0 386--400, 2016.

\bibitem[Nguyen et~al.(2016)Nguyen, Li, Meng, Kath, Nsoesie, Li, and
  Wen]{nguyen2016building}
Quynh~C Nguyen, Dapeng Li, Hsien-Wen Meng, Suraj Kath, Elaine Nsoesie, Feifei
  Li, and Ming Wen.
\newblock Building a national neighborhood dataset from geotagged twitter data
  for indicators of happiness, diet, and physical activity.
\newblock \emph{JMIR public health and surveillance}, 2\penalty0 (2):\penalty0
  e5869, 2016.

\bibitem[Nicoletti et~al.(2022)Nicoletti, Sirenko, and
  Verma]{nicoletti2022disadvantaged}
Leonardo Nicoletti, Mikhail Sirenko, and Trivik Verma.
\newblock Disadvantaged communities have lower access to urban infrastructure.
\newblock \emph{arXiv preprint arXiv:2203.13784}, 2022.

\bibitem[Oliver et~al.(2020)Oliver, Lepri, Sterly, Lambiotte, Deletaille,
  De~Nadai, Letouz{\'e}, Salah, Benjamins, Cattuto, et~al.]{oliver2020mobile}
Nuria Oliver, Bruno Lepri, Harald Sterly, Renaud Lambiotte, S{\'e}bastien
  Deletaille, Marco De~Nadai, Emmanuel Letouz{\'e}, Albert~Ali Salah, Richard
  Benjamins, Ciro Cattuto, et~al.
\newblock Mobile phone data for informing public health actions across the
  covid-19 pandemic life cycle.
\newblock \emph{Science advances}, 6\penalty0 (23):\penalty0 eabc0764, 2020.

\bibitem[Pan et~al.(2013)Pan, Ghoshal, Krumme, Cebrian, and Pentland]{pan2013}
W.~Pan, G.~Ghoshal, C.~Krumme, M.~Cebrian, and A.~Pentland.
\newblock Urban characteristics attributable to density-driven tie formation.
\newblock \emph{Nature Communications}, 4\penalty0 (1):\penalty0 1--7, 2013.

\bibitem[Pangallo et~al.(2022)Pangallo, Aleta, Chanona, Pichler,
  Mart{\'\i}n-Corral, Chinazzi, Lafond, Ajelli, Moro, Moreno,
  et~al.]{pangallo2022unequal}
Marco Pangallo, Alberto Aleta, R~Chanona, Anton Pichler, David
  Mart{\'\i}n-Corral, Matteo Chinazzi, Fran{\c{c}}ois Lafond, Marco Ajelli,
  Esteban Moro, Yamir Moreno, et~al.
\newblock The unequal effects of the health-economy tradeoff during the
  covid-19 pandemic.
\newblock \emph{arXiv preprint arXiv:2212.03567}, 2022.

\bibitem[Paolotti et~al.(2014)Paolotti, Carnahan, Colizza, Eames, Edmunds,
  Gomes, Koppeschaar, Rehn, Smallenburg, Turbelin, et~al.]{paolotti2014web}
Daniela Paolotti, Annasara Carnahan, Vittoria Colizza, Ken Eames, John Edmunds,
  Gabriel Gomes, C~Koppeschaar, Moa Rehn, Ronald Smallenburg, Cl{\'e}ment
  Turbelin, et~al.
\newblock Web-based participatory surveillance of infectious diseases: the
  influenzanet participatory surveillance experience.
\newblock \emph{Clinical Microbiology and Infection}, 20\penalty0 (1):\penalty0
  17--21, 2014.

\bibitem[Papachristos \& Bastomski(2018)Papachristos and
  Bastomski]{PapachristosConnectedCrimeEnduring2018a}
Andrew~V. Papachristos and Sara Bastomski.
\newblock Connected in {Crime}: {The} {Enduring} {Effect} of {Neighborhood}
  {Networks} on the {Spatial} {Patterning} of {Violence}.
\newblock \emph{American Journal of Sociology}, 124\penalty0 (2):\penalty0
  517--568, September 2018.
\newblock ISSN 0002-9602.
\newblock \doi{10.1086/699217}.
\newblock URL \url{https://www.journals.uchicago.edu/doi/abs/10.1086/699217}.
\newblock Publisher: The University of Chicago Press.

\bibitem[Pappalardo et~al.(2015)Pappalardo, Simini, Rinzivillo, Pedreschi,
  Giannotti, and Barab{\'a}si]{pappalardo2015returners}
Luca Pappalardo, Filippo Simini, Salvatore Rinzivillo, Dino Pedreschi, Fosca
  Giannotti, and Albert-L{\'a}szl{\'o} Barab{\'a}si.
\newblock Returners and explorers dichotomy in human mobility.
\newblock \emph{Nature communications}, 6\penalty0 (1):\penalty0 1--8, 2015.

\bibitem[Paul \& Dredze(2011)Paul and Dredze]{paul2011you}
Michael Paul and Mark Dredze.
\newblock You are what you tweet: Analyzing twitter for public health.
\newblock In \emph{Proceedings of the International AAAI Conference on Web and
  Social Media}, volume~5, pp.\  265--272, 2011.

\bibitem[Perkins et~al.(2014)Perkins, Garcia, Paz-Sold{\'a}n, Stoddard,
  Reiner~Jr, Vazquez-Prokopec, Bisanzio, Morrison, Halsey, Kochel,
  et~al.]{perkins2014theory}
T~Alex Perkins, Andres~J Garcia, Valerie~A Paz-Sold{\'a}n, Steven~T Stoddard,
  Robert~C Reiner~Jr, Gonzalo Vazquez-Prokopec, Donal Bisanzio, Amy~C Morrison,
  Eric~S Halsey, Tadeusz~J Kochel, et~al.
\newblock Theory and data for simulating fine-scale human movement in an urban
  environment.
\newblock \emph{Journal of the Royal Society Interface}, 11\penalty0
  (99):\penalty0 20140642, 2014.

\bibitem[Perra et~al.(2011)Perra, Balcan, Gon{\c{c}}alves, and
  Vespignani]{perra2011towards}
Nicola Perra, Duygu Balcan, Bruno Gon{\c{c}}alves, and Alessandro Vespignani.
\newblock Towards a characterization of behavior-disease models.
\newblock \emph{PloS one}, 6\penalty0 (8):\penalty0 e23084, 2011.

\bibitem[Perry(2013)]{PerryPredictivePolicingRole2013}
Walt~L. Perry.
\newblock \emph{Predictive {Policing}: {The} {Role} of {Crime} {Forecasting} in
  {Law} {Enforcement} {Operations}}.
\newblock Rand Corporation, September 2013.
\newblock ISBN 978-0-8330-8155-1.

\bibitem[Peterson \& Krivo(2009)Peterson and
  Krivo]{PetersonSegregatedSpatialLocations2009}
Ruth~D. Peterson and Lauren~J. Krivo.
\newblock Segregated {Spatial} {Locations}, {Race}-{Ethnic} {Composition}, and
  {Neighborhood} {Violent} {Crime}.
\newblock \emph{The ANNALS of the American Academy of Political and Social
  Science}, 623\penalty0 (1):\penalty0 93--107, May 2009.
\newblock ISSN 0002-7162, 1552-3349.
\newblock \doi{10.1177/0002716208330490}.
\newblock URL \url{http://journals.sagepub.com/doi/10.1177/0002716208330490}.

\bibitem[Phillips et~al.(2021)Phillips, Levy, Sampson, Small, and
  Wang]{phillips2021social}
Nolan~E Phillips, Brian~L Levy, Robert~J Sampson, Mario~L Small, and Ryan~Q
  Wang.
\newblock The social integration of american cities: Network measures of
  connectedness based on everyday mobility across neighborhoods.
\newblock \emph{Sociological Methods \& Research}, 50\penalty0 (3):\penalty0
  1110--1149, 2021.

\bibitem[Piatkowska \& Lantz(2021)Piatkowska and
  Lantz]{PiatkowskaTemporalClusteringHate2021}
Sylwia~J Piatkowska and Brendan Lantz.
\newblock Temporal {Clustering} of {Hate} {Crimes} in the {Aftermath} of the
  {Brexit} {Vote} and {Terrorist} {Attacks}: {A} {Comparison} of {Scotland} and
  {England} and {Wales}.
\newblock \emph{The British Journal of Criminology}, 61\penalty0 (3):\penalty0
  648--669, April 2021.
\newblock ISSN 0007-0955, 1464-3529.
\newblock \doi{10.1093/bjc/azaa090}.
\newblock URL \url{https://academic.oup.com/bjc/article/61/3/648/6087651}.

\bibitem[Pickett \& Wilkinson(2015)Pickett and Wilkinson]{pickett2015}
Kate~E Pickett and Richard~G Wilkinson.
\newblock Income inequality and health: A causal review.
\newblock \emph{Social Science \& Medicine}, 128:\penalty0 316--326, 2015.

\bibitem[Poletto et~al.(2020)Poletto, Scarpino, and
  Volz]{poletto2020applications}
Chiara Poletto, Samuel~V Scarpino, and Erik~M Volz.
\newblock Applications of predictive modelling early in the covid-19 epidemic.
\newblock \emph{The Lancet Digital Health}, 2\penalty0 (10):\penalty0
  e498--e499, 2020.

\bibitem[Polgreen et~al.(2008)Polgreen, Chen, Pennock, Nelson, and
  Weinstein]{polgreen2008using}
Philip~M Polgreen, Yiling Chen, David~M Pennock, Forrest~D Nelson, and Robert~A
  Weinstein.
\newblock Using internet searches for influenza surveillance.
\newblock \emph{Clinical infectious diseases}, 47\penalty0 (11):\penalty0
  1443--1448, 2008.

\bibitem[Purves(2022)]{PurvesFairnessAlgorithmicPolicing2022}
Duncan Purves.
\newblock Fairness in {Algorithmic} {Policing}.
\newblock \emph{Journal of the American Philosophical Association}, pp.\
  1--21, March 2022.
\newblock ISSN 2053-4477, 2053-4485.
\newblock \doi{10.1017/apa.2021.39}.
\newblock URL
  \url{https://www.cambridge.org/core/product/identifier/S2053447721000397/type/journal_article}.

\bibitem[Quetelet(1831)]{QueteletResearchPropensityCrime1831}
Adolphe Quetelet.
\newblock \emph{Research on the {Propensity} for {Crime} at {Different}
  {Ages}}.
\newblock Anderson, Cincinnati, Ohio, 1831.

\bibitem[Quillian(2014)]{quillian2014}
Lincoln Quillian.
\newblock Does segregation create winners and losers? residential segregation
  and inequality in educational attainment.
\newblock \emph{Social Problems}, 61\penalty0 (3):\penalty0 402--426, 2014.

\bibitem[Rader et~al.(2020)Rader, Scarpino, Nande, Hill, Adlam, Reiner, Pigott,
  Gutierrez, Zarebski, Shrestha, et~al.]{rader2020crowding}
Benjamin Rader, Samuel~V Scarpino, Anjalika Nande, Alison~L Hill, Ben Adlam,
  Robert~C Reiner, David~M Pigott, Bernardo Gutierrez, Alexander~E Zarebski,
  Munik Shrestha, et~al.
\newblock Crowding and the shape of covid-19 epidemics.
\newblock \emph{Nature medicine}, 26\penalty0 (12):\penalty0 1829--1834, 2020.

\bibitem[Reid et~al.(2016)Reid, Hamidi, Grace, and Wei]{ewing2016}
E.~Reid, S.~Hamidi, J.B. Grace, and Y.D. Wei.
\newblock Does urban sprawl hold down upward mobility.
\newblock \emph{Landscape and Urban Planning}, 148:\penalty0 80--88, 2016.

\bibitem[Ribeiro et~al.(2018)Ribeiro, Alves, Martins, Lenzi, and
  Perc]{Ribeirodynamicalstructurepolitical2018b}
Haroldo~V Ribeiro, Luiz G~A Alves, Alvaro~F Martins, Ervin~K Lenzi, and Matjaž
  Perc.
\newblock The dynamical structure of political corruption networks.
\newblock \emph{Journal of Complex Networks}, 6\penalty0 (6):\penalty0
  989--1003, December 2018.
\newblock ISSN 2051-1329.
\newblock \doi{10.1093/comnet/cny002}.
\newblock URL \url{https://doi.org/10.1093/comnet/cny002}.

\bibitem[Richardson et~al.(2019)Richardson, Schultz, and
  Crawford]{RichardsonDirtyDataBad2019}
Rashida Richardson, Jason Schultz, and Kate Crawford.
\newblock Dirty {Data}, {Bad} {Predictions}: {How} {Civil} {Rights}
  {Violations} {Impact} {Police} {Data}, {Predictive} {Policing} {Systems}, and
  {Justice}.
\newblock {SSRN} {Scholarly} {Paper} ID 3333423, Social Science Research
  Network, Rochester, NY, February 2019.
\newblock URL \url{https://papers.ssrn.com/abstract=3333423}.

\bibitem[Rocha et~al.(2015)Rocha, Thorson, and Lambiotte]{rocha2015non}
Luis~EC Rocha, Anna~E Thorson, and Renaud Lambiotte.
\newblock The non-linear health consequences of living in larger cities.
\newblock \emph{Journal of Urban Health}, 92\penalty0 (5):\penalty0 785--799,
  2015.

\bibitem[Rogers et~al.(2002)Rogers, Randolph, Snow, and
  Hay]{rogers2002satellite}
David~J Rogers, Sarah~E Randolph, Robert~W Snow, and Simon~I Hay.
\newblock Satellite imagery in the study and forecast of malaria.
\newblock \emph{Nature}, 415\penalty0 (6872):\penalty0 710--715, 2002.

\bibitem[Ross(1916)]{ross1916application}
Ronald Ross.
\newblock An application of the theory of probabilities to the study of a
  priori pathometry.—part i.
\newblock \emph{Proceedings of the Royal Society of London. Series A,
  Containing papers of a mathematical and physical character}, 92\penalty0
  (638):\penalty0 204--230, 1916.

\bibitem[Rosés et~al.(2020)Rosés, Kadar, Gerritsen, and
  Rouly]{RosesSimulatingOffenderMobility2020}
Raquel Rosés, Cristina Kadar, Charlotte Gerritsen, and Chris Rouly.
\newblock Simulating {Offender} {Mobility}: {Modeling} {Activity} {Nodes} from
  {Large}-{Scale} {Human} {Activity} {Data}.
\newblock \emph{Journal of Artificial Intelligence Research}, 68:\penalty0
  541--570, July 2020.
\newblock ISSN 1076-9757.
\newblock \doi{10.1613/jair.1.11831}.
\newblock URL \url{https://www.jair.org/index.php/jair/article/view/11831}.

\bibitem[Rosés et~al.(2021)Rosés, Kadar, and
  Malleson]{Rosesdatadrivenagentbasedsimulation2021}
Raquel Rosés, Cristina Kadar, and Nick Malleson.
\newblock A data-driven agent-based simulation to predict crime patterns in an
  urban environment.
\newblock \emph{Computers, Environment and Urban Systems}, 89:\penalty0 101660,
  September 2021.
\newblock ISSN 01989715.
\newblock \doi{10.1016/j.compenvurbsys.2021.101660}.
\newblock URL
  \url{https://linkinghub.elsevier.com/retrieve/pii/S0198971521000673}.

\bibitem[Rvachev \& Longini~Jr(1985)Rvachev and
  Longini~Jr]{rvachev1985mathematical}
Leonid~A Rvachev and Ira~M Longini~Jr.
\newblock A mathematical model for the global spread of influenza.
\newblock \emph{Mathematical biosciences}, 75\penalty0 (1):\penalty0 3--22,
  1985.

\bibitem[Sadilek \& Kautz(2013)Sadilek and Kautz]{sadilek2013modeling}
Adam Sadilek and Henry Kautz.
\newblock Modeling the impact of lifestyle on health at scale.
\newblock In \emph{Proceedings of the sixth ACM international conference on Web
  search and data mining}, pp.\  637--646, 2013.

\bibitem[Sadilek et~al.(2012)Sadilek, Kautz, and
  Silenzio]{sadilek2012predicting}
Adam Sadilek, Henry Kautz, and Vincent Silenzio.
\newblock Predicting disease transmission from geo-tagged micro-blog data.
\newblock In \emph{Twenty-Sixth AAAI Conference on Artificial Intelligence},
  2012.

\bibitem[Salath{\'e}(2018)]{salathe2018digital}
Marcel Salath{\'e}.
\newblock Digital epidemiology: what is it, and where is it going?
\newblock \emph{Life sciences, society and policy}, 14\penalty0 (1):\penalty0
  1--5, 2018.

\bibitem[Salathe et~al.(2012)Salathe, Bengtsson, Bodnar, Brewer, Brownstein,
  Buckee, Campbell, Cattuto, Khandelwal, Mabry, et~al.]{salathe2012digital}
Marcel Salathe, Linus Bengtsson, Todd~J Bodnar, Devon~D Brewer, John~S
  Brownstein, Caroline Buckee, Ellsworth~M Campbell, Ciro Cattuto, Shashank
  Khandelwal, Patricia~L Mabry, et~al.
\newblock Digital epidemiology.
\newblock 2012.

\bibitem[Sampson \& Levy(2022)Sampson and
  Levy]{SampsonEnduringNeighborhoodEffect2022}
Robert Sampson and Brian Levy.
\newblock The {Enduring} {Neighborhood} {Effect}, {Everyday} {Urban}
  {Mobility}, and {Violence} in {Chicago}.
\newblock \emph{University of Chicago Law Review}, 89\penalty0 (2), March 2022.
\newblock ISSN 0041-9494.
\newblock URL \url{https://chicagounbound.uchicago.edu/uclrev/vol89/iss2/2}.

\bibitem[Sampson(2012)]{SampsonGreatAmericancity2012}
Robert~J. Sampson.
\newblock \emph{Great {American} city: {Chicago} and the enduring neighborhood
  effect}.
\newblock Univ. of Chicago Press, Chicago, Ill., 2012.
\newblock ISBN 978-0-226-05568-8 978-0-226-73456-9.

\bibitem[Sampson(2019)]{sampson2019}
Robert~J Sampson.
\newblock Neighbourhood effects and beyond: Explaining the paradoxes of
  inequality in the changing american metropolis.
\newblock \emph{Urban Studies}, 56\penalty0 (1):\penalty0 3--32, 2019.

\bibitem[Sarker et~al.(2019)Sarker, Gonzalez-Hernandez, Ruan, and
  Perrone]{SarkerMachineLearningNatural2019}
Abeed Sarker, Graciela Gonzalez-Hernandez, Yucheng Ruan, and Jeanmarie Perrone.
\newblock Machine {Learning} and {Natural} {Language} {Processing} for
  {Geolocation}-{Centric} {Monitoring} and {Characterization} of
  {Opioid}-{Related} {Social} {Media} {Chatter}.
\newblock \emph{JAMA Network Open}, 2\penalty0 (11):\penalty0 e1914672,
  November 2019.
\newblock ISSN 2574-3805.
\newblock \doi{10.1001/jamanetworkopen.2019.14672}.
\newblock URL
  \url{https://jamanetwork.com/journals/jamanetworkopen/fullarticle/2753983}.

\bibitem[Sassen(2019)]{sassen2019}
S.~Sassen.
\newblock \emph{Cities in a World Economy}.
\newblock AGE Publications Inc., 2019.

\bibitem[Sattenspiel \& Dietz(1995)Sattenspiel and
  Dietz]{sattenspiel1995structured}
Lisa Sattenspiel and Klaus Dietz.
\newblock A structured epidemic model incorporating geographic mobility among
  regions.
\newblock \emph{Mathematical biosciences}, 128\penalty0 (1-2):\penalty0 71--91,
  1995.

\bibitem[Schl{\"a}pfer et~al.(2014)Schl{\"a}pfer, Bettencourt, Grauwin,
  Raschke, Claxton, Smoreda, West, and Ratti]{schlapfer2014scaling}
Markus Schl{\"a}pfer, Lu{\'\i}s~MA Bettencourt, S{\'e}bastian Grauwin, Mathias
  Raschke, Rob Claxton, Zbigniew Smoreda, Geoffrey~B West, and Carlo Ratti.
\newblock The scaling of human interactions with city size.
\newblock \emph{Journal of the Royal Society Interface}, 11\penalty0
  (98):\penalty0 20130789, 2014.

\bibitem[Shalaginov et~al.(2017)Shalaginov, Johnsen, and
  Franke]{ShalaginovCybercrimeinvestigations2017}
Andrii Shalaginov, Jan~William Johnsen, and Katrin Franke.
\newblock Cyber crime investigations in the era of big data.
\newblock In \emph{2017 {IEEE} {International} {Conference} on {Big} {Data}
  ({Big} {Data})}, pp.\  3672--3676, December 2017.
\newblock \doi{10.1109/BigData.2017.8258362}.

\bibitem[Shaman \& Karspeck(2012)Shaman and Karspeck]{shaman2012forecasting}
Jeffrey Shaman and Alicia Karspeck.
\newblock Forecasting seasonal outbreaks of influenza.
\newblock \emph{Proceedings of the National Academy of Sciences}, 109\penalty0
  (50):\penalty0 20425--20430, 2012.

\bibitem[Shaman et~al.(2013)Shaman, Karspeck, Yang, Tamerius, and
  Lipsitch]{shaman2013real}
Jeffrey Shaman, Alicia Karspeck, Wan Yang, James Tamerius, and Marc Lipsitch.
\newblock Real-time influenza forecasts during the 2012--2013 season.
\newblock \emph{Nature communications}, 4\penalty0 (1):\penalty0 1--10, 2013.

\bibitem[Shaw \& McKay(1942)Shaw and McKay]{ShawJuveniledelinquencyurban1942}
C.~R. Shaw and H.~D. McKay.
\newblock \emph{Juvenile delinquency and urban areas}.
\newblock Juvenile delinquency and urban areas. University of Chicago Press,
  Chicago, IL, US, 1942.
\newblock Pages: xxxii, 451.

\bibitem[Shelton et~al.(2015)Shelton, Poorthuis, and Zook]{shelton2015social}
Taylor Shelton, Ate Poorthuis, and Matthew Zook.
\newblock Social media and the city: Rethinking urban socio-spatial inequality
  using user-generated geographic information.
\newblock \emph{Landscape and urban planning}, 142:\penalty0 198--211, 2015.

\bibitem[Shiode et~al.(2015)Shiode, Shiode, Rod-Thatcher, Rana, and
  Vinten-Johansen]{shiode2015mortality}
Narushige Shiode, Shino Shiode, Elodie Rod-Thatcher, Sanjay Rana, and Peter
  Vinten-Johansen.
\newblock The mortality rates and the space-time patterns of john snow’s
  cholera epidemic map.
\newblock \emph{International journal of health geographics}, 14\penalty0
  (1):\penalty0 1--15, 2015.

\bibitem[Sigalo et~al.(2022)Sigalo, St~Jean, Frias-Martinez,
  et~al.]{sigalo2022using}
Nekabari Sigalo, Beth St~Jean, Vanessa Frias-Martinez, et~al.
\newblock Using social media to predict food deserts in the united states:
  Infodemiology study of tweets.
\newblock \emph{JMIR Public Health and Surveillance}, 8\penalty0 (7):\penalty0
  e34285, 2022.

\bibitem[Simini et~al.(2012)Simini, Gonz{\'a}lez, Maritan, and
  Barab{\'a}si]{simini2012universal}
Filippo Simini, Marta~C Gonz{\'a}lez, Amos Maritan, and Albert-L{\'a}szl{\'o}
  Barab{\'a}si.
\newblock A universal model for mobility and migration patterns.
\newblock \emph{Nature}, 484\penalty0 (7392):\penalty0 96--100, 2012.

\bibitem[Singh et~al.(2015)Singh, Bozkaya, and Pentland]{singh2015money}
Vivek~Kumar Singh, Burcin Bozkaya, and Alex Pentland.
\newblock Money walks: implicit mobility behavior and financial well-being.
\newblock \emph{PloS one}, 10\penalty0 (8):\penalty0 e0136628, 2015.

\bibitem[Smolinski et~al.(2015)Smolinski, Crawley, Baltrusaitis, Chunara,
  Olsen, W{\'o}jcik, Santillana, Nguyen, and Brownstein]{smolinski2015flu}
Mark~S Smolinski, Adam~W Crawley, Kristin Baltrusaitis, Rumi Chunara,
  Jennifer~M Olsen, Oktawia W{\'o}jcik, Mauricio Santillana, Andre Nguyen, and
  John~S Brownstein.
\newblock Flu near you: crowdsourced symptom reporting spanning 2 influenza
  seasons.
\newblock \emph{American journal of public health}, 105\penalty0 (10):\penalty0
  2124--2130, 2015.

\bibitem[Song et~al.(2019)Song, Bernasco, Liu, Xiao, Zhou, and
  Liao]{SongCrimeFeedsLegal2019}
Guangwen Song, Wim Bernasco, Lin Liu, Luzi Xiao, Suhong Zhou, and Weiwei Liao.
\newblock Crime {Feeds} on {Legal} {Activities}: {Daily} {Mobility} {Flows}
  {Help} to {Explain} {Thieves}’ {Target} {Location} {Choices}.
\newblock \emph{Journal of Quantitative Criminology}, 35\penalty0 (4):\penalty0
  831--854, December 2019.
\newblock ISSN 1573-7799.
\newblock \doi{10.1007/s10940-019-09406-z}.
\newblock URL \url{https://doi.org/10.1007/s10940-019-09406-z}.

\bibitem[Spinsanti et~al.(2013)Spinsanti, Berlingerio, and
  Pappalardo]{spinsanti2013mobility}
L.~Spinsanti, M.~Berlingerio, and L.~Pappalardo.
\newblock \emph{Mobility and Geo-Social Networks}, pp.\  315–333.
\newblock Cambridge University Press, 2013.
\newblock \doi{10.1017/CBO9781139128926.017}.

\bibitem[Strano et~al.(2021)Strano, Simini, De~Nadai, Esch, and
  Marconcini]{strano2021agglomeration}
Emanuele Strano, Filippo Simini, Marco De~Nadai, Thomas Esch, and Mattia
  Marconcini.
\newblock The agglomeration and dispersion dichotomy of human settlements on
  earth.
\newblock \emph{Scientific reports}, 11\penalty0 (1):\penalty0 1--10, 2021.

\bibitem[Tatem(2009)]{tatem2009worldwide}
Andrew~J Tatem.
\newblock The worldwide airline network and the dispersal of exotic species:
  2007--2010.
\newblock \emph{Ecography}, 32\penalty0 (1):\penalty0 94--102, 2009.

\bibitem[Tatem(2017)]{tatem2017worldpop}
Andrew~J Tatem.
\newblock Worldpop, open data for spatial demography.
\newblock \emph{Scientific data}, 4\penalty0 (1):\penalty0 1--4, 2017.

\bibitem[Tatem et~al.(2009)Tatem, Qiu, Smith, Sabot, Ali, and
  Moonen]{tatem2009use}
Andrew~J Tatem, Youliang Qiu, David~L Smith, Oliver Sabot, Abdullah~S Ali, and
  Bruno Moonen.
\newblock The use of mobile phone data for the estimation of the travel
  patterns and imported plasmodium falciparum rates among zanzibar residents.
\newblock \emph{Malaria journal}, 8\penalty0 (1):\penalty0 1--12, 2009.

\bibitem[Tita \& Greenbaum(2009)Tita and
  Greenbaum]{TitaCrimeNeighborhoodsUnits2009}
George~E. Tita and Robert~T. Greenbaum.
\newblock Crime, {Neighborhoods}, and {Units} of {Analysis}: {Putting} {Space}
  in {Its} {Place}.
\newblock In David Weisburd, Wim Bernasco, and Gerben~J.N. Bruinsma (eds.),
  \emph{Putting {Crime} in its {Place}}, pp.\  145--170. Springer New York, New
  York, NY, 2009.
\newblock ISBN 978-1-4419-0973-2 978-0-387-09688-9.
\newblock \doi{10.1007/978-0-387-09688-9_7}.
\newblock URL \url{http://link.springer.com/10.1007/978-0-387-09688-9_7}.

\bibitem[Tizzoni et~al.(2015)Tizzoni, Sun, Benusiglio, Karsai, and
  Perra]{tizzoni2015scaling}
Michele Tizzoni, Kaiyuan Sun, Diego Benusiglio, M{\'a}rton Karsai, and Nicola
  Perra.
\newblock The scaling of human contacts and epidemic processes in
  metapopulation networks.
\newblock \emph{Scientific reports}, 5\penalty0 (1):\penalty0 1--11, 2015.

\bibitem[Tizzoni et~al.(2020)Tizzoni, Panisson, Paolotti, and
  Cattuto]{tizzoni2020impact}
Michele Tizzoni, Andr{\'e} Panisson, Daniela Paolotti, and Ciro Cattuto.
\newblock The impact of news exposure on collective attention in the united
  states during the 2016 zika epidemic.
\newblock \emph{PLoS computational biology}, 16\penalty0 (3):\penalty0
  e1007633, 2020.

\bibitem[Tizzoni et~al.(2022)Tizzoni, Nsoesie, Gauvin, Karsai, Perra, and
  Bansal]{tizzoni2022addressing}
Michele Tizzoni, Elaine~O Nsoesie, Laetitia Gauvin, M{\'a}rton Karsai, Nicola
  Perra, and Shweta Bansal.
\newblock Addressing the socioeconomic divide in computational modeling for
  infectious diseases.
\newblock \emph{Nature Communications}, 13\penalty0 (1):\penalty0 1--7, 2022.

\bibitem[Tollenaar \& van~der Heijden(2013)Tollenaar and van~der
  Heijden]{TollenaarWhichmethodpredicts2013}
N.~Tollenaar and P.~G.~M. van~der Heijden.
\newblock Which method predicts recidivism best?: a comparison of statistical,
  machine learning and data mining predictive models: \textit{{Which} {Method}
  {Predicts} {Recidivism} {Best}?}
\newblock \emph{Journal of the Royal Statistical Society: Series A (Statistics
  in Society)}, 176\penalty0 (2):\penalty0 565--584, February 2013.
\newblock ISSN 09641998.
\newblock \doi{10.1111/j.1467-985X.2012.01056.x}.
\newblock URL
  \url{https://onlinelibrary.wiley.com/doi/10.1111/j.1467-985X.2012.01056.x}.

\bibitem[Tovanich et~al.(2021)Tovanich, Centellegher, Seghouani, Gladstone,
  Matz, and Lepri]{tovanich2021inferring}
Natkamon Tovanich, Simone Centellegher, Nac{\'e}ra~Bennacer Seghouani, Joe
  Gladstone, Sandra Matz, and Bruno Lepri.
\newblock Inferring psychological traits from spending categories and dynamic
  consumption patterns.
\newblock \emph{EPJ Data Science}, 10\penalty0 (1):\penalty0 24, 2021.

\bibitem[Traunmueller et~al.(2014)Traunmueller, Quattrone, and
  Capra]{TraunmuellerMiningMobilePhone2014}
Martin Traunmueller, Giovanni Quattrone, and Licia Capra.
\newblock Mining {Mobile} {Phone} {Data} to {Investigate} {Urban} {Crime}
  {Theories} at {Scale}.
\newblock In Luca~Maria Aiello and Daniel McFarland (eds.), \emph{Social
  {Informatics}: 6th {International} {Conference}, {SocInfo} 2014, {Barcelona},
  {Spain}, {November} 11-13, 2014. {Proceedings}}, Lecture {Notes} in
  {Computer} {Science}, pp.\  396--411. Springer International Publishing,
  Cham, 2014.
\newblock ISBN 978-3-319-13734-6.
\newblock \doi{10.1007/978-3-319-13734-6_29}.
\newblock URL \url{https://doi.org/10.1007/978-3-319-13734-6_29}.

\bibitem[Troitzsch(2017)]{TroitzschCanagentbasedsimulation2017a}
Klaus~G. Troitzsch.
\newblock Can agent-based simulation models replicate organised crime?
\newblock \emph{Trends in Organized Crime}, 20\penalty0 (1):\penalty0 100--119,
  June 2017.
\newblock ISSN 1936-4830.
\newblock \doi{10.1007/s12117-016-9298-8}.
\newblock URL \url{https://doi.org/10.1007/s12117-016-9298-8}.

\bibitem[Tucker et~al.(2021)Tucker, O’Brien, Ciomek, Castro, Wang, and
  Phillips]{TuckerWhoTweetsWhere2021}
Riley Tucker, Daniel~T. O’Brien, Alexandra Ciomek, Edgar Castro, Qi~Wang, and
  Nolan~Edward Phillips.
\newblock Who ‘{Tweets}’ {Where} and {When}, and {How} {Does} it {Help}
  {Understand} {Crime} {Rates} at {Places}? {Measuring} the {Presence} of
  {Tourists} and {Commuters} in {Ambient} {Populations}.
\newblock \emph{Journal of Quantitative Criminology}, 37\penalty0 (2):\penalty0
  333--359, June 2021.
\newblock ISSN 1573-7799.
\newblock \doi{10.1007/s10940-020-09487-1}.
\newblock URL \url{https://doi.org/10.1007/s10940-020-09487-1}.

\bibitem[Umar et~al.(2020)Umar, Johnson, and
  Cheshire]{UmarAssessingSpatialConcentration2020}
Faisal Umar, Shane~D. Johnson, and James~A. Cheshire.
\newblock Assessing the {Spatial} {Concentration} of {Urban} {Crime}: {An}
  {Insight} from {Nigeria}.
\newblock \emph{Journal of Quantitative Criminology}, January 2020.
\newblock ISSN 0748-4518, 1573-7799.
\newblock \doi{10.1007/s10940-019-09448-3}.
\newblock URL \url{http://link.springer.com/10.1007/s10940-019-09448-3}.

\bibitem[Viboud \& Santillana(2020)Viboud and Santillana]{viboud2020fitbit}
Cecile Viboud and Mauricio Santillana.
\newblock Fitbit-informed influenza forecasts.
\newblock \emph{The Lancet Digital Health}, 2\penalty0 (2):\penalty0 e54--e55,
  2020.

\bibitem[Wang et~al.(2016)Wang, Kifer, Graif, and
  Li]{WangCrimeRateInference2016}
Hongjian Wang, Daniel Kifer, Corina Graif, and Zhenhui Li.
\newblock Crime {Rate} {Inference} with {Big} {Data}.
\newblock In \emph{Proceedings of the 22nd {ACM} {SIGKDD} {International}
  {Conference} on {Knowledge} {Discovery} and {Data} {Mining}}, {KDD} '16, pp.\
   635--644, New York, NY, USA, August 2016. Association for Computing
  Machinery.
\newblock ISBN 978-1-4503-4232-2.
\newblock \doi{10.1145/2939672.2939736}.
\newblock URL \url{https://doi.org/10.1145/2939672.2939736}.

\bibitem[Wang et~al.(2019)Wang, Yao, Kifer, Graif, and
  Li]{WangNonStationaryModelCrime2019}
Hongjian Wang, Huaxiu Yao, Daniel Kifer, Corina Graif, and Zhenhui Li.
\newblock Non-{Stationary} {Model} for {Crime} {Rate} {Inference} {Using}
  {Modern} {Urban} {Data}.
\newblock \emph{IEEE Transactions on Big Data}, 5\penalty0 (2):\penalty0
  180--194, June 2019.
\newblock ISSN 2332-7790, 2372-2096.
\newblock \doi{10.1109/TBDATA.2017.2786405}.
\newblock URL \url{https://ieeexplore.ieee.org/document/8234616/}.

\bibitem[Wang \& Gerber(2015)Wang and Gerber]{WangUsingTwitterNextPlace2015}
Mingjun Wang and Matthew~S. Gerber.
\newblock Using {Twitter} for {Next}-{Place} {Prediction}, with an
  {Application} to {Crime} {Prediction}.
\newblock In \emph{2015 {IEEE} {Symposium} {Series} on {Computational}
  {Intelligence}}, pp.\  941--948, December 2015.
\newblock \doi{10.1109/SSCI.2015.138}.

\bibitem[Wang et~al.(2018)Wang, Phillips, Small, and Sampson]{wang2018}
Qi~Wang, Nolan~Edward Phillips, Mario~L Small, and Robert~J Sampson.
\newblock Urban mobility and neighborhood isolation in america’s 50 largest
  cities.
\newblock \emph{Proc. Natl Acad. Sci. USA}, 115:\penalty0 7735–7740, 2018.

\bibitem[Wang et~al.(2012)Wang, Gerber, and
  Brown]{WangAutomaticCrimePrediction2012a}
Xiaofeng Wang, Matthew~S. Gerber, and Donald~E. Brown.
\newblock Automatic {Crime} {Prediction} {Using} {Events} {Extracted} from
  {Twitter} {Posts}.
\newblock In Shanchieh~Jay Yang, Ariel~M. Greenberg, and Mica Endsley (eds.),
  \emph{Social {Computing}, {Behavioral} - {Cultural} {Modeling} and
  {Prediction}}, Lecture {Notes} in {Computer} {Science}, pp.\  231--238,
  Berlin, Heidelberg, 2012. Springer.
\newblock ISBN 978-3-642-29047-3.
\newblock \doi{10.1007/978-3-642-29047-3_28}.

\bibitem[Weber et~al.(2018)Weber, Seaman, Stewart, Bird, Tatem, McKee, Bhaduri,
  Moehl, and Reith]{weber2018census}
Eric~M Weber, Vincent~Y Seaman, Robert~N Stewart, Tomas~J Bird, Andrew~J Tatem,
  Jacob~J McKee, Budhendra~L Bhaduri, Jessica~J Moehl, and Andrew~E Reith.
\newblock Census-independent population mapping in northern nigeria.
\newblock \emph{Remote Sensing of Environment}, 204:\penalty0 786--798, 2018.

\bibitem[Weisburd(2015)]{WeisburdLawCrimeConcentration2015}
David Weisburd.
\newblock The {Law} of {Crime} {Concentration} and the {Criminology} of
  {Place}.
\newblock \emph{Criminology}, 53\penalty0 (2):\penalty0 133--157, May 2015.
\newblock ISSN 00111384.
\newblock \doi{10.1111/1745-9125.12070}.
\newblock URL \url{http://doi.wiley.com/10.1111/1745-9125.12070}.

\bibitem[Wesolowski et~al.(2012)Wesolowski, Eagle, Tatem, Smith, Noor, Snow,
  and Buckee]{wesolowski2012quantifying}
Amy Wesolowski, Nathan Eagle, Andrew~J Tatem, David~L Smith, Abdisalan~M Noor,
  Robert~W Snow, and Caroline~O Buckee.
\newblock Quantifying the impact of human mobility on malaria.
\newblock \emph{Science}, 338\penalty0 (6104):\penalty0 267--270, 2012.

\bibitem[Wesolowski et~al.(2016)Wesolowski, Buckee, Eng{\o}-Monsen, and
  Metcalf]{wesolowski2016connecting}
Amy Wesolowski, Caroline~O Buckee, Kenth Eng{\o}-Monsen, and Charlotte
  Jessica~Eland Metcalf.
\newblock Connecting mobility to infectious diseases: the promise and limits of
  mobile phone data.
\newblock \emph{The Journal of infectious diseases}, 214\penalty0
  (suppl\_4):\penalty0 S414--S420, 2016.

\bibitem[Wiedermann et~al.(2022)Wiedermann, Rose, Maier, Kolb, Hinrichs, and
  Brockmann]{wiedermann2022evidence}
Marc Wiedermann, Annika~H Rose, Benjamin~F Maier, Jakob~J Kolb, David Hinrichs,
  and Dirk Brockmann.
\newblock Evidence for positive long-and short-term effects of vaccinations
  against covid-19 in wearable sensor metrics--insights from the german corona
  data donation project.
\newblock \emph{arXiv preprint arXiv:2204.02846}, 2022.

\bibitem[Wilkinson \& Pickett(2006)Wilkinson and Pickett]{wilkinson2006}
Richard~G Wilkinson and Kate~E Pickett.
\newblock Income inequality and population health: A review and explanation of
  the evidence.
\newblock \emph{Social Science \& Medicine}, 62\penalty0 (7):\penalty0
  1768--1784, 2006.

\bibitem[Wo et~al.(2022)Wo, Rogers, Berg, and
  Koylu]{WoRecreatingHumanMobility2022}
James~C. Wo, Ethan~M. Rogers, Mark~T. Berg, and Caglar Koylu.
\newblock Recreating {Human} {Mobility} {Patterns} {Through} the {Lens} of
  {Social} {Media}: {Using} {Twitter} to {Model} the {Social} {Ecology} of
  {Crime}.
\newblock \emph{Crime \& Delinquency}, pp.\  00111287221106946, June 2022.
\newblock ISSN 0011-1287.
\newblock \doi{10.1177/00111287221106946}.
\newblock URL \url{https://doi.org/10.1177/00111287221106946}.
\newblock Publisher: SAGE Publications Inc.

\bibitem[Woolgar(1989)]{WoolgarWhynotsociology1989}
S.~Woolgar.
\newblock Why not a sociology of machines? {An} evaluation of prospects for an
  association between sociology and artificial intelligence.
\newblock In \emph{Intelligent {Systems} in a {Human} {Context}: {Development},
  {Implications}, and {Applications}}, pp.\  53--70. Oxford University Press,
  Inc., USA, January 1989.
\newblock ISBN 978-0-19-853736-6.

\bibitem[Xu et~al.(2019)Xu, Belyi, Santi, and Ratti]{xu2019quantifying}
Yang Xu, Alexander Belyi, Paolo Santi, and Carlo Ratti.
\newblock Quantifying segregation in an integrated urban physical-social space.
\newblock \emph{Journal of the Royal Society Interface}, 16\penalty0
  (160):\penalty0 20190536, 2019.

\bibitem[Yabe et~al.(2022)Yabe, Bueno, Dong, Pentland, and
  Moro]{yabe2022behavioral}
Takahiro Yabe, Bernardo Garcia~Bulle Bueno, Xiaowen Dong, AlexSandy' Pentland,
  and Esteban Moro.
\newblock Behavioral changes during the pandemic worsened income diversity of
  urban encounters.
\newblock \emph{arXiv preprint arXiv:2207.06895}, 2022.

\bibitem[Yan(2004)]{YanSeasonalityPropertyCrime2004}
Y.~Y. Yan.
\newblock Seasonality of {Property} {Crime} in {Hong} {Kong}.
\newblock \emph{British Journal of Criminology}, 44\penalty0 (2):\penalty0
  276--283, March 2004.
\newblock ISSN 0007-0955, 1464-3529.
\newblock \doi{10.1093/bjc/44.2.276}.
\newblock URL
  \url{https://academic.oup.com/bjc/article-lookup/doi/10.1093/bjc/44.2.276}.

\bibitem[Yang et~al.(2018)Yang, Heaney, Tonon, Wang, and
  Cudré-Mauroux]{YangCrimeTelescopecrimehotspot2018}
Dingqi Yang, Terence Heaney, Alberto Tonon, Leye Wang, and Philippe
  Cudré-Mauroux.
\newblock {CrimeTelescope}: crime hotspot prediction based on urban and social
  media data fusion.
\newblock \emph{World Wide Web}, 21\penalty0 (5):\penalty0 1323--1347,
  September 2018.
\newblock ISSN 1573-1413.
\newblock \doi{10.1007/s11280-017-0515-4}.
\newblock URL \url{https://doi.org/10.1007/s11280-017-0515-4}.

\bibitem[Ye et~al.(2015)Ye, Xu, Lee, Zhu, and Wu]{YeSpacetimeinteraction2015}
Xinyue Ye, Xiao Xu, Jay Lee, Xinyan Zhu, and Ling Wu.
\newblock Space–time interaction of residential burglaries in {Wuhan},
  {China}.
\newblock \emph{Applied Geography}, 60:\penalty0 210--216, June 2015.
\newblock ISSN 0143-6228.
\newblock \doi{10.1016/j.apgeog.2014.11.022}.
\newblock URL
  \url{https://www.sciencedirect.com/science/article/pii/S014362281400277X}.

\bibitem[Yeh et~al.(2020)Yeh, Perez, Driscoll, Azzari, Tang, Lobell, Ermon, and
  Burke]{yeh2020using}
Christopher Yeh, Anthony Perez, Anne Driscoll, George Azzari, Zhongyi Tang,
  David Lobell, Stefano Ermon, and Marshall Burke.
\newblock Using publicly available satellite imagery and deep learning to
  understand economic well-being in africa.
\newblock \emph{Nature communications}, 11\penalty0 (1):\penalty0 1--11, 2020.

\bibitem[Yuan et~al.(2013)Yuan, Nsoesie, Lv, Peng, Chunara, and
  Brownstein]{yuan2013monitoring}
Qingyu Yuan, Elaine~O Nsoesie, Benfu Lv, Geng Peng, Rumi Chunara, and John~S
  Brownstein.
\newblock Monitoring influenza epidemics in china with search query from baidu.
\newblock \emph{PloS one}, 8\penalty0 (5):\penalty0 e64323, 2013.

\bibitem[Zachreson et~al.(2018)Zachreson, Fair, Cliff, Harding, Piraveenan, and
  Prokopenko]{zachreson2018urbanization}
Cameron Zachreson, Kristopher~M Fair, Oliver~M Cliff, Nathan Harding, Mahendra
  Piraveenan, and Mikhail Prokopenko.
\newblock Urbanization affects peak timing, prevalence, and bimodality of
  influenza pandemics in australia: Results of a census-calibrated model.
\newblock \emph{Science advances}, 4\penalty0 (12):\penalty0 eaau5294, 2018.

\bibitem[Zhang et~al.(2015)Zhang, Gioannini, Paolotti, Perra, Perrotta,
  Quaggiotto, Tizzoni, and Vespignani]{zhang2015social}
Qian Zhang, Corrado Gioannini, Daniela Paolotti, Nicola Perra, Daniela
  Perrotta, Marco Quaggiotto, Michele Tizzoni, and Alessandro Vespignani.
\newblock Social data mining and seasonal influenza forecasts: the fluoutlook
  platform.
\newblock In \emph{Joint European Conference on Machine Learning and Knowledge
  Discovery in Databases}, pp.\  237--240. Springer, 2015.

\bibitem[Zhang et~al.(2017)Zhang, Perra, Perrotta, Tizzoni, Paolotti, and
  Vespignani]{zhang2017forecasting}
Qian Zhang, Nicola Perra, Daniela Perrotta, Michele Tizzoni, Daniela Paolotti,
  and Alessandro Vespignani.
\newblock Forecasting seasonal influenza fusing digital indicators and a
  mechanistic disease model.
\newblock In \emph{Proceedings of the 26th international conference on world
  wide web}, pp.\  311--319, 2017.

\bibitem[Zheng et~al.(2010)Zheng, Xie, Ma, et~al.]{zheng2010geolife}
Yu~Zheng, Xing Xie, Wei-Ying Ma, et~al.
\newblock Geolife: A collaborative social networking service among user,
  location and trajectory.
\newblock \emph{IEEE Data Eng. Bull.}, 33\penalty0 (2):\penalty0 32--39, 2010.

\end{thebibliography}
\bibliographystyle{iclr2021_conference}

\end{document}